\documentclass[12pt,a4paper]{article} 
 
\usepackage{amsfonts}
\usepackage{amsmath}
\usepackage{amssymb}
\usepackage{amsthm}
\usepackage{authblk}
\usepackage{bbm}
\usepackage{booktabs}
\usepackage[margin=5pt,font=small,labelfont=bf,labelsep=colon, skip=2pt, parskip=5pt]{caption}
\usepackage{epsfig}
\usepackage{enumerate}
\usepackage{float}
\usepackage{graphicx}
\usepackage{indentfirst}
\usepackage{latexsym}
\usepackage{lscape}
\usepackage{mathabx}
\usepackage{multirow}
\usepackage{multicol}
\usepackage{rotate}
\usepackage{subfig}
\usepackage{tabularx}
\usepackage{tikz}
\usepackage{url}
\usepackage{verbatim}
\usepackage{colortbl}

\usepackage[authoryear,round]{natbib}

\newlength{\defaultbaselineskip}
\definecolor{asparagus}{rgb}{0.55, 0.71, 0.0}

\begin{document}

\begin{titlepage}
\title{\bf Modeling and evaluating conditional quantile dynamics in  VaR forecasts}
\author{Fabrizio Cipollini$^{1}$, Giampiero M. Gallo$^{2,}$\footnote{Corresponding author: {\tt giampiero.gallo@nyu.edu}. The views expressed in the article are those of the authors and do not involve the responsibility of the Corte dei conti.} ~and Alessandro Palandri$^{1}$}
\affil{$^{1}$Dipartimento di Statistica, Informatica, Applicazioni (DiSIA)\\ {\it G. Parenti}, Universit\`a di Firenze}
\affil{$^{2}$Corte dei conti, New York University in Florence, and CRENoS}
\date{This Version: \today}
\end{titlepage}

\maketitle\thispagestyle{empty}

\renewcommand{\arraystretch}{1.5}
\setlength{\baselineskip}{6.5mm} 
\begin{abstract}
We focus on the time-varying modeling of VaR at a given coverage $\tau$, assessing whether the quantiles of the distribution of the returns standardized by their conditional means and standard deviations exhibit predictable dynamics. 
Models are evaluated via simulation, determining the merits of the asymmetric Mean Absolute Deviation as a loss function to rank forecast performances. The empirical application on the Fama--French 25 value--weighted portfolios with a moving forecast window shows  substantial improvements in forecasting conditional quantiles by keeping the predicted quantile unchanged unless the empirical frequency of violations falls outside a data-driven interval around $\tau$.
\end{abstract}
\vspace{2cm}
\textcolor{white}{space}

\pagebreak
\setcounter{page}{1}

\section{Introduction}\label{intro}

Value at Risk (VaR), which measures the portfolio loss that will be exceeded with probability $\tau$ over a given time period, is a well-known and solidly established measure for risk management. More formally, VaR forecasting is tantamount to forecasting the $\tau$-quantile of future portfolio returns. {As it is an unobserved object, a large literature exists about how it is measured and how it can be forecast one or more periods in advance, cf. \citet{Duffie:Pan:97}, \citet{Embrechts:Resnick:Samarodnitsky:99}, \citet{Embrechts:Kluppelberg:Mikosch:97}, and \citet{Jorion:07}.}

A first stream of approaches, say Historical Simulations (HS) and Extreme Value Theory (EVT), assumed that future values of the quantile at short horizons could be well represented by the recent past behavior of returns; as an example, the VaR in the HS  is calculated as the empirical quantile of the past distribution of returns over a rolling window of fixed size (a sort of realized measure).

On the wake of the success of GARCH, Stochastic Volatility and Realized Variance models in forecasting conditional variances, VaR approaches have evolved by adopting a Stepwise Distribution Modeling (SDM): the standard reference here is the book by \citet{Christoffersen:2012}. The idea is that a simple location-scale decomposition of the portfolio returns allows for the separate modeling of the conditional mean (e.g. ARMA), the conditional variance (e.g. GARCH) and the $\tau$-quantile of the returns {\it innovation} term thus obtained. {While existing SDM approaches have fully exploited the forecastability of the conditional variance as a major driver of quantile dynamics, they rely on procedures devised for the original returns to deliver the forecasts of the {\it innovation} quantiles (e.g. HS, EVT, fitting parametric distributions to all data values, etc).}

As an alternative to the SDM approach, \citet{Engle:Manganelli:2004} suggest for their CAViaR several dynamic specifications of the original return quantiles. In particular, the {\it Adaptive} parameterization may capture shifts in the quantiles not solely driven by time-varying variances, while other CAViaR specifications are driven by forcing terms carrying no autonomous quantile information. 

Modeling time--varying quantiles separately from volatility dynamics is not present in the literature.
In this paper, we adhere to the SDM framework, and we fill the existing gap by suggesting suitable dynamics for the quantiles of the innovations. Rather than referring to generic time--varying quantiles, in presenting our modeling approaches we focus on new objects of interest, making a distinction between \textit{unconditional}, \textit{conditional}  and \textit{actual} quantiles.
This allows us to highlight that, in general, the conditional expectation of the actual quantile and the conditional quantile do not coincide.\footnote{This is a major difference relative to what is found in the literature on volatility, where the conditional expectation of the squared-return (actual) equals the conditional variance.} Furthermore, our distinction of quantiles allows us to qualify both the existing and the specifications proposed here as either Direct- or Indirect-Dynamics approaches. Direct-Dynamics refers to all specifications where the conditional quantile adjustment depends on the past discrepancies between actual- and conditional-quantiles\footnote{{To visualize the logic behind Direct-Dynamics specifications, an example would be a GARCH for the conditional variance of zero--mean returns: $\sigma_{t}^{2}=\omega+(\alpha+\beta)\sigma_{t-1}^{2} +\alpha(r_{t-1}^{2} -\sigma_{t-1}^{2})$.}}; Indirect-Dynamics, first introduced in the {\it Adaptive} CAViaR, adjust the conditional quantiles depending on the differences between past nominal and empirical rejection frequencies. The value of such distinction is not a mere matter of taxonomy, since both our simulation study and our empirical analysis point to an overall superiority of forecasts generated by Indirect-Dynamics specifications. 

A problem arises when evaluating competing forecasts, namely the lack of a consistent measure of goodness--of--fit, in the sense of \cite{Hansen:Lunde:2005b} and \cite{Patton:2011}. We face the infeasibility of a root-mean-squared-error (RMSE) measuring the distance between forecasts and the unobservable underlying conditional quantiles for which no unbiased proxies exist; the results of  a simulation study show that, despite not possessing ideal theoretical properties, the MAD$_{\tau}$ induces  model rankings very similar to the consistent but infeasible RMSE.

The empirical application is conducted on the daily returns of the Fama--French 25 value-weighted portfolios from January 2010 to December 2020. The sample is split into six 5-year in-sample periods: model parameters are estimated in each period and out-of-sample forecasts are generated for the ensuing 1-year period. In--sample, contrary to conditional volatilities which exhibit strong persistence in every sub-period, we find that conditional quantiles alternate time-variation to constancy. Furthermore, in the case of the 1\% VaR's, filtered innovations are more prone to display \textit{anti-clustering} (perhaps as a result of GARCH filtering),  that is, the probability of two consecutive violations is lower than under independence. We model such a feature by allowing model dynamics to capture negative lag-1 correlations while imposing, when needed, suitable constraints to avoid unwanted explosive behavior.

Among the novel specifications introduced in the paper, a significant departure from standard autoregressive modeling is achieved by the \textit{Test Tracking} (TT),  where a friction is inserted in the updating mechanism by keeping existing quantile predictions, unless in contrast with the empirical evidence. Accordingly, as if guided by a test, TT adjusts the predictions only when the empirical frequency of the violations either falls below (respectively, exceeds) a lower (respectively, an upper) threshold. Empirically, we find this to be the only model specification to consistently improve upon constant VaR forecasts \citep[in terms of coverage and independence tests,  see][and of our MAD$_{\tau}$]{Christoffersen:2012}.

The structure of the paper is the following: in Section \ref{sec:newmod} we put forth the definitions of actual, conditional and unconditional quantiles, with a simple illustrative example in the context of a mixture of distributions. Section \ref{sec:ModelQDyn} sets up existing and new models within our Direct-- and Indirect--Dynamics categories with Section \ref{sec:estima} outlining the estimation strategy. In Section \ref{sec:QTrackEval}, the ability of the specifications in Section \ref{sec:ModelQDyn} to track conditional $\tau$--quantiles and the merits of MAD$_{\tau}$ as a measure of goodness--of--fit are investigated via a long Monte Carlo simulation. Section \ref{sec:EmpAnalysis} presents the results of the empirical application and Section \ref{sec:concl} concludes.

\section{Actual, Conditional and Unconditional Quantiles}
\label{sec:newmod}
The one-period conditional Value at Risk (VaR) of an investment is the percentage loss of the corresponding value that will be exceeded (as a negative value) with probability $\tau$:
\begin{equation*}
	\mathbb{P}_{t-1}\left(r_{t}<VaR_{t}\right)=\tau
\end{equation*}
where $\mathbb{P}_{t-1}(\cdot)$ is the probability of the event evaluated at $(t-1)$, $r_{t}$ the log-return\footnote{Recall that it is always possible to map the VaR of the log-returns to that of the corresponding net-returns from the monotonic transformation $\exp(VaR_{t})-1$.} and $VaR_{t}$ its $\tau$-quantile. 

The Stepwise Distribution Modeling (SDM) of VaR, followed in this paper, moves from the location-scale decomposition of the return $r_{t}=\mu_{t}+\sigma_{t}z_{t}$, where $\mu_{t}$ is the conditional mean, $\sigma_{t}$ the conditional volatility and $z_{t}$ the standardized innovation:
\begin{equation*}
	\mathbb{P}_{t-1}\left(\frac{r_{t}-\mu_{t}}{\sigma_{t}} < \frac{VaR_{t}-\mu_{t}}{\sigma_{t}}\right)=\tau \quad \Longleftrightarrow\quad \mathbb{P}_{t-1}\left(z_{t}<c_{t}\right)=\tau
\end{equation*}
where $c_{t}$ is the $\tau$-quantile of $z_{t}$. The step-by-step modeling of $\mu_{t}$, $\sigma_{t}$ and $c_{t}$ allows to reconstruct the VaR as $VaR_{t}=\mu_{t}+\sigma_{t}c_{t}$. Compared to other unconditional approaches, SDM allows to factor the forecastability of the conditional volatility (GARCH, Stochastic Volatility or Realized Variance models) into the VaR calculations. Furthermore, SDM allows to investigate whether $VaR_{t}$ movements may be explained by the return's first two conditional moments alone ($VaR_{t}=\mu_{t}+\sigma_{t}c$) or time-varying quantiles are also needed ($VaR_{t}=\mu_{t}+\sigma_{t}c_{t}$).

When the prediction of the quantiles is in question, it is advisable to recognize that there are differences between what influences the distribution of $z_t$ at time $t$ and what allows it to be predicted on the basis of what is known at time $t-1$. To that end, let us assume that the standardized innovation $z_{t}$ is the observable variable of interest, and let its cumulative distribution function $F(z_t|\omega_{t})$ depend on the realization of a generic vector of contemporary latent random factors $\omega_{t}$. As customary, then, we are interested in the possibility that some observable variables, $x_{t-1}$, belonging to the information set at time $t-1$, may be relevant in predicting the $\omega_{t}$ one-step ahead, given their role in determining the shape of the distribution of $z_t$: thus, the conditional probability density function $g(\omega_{t}|x_{t-1})$ is different from its unconditional pdf $f(\omega_{t})=\int g(\omega_{t}|x_{t-1})h(x_{t-1})dx_{t-1}$, where  $h(x_{t-1})$ is the pdf of $x_{t-1}$. 

It is therefore convenient to represent the conditional distribution $F(z_t|x_{t-1})$ of $z_t$ conditional on $x_{t-1}$,  as:
\begin{equation*}
F(z_t|x_{t-1}) = \int F(z_t|\omega_{t})g(\omega_{t}|x_{t-1})d\omega_{t} =  \mathbb{E}_{t-1}\left[F(z_t|\omega_{t})\right]
\end{equation*}
where, from now on, $\mathbb{E}_{t-1}$ indicates taking expectations conditional on $x_{t-1}$. To complete the definitions, the unconditional distribution $F(z_t)$ is given by:
\begin{eqnarray*}
F(z_t) &=& \int F(z_t|x_{t-1}) h(x_{t-1})dx_{t-1} = \mathbb{E}\left[F(z_t|x_{t-1})\right]\\ 
&=& \int F(z_t|\omega_t)f(\omega_{t})d\omega_{t} = \mathbb{E}\left[F(z_t|\omega_t)\right]
\end{eqnarray*}
In what follows, we will call $F(z_t|\omega_{t})$ the \textit{actual} distribution, $F(z_t|x_{t-1})$ the \textit{conditional} distribution, and $F(z_t)$ the \textit{unconditional} distribution of the variable of interest $z_t$. 

As an illustrative example of the above setup, let us consider a case where the $\omega_t$ are  non--negative random weights $\left\{\omega_{i,t}\right\}_{i=1}^{N}$ (summing to one at each $t$) for a mixture of distributions  $\left\{P_{i}(z_t)\right\}_{i=1}^{N}$. Then:
\begin{eqnarray}\label{eq:mixture}
 F(z_t|\omega_t)  &=& \sum_{i=1}^{N}\omega_{i,t}P_{i}(z_t),\\
 F(z_t|x_{t-1}) &=& \sum_{i=1}^{N}\mathbb{E}_{t-1}\left[\omega_{i,t}\right]P_{i}(z_t),\\ 
 F(z_t) &=& \sum_{i=1}^{N}\mathbb{E}\left[\omega_{i,t}\right]P_{i}(z_t)
\end{eqnarray}
In the literature on conditional heteroskedasticity, similar concepts are familiar: given $r_{t}=\sigma_{t}z_{t}$, with $z_{t}\sim(0,1)$,  the {\it actual} squared-return is  $\sigma_{t}^{2}z_{t}^{2}$, the {\it conditional} squared-return is $\mathbb{E}_{t-1}[r_{t}^2]=\sigma_{t}^{2}$ and $\mathbb{E}[r_{t}^2]=\sigma^{2}$ is the {\it unconditional} squared-return. Furthermore, assuming that $\sigma_{t}^{2}$ follows a GARCH(1,1) process, it follows that $\mathbb{E}_{t-1}[r_{t}^2]=\mathbb{E}[r_{t}^2|x_{t-1}]$ where $x_{t-1}'=(r_{t-1}^2,\sigma_{t-1}^{2})$.

As it is relevant in what follows, we extend the definitions to the corresponding actual, conditional and unconditional quantiles, as: 
\begin{itemize}

	\item the unobservable \textit{actual} $\tau$-quantile $q_{t}$ satisfies $F(q_{t}|\omega_t)=\tau$; 
	\item the \textit{conditional} $\tau$-quantile $c_{t}$ satisfies $F(c_{t}|x_{t-1})=\tau$, and it is the main object of interest for time $t-1$ predictions; 
	\item the \textit{unconditional} $\tau$-quantile $c$ satisfies $F(c)=\tau$.
\end{itemize}

Given the same $\tau$, the actual quantile $q_{t}(\tau) \equiv F^{-1}(\tau|\omega_t)$ may be written in terms of $c_t(\tau)\equiv F^{-1}(\tau|x_{t-1})$:
$$
q_{t} = c_{t} + \left[F^{-1}(\tau|\omega_t)-F^{-1}(\tau|x_{t-1})\right].
$$ 
Although $\mathbb{E}_{t-1}\left[F(z_t|\omega_t)\right]= F(z_t|x_{t-1})$, in general  $\mathbb{E}_{t-1}\left[F^{-1}(\tau|\omega_t)\right]\ne F^{-1}(\tau|x_{t-1})$, so that when 
forming expectations of the actual quantile $q_t$ conditional on $x_{t-1}$,  a bias $b_{t}$ term for $c_t$ arises, that is,
\begin{equation}\label{eq:real_bias}
\mathbb{E}_{t-1}\left[q_{t}\right] = c_{t} + b_{t}
\end{equation}
where $b_{t}=\mathbb{E}_{t-1}\left[F^{-1}(\tau|\omega_t)\right]-F^{-1}(\tau|x_{t-1})$.

\begin{figure}[ht]
\captionsetup{singlelinecheck=off}
\caption[ac]{
Mixture of distributions example (cf. Appendix \ref{example}).  (\textbf{a}) the actual pdf  $f(z_t|\omega_{t}=0.2)$; (\textbf{b}) the conditional pdf $f(z_t|x_{t-1}=1)$; (\textbf{c}) the unconditional pdf $f(z_t)$.  Blue and red vertical lines identify the $5\%$ and $1\%$ quantiles, respectively.}\vspace{-0.25cm}
\label{fig:figure00}
\begin{center}
\subfloat[Actual $f(z_t|\omega_{t}=0.2)$]{\includegraphics[width=0.32\textwidth]{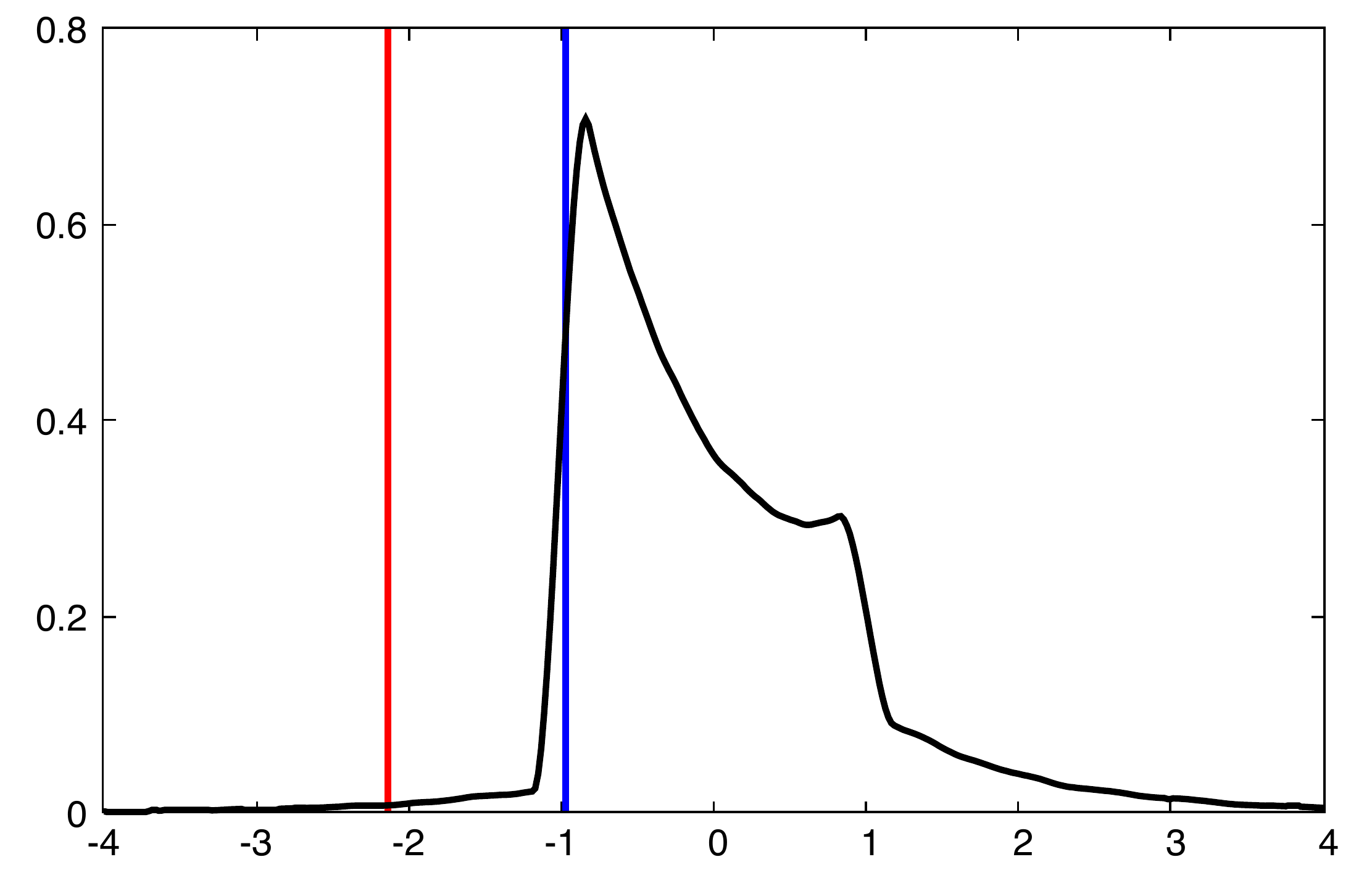}}
\subfloat[Condit. $f(z_t|x_{t-1}=1)$]{\includegraphics[width=0.32\textwidth]{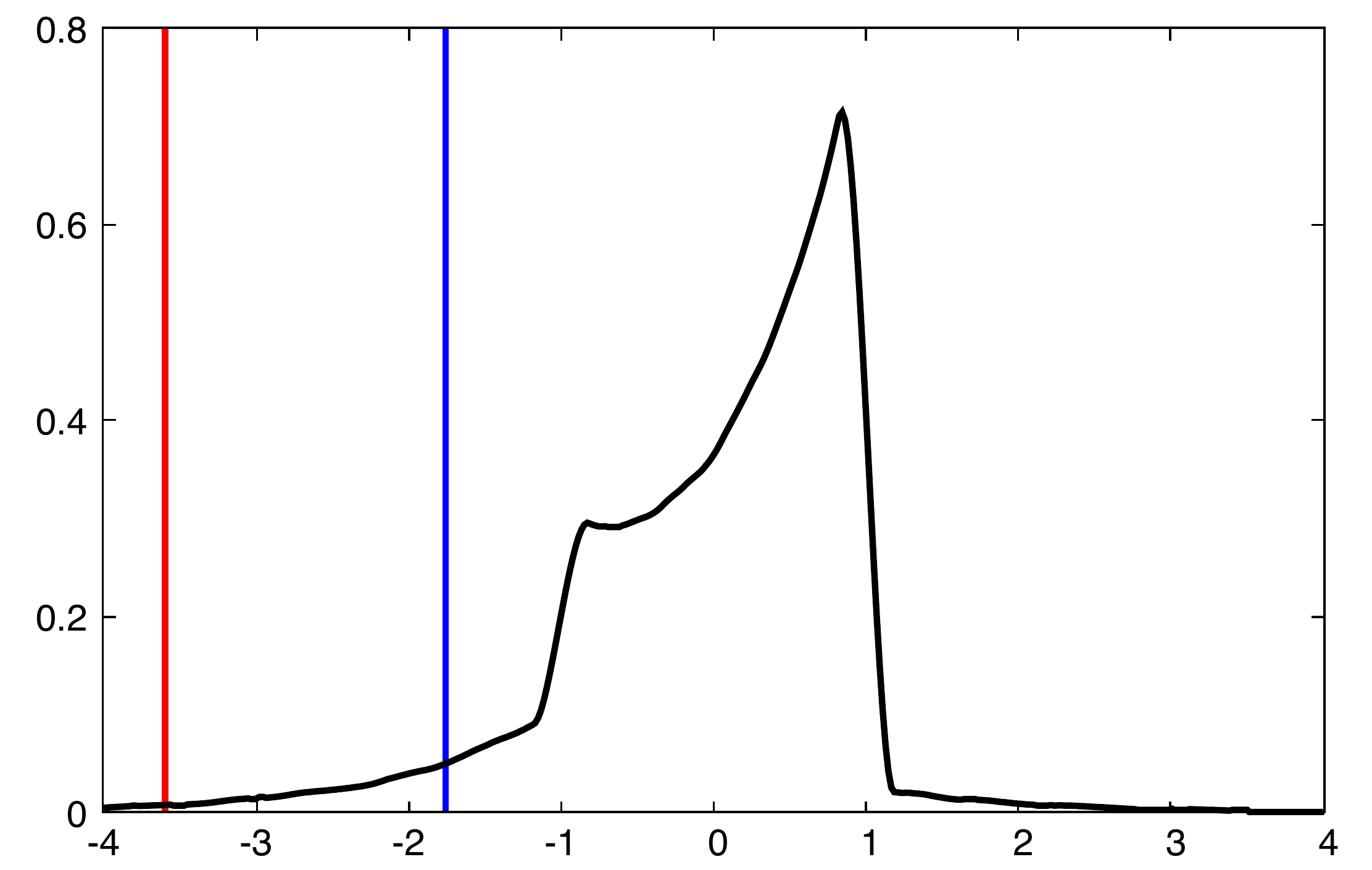}}
\subfloat[Unconditional $f(z_t)$]{\includegraphics[width=0.32\textwidth]{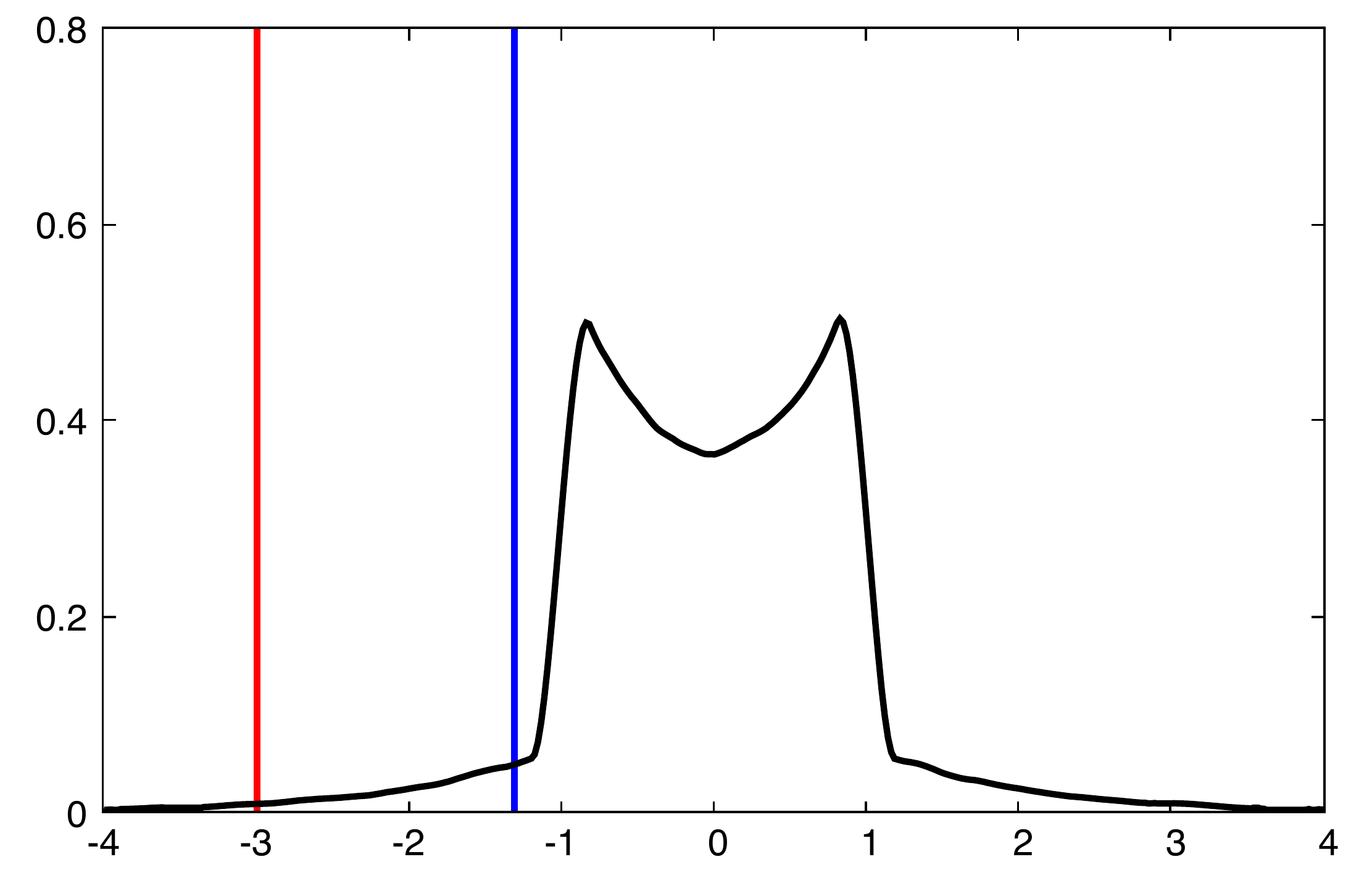}}
\end{center}
\end{figure}

For the mixture example, with details available in Appendix \ref{example}, a graphical appraisal of the differences arising among the probability density functions derived from the distributions, namely, $f(z_t|\omega_t)$, $f(z_t|x_{t-1})$, and $f(z_t)$, with the corresponding quantiles is provided in Figure \ref{fig:figure00}.\footnote{In the case examined, the easiest assessment of the magnitude of the bias in Equation (\ref{eq:real_bias}) is when closed-form solutions to the quantiles of interest exist, i.e.,  in the chosen design, when $\tau<0.009$: conditional on $x_{t-1}=1$, $\mathbb{E}_{t-1}[q_{t}(\tau)]=1+1.2\ln2+\ln5+\ln\tau$, $c_{t}(\tau)=1+\ln5+\ln\tau$ and $b_{t}(\tau)=1.2\ln2$ (in other cases, not always a constant). When $\tau=0.005$, we thus have $\mathbb{E}_{t-1}[q_{t}(0.005)]\approx-1.8571$, $c_{t}(0.005)\approx-2.6889$ and $b_{t}(0.005)\approx0.8318$.}

\section{Modeling Quantile Dynamics }
\label{sec:ModelQDyn}

In order to specify the dynamics of $c_t$, in the literature one encounters methods which we classify as \textit{Direct-Dynamics} models, in that they provide autoregressive schemes involving past $q_t$'s. Given their unobservability, they resort to some form of proxy from past observable time series. A separate strain falls under the category of \textit{Indirect-Dynamics}, where specifications exploit the information that comes from past violations of the conditional quantiles and amounts to providing an updating rule over current prediction. In what follows, we provide a taxonomy of existing methods within the framework we suggested and advance some novel specifications in both classes of dynamics.

\subsection{Direct-Dynamics}\label{ddyn}
Direct-Dynamics consist of a specification where the conditional quantiles $c_{t}$ depend on lagged actual quantiles $\{q_{t-i}\}_{i=1}^{\infty}$. Since the actual quantiles are not directly observable, a feasible direct-dynamics specification of $c_{t}$ requires the lagged $q_{t}$'s to be estimated or proxied, as done by the popular historical simulations \citep{Christoffersen:2012},  from the time series of past observable variables.

\subsubsection{Historical Simulations}

The (baseline) Historical Simulation \citep[HS; among others, cf.][]{Hendricks:1996} estimates $q_{t-1}$  over a window of the past $N$ standardized observations $\{z_{t-n}\}_{n=1}^{N}$. Linear interpolation is used in the estimate, $\widehat{q}_{t-1}$, whenever the empirical quantile falls between two observations.\footnote{Specifically,
\begin{equation*}
\widehat{q}_{t-1} = z^{(l)}+(\tau N -\lfloor\tau N\rfloor)(z^{(h)}-z^{(l)})
\end{equation*}
where $\lfloor~\rfloor$ is the floor function, $l=\lfloor\tau N\rfloor+1$, $h=\lfloor\tau N\rfloor+2$ and $z^{(i)}$ is the $i$-th of the $\{z_{t+1-n}\}_{n=1}^{N}$ sorted in ascending order.} As with other non-parametric methods, $N$ regulates the trade--off between the variance and the bias in $\widehat{q}_{t-1}$ by giving more or less weight to the most recent and presumably most informative observations. Generally adopted window lengths are $N=250$ and $N=1000$, leading to the HS250 and HS1000 estimates, respectively, and we adhere to this choice.

The Weighted Historical Simulation \citep[WHS;][]{Bodoukh:Richardson:Whitelaw:1998}  is an extension which assigns relatively more weight to the most recent observations.\footnote{In particular, given $\lambda\in(0,1)$, the weight $\lambda_{r}$ associated to $z_{r}$, with $r\in[t-N,t-1]$, is given by $\lambda_{r} = (1-\lambda^{N})^{-1}(1-\lambda)\lambda^{t-1-r}$. The observations $z_{r}$ are then sorted in ascending order along with their weights $\lambda_{r}$. From the resulting sort, let $x$ define the estimated probabilities $\widehat{p}_{l}$ and $\widehat{p}_{h}$ according to $\widehat{p}_{l}=\sum_{i=1}^{x}\lambda^{(i)}\le\tau$ and $\widehat{p}_{h}=\sum_{i=1}^{x+1}\lambda^{(i)}>\tau$. Letting $z^{(l)}$ and $z^{(h)}$ be the sorted observations associated to the probabilities $\widehat{p}_{l}$ and $\widehat{p}_{h}$, respectively, the WHS estimated quantile is given by:
\begin{equation*}
\widehat{q}_{t-1} = z^{(l)}+\frac{\tau-p_{l}}{p_{h}-p_{l}}\cdot(z^{(h)}-z^{(l)})
\end{equation*}
Although weighting the observations by $\lambda_{r}$ makes the choice of $N$ less crucial, WHS neither specifies $N$ nor the parameter $\lambda$.} In our implementation of WHS, we make use of all available data by setting $N=t-1$ while for $\lambda$ we adopt the standard values of $0.95$ and $0.99$ in WHS95 and WHS99, respectively.

When forecasting the conditional quantile $c_{t}$, the usual implementation of these approaches is based on $\widehat{c}_{t}=\widehat{q}_{t-1}$, which stems from the following two assumptions: first, the actual quantiles follow a random walk, so  that $\mathbb{E}_{t-1}\left[q_{t}\right]=q_{t-1}$; second, the bias $b_{t}$ in Equation (\ref{eq:real_bias}) is zero ($c_{t}=q_{t-1}$).

\subsubsection{Generalized Autoregressive Conditional Quantiles}
Inspired by the successful stationary GARCH parameterizations, a class of Generalized Autoregressive Conditional Quantiles (GARCQ) can be devised for the conditional quantiles $c_{t}$. In particular, the GARCQ(1,1) dynamics is given by:
\begin{equation}\label{eq:garcq}
c_{t} = \omega + \alpha q_{t-1} + \beta c_{t-1}
\end{equation}
with $\alpha\ge 0$, $\beta\ge 0$ and $(\alpha+\beta)<1$.\footnote{Although the empirical analysis will focus on the GARCQ(1,1), it is straightforward to consider additional lags or altogether different parameterizations. For example, a log-GARCH(1,1) analogue \citep{Geweke:1986} could be parameterized as $\ln|c_{t}|=\omega+\alpha\ln|q_{t-1}|+\beta\ln|c_{t-1}|$ and $c_{t}=-|c_{t}|$ with $\alpha\ge0$, $\beta\ge 0$ and $(\alpha+\beta)<1$. } To make it operational, $q_{t-1}$ is chosen as the $\widehat{q}_{t-1}$ in WHS99. 

\subsection{Indirect-Dynamics}\label{idyn}
Within the class of Indirect-Dynamics we comprise specifications of conditional quantiles $c_{t}$ that depend on lagged empirical rejection frequencies. The idea behind these modeling approaches is to update $c_{t}$ by {\it adjusting} the previous  $c_{t-1}$ on the basis the discrepancy between the nominal $\tau$ and the empirical frequency $\widehat p_{t-1}$ of quantile violations accrued up to time $(t-1)$. 

The approach is derived from the fact that  actual quantiles $q_{t-1}$ and probabilities $F(c_{t-1}|\omega_{t-1})$ can be substituted for one another. As a matter of fact, assuming strictly positive differentiability of $F(c_{t-1}|\omega_{t-1})$, we can resort to the mean-value theorem to express
$ F(q_{t-1}|\omega_{t-1})=F(c_{t-1}|\omega_{t-1})+F'(q_{t-1}^{*}|\omega_{t-1})\cdot(q_{t-1}-c_{t-1}),$ 
where $q_{t-1}^{*}$ is a convex combination of $q_{t-1}$ and $c_{t-1}$.\footnote{By the definitions of quantiles, $F'(q_{t-1}^{*}|\omega_{t-1})>0$.} Defining $\theta_{t-1}=\left[F'(q_{t-1}^{*}|\omega_{t-1})\right]^{-1}$ and rearranging terms yields:
\begin{equation}\label{eq:di-ind}
q_{t-1}-c_{t-1} = \theta_{t-1}\left[\tau-F(c_{t-1}|\omega_{t-1})\right].
\end{equation}
Therefore, while Direct-Dynamics relies on the LHS of Equation (\ref{eq:di-ind}), with the need to choose a window of observations to make it operational (as seen before), Indirect--Dynamics uses $F(c_{t-1}|\omega_{t-1})$, with two main advantages: first, it depends on $c_{t-1}$ which is model generated and readily available; second, an unbiased proxy of $F(c_{t-1}|\omega_{t-1})$ exists in the form of the indicator function $d_{t-1} = \mathbbm{1}_{[z_{t-1}<c_{t-1}]}$.\footnote{Recall that $\mathbb{E}_{t-2} \left[\mathbbm{1}_{[z_{t-1}<c_{t-1}]} \right] = F(c_{t-1}|x_{t-2})\equiv \tau.$}

\subsubsection{CAViaR}
Indirect-Dynamics specification was popularized by the Conditional Autoregressive VaR (CAViaR) of \citet{Engle:Manganelli:2004} where an autoregressive dynamics of the conditional quantile $c_{t}$ is assumed. In the Adaptive CAViaR we have:
\begin{equation}\label{eq:caviar}
c_{t} = c_{t-1} - \alpha\left\{\left[1+\exp\left(G\left[z_{t-1}-c_{t-1}\right]\right)\right]^{-1}-\tau\right\},
\end{equation}
with $\alpha>0$. As a proxy of $F(c_{t-1}|\omega_{t-1})$, the CAViaR employs a logistic approximation (with $G$ set equal to $10$) to the indicator function $d_{t-1} = \mathbbm{1}_{[z_{t-1}<c_{t-1}]}$, with the main advantage to be continuous with respect to $c_{t-1}$, which somewhat simplifies the estimation procedure. On the downside, the logistic approximation introduces a bias in the probability of exceeding the quantile. Specifically, since $c_{t-1}$ is such that $F(c_{t-1}|x_{t-2})=\tau$, it follows that $\mathbb{E}_{t-2}\left\{\left[1+\exp\left(G\left[z_{t-1}-c_{t-1}\right]\right)\right]^{-1}\right\}\ne\tau$, with the magnitude of the bias being inversely related to $G$.

Although this CAViaR parameterization was originally specified to model the quantiles of {\it non-standardized} observations\footnote{In \citet{Engle:Manganelli:2004} there are other CAViaR specifications: the Symmetric Absolute Value, the Asymmetric Slope and the Indirect GARCH(1,1) are parameterizations capturing changes in the quantiles that are exclusively driven by time-varying conditional variances. In particular, the conditional expectations in $(t-1)$ of the forcing terms in $t$ are: $\sigma_{t}\mathbb{E}_{t-1}|z_{t}|$, $\sigma_{t}\mathbb{E}_{t-1}[z_{t}\mathbbm{1}_{[z_{t}\lessgtr 0]}]$ and $\sigma_{t}^{2}$ for the Symmetric Absolute Value, Asymmetric Slope and Indirect GARCH(1,1), respectively. Given that the factors multiplying the functions of the conditional variance are generally uninformative about the $\tau$-quantile of $z_{t}$, the CAViaR specifications employing such forcing terms should be regarded as GARCH processes rescaled by a constant value of the $z_{t}$ quantile of interest, and thus will not be considered in what follows. Simulations studies, not presented here, have confirmed that these parameterizations cannot track time-varying quantiles of $z_{t}$.}, the specification of Equation (\ref{eq:caviar}) may equally well be applied to modeling the quantiles of the standardized innovations $z_{t}$. Such CAViaR specification is neither mean-reverting ($c_{t-1}$ has a unit coefficient) nor does it provide a value for the unconditional quantile. 

\subsubsection{QPI: Quantile-Probability Indicator}
Switching from actual quantiles to empirical frequencies by substituting Equation (\ref{eq:di-ind}) in the GARCQ(1,1) of Equation (\ref{eq:garcq}) and reparameterizing, leads to a specification with a time-varying parameter that maps probabilities to quantiles:
\begin{equation*}
c_{t} = \omega + \alpha_{t-1}\left[\tau-F(c_{t-1}|\omega_{t-1})\right]+\beta c_{t-1}
\end{equation*}
Setting $\alpha_{t-1}=\alpha$ and replacing the unobservable $F(c_{t-1}|\omega_{t-1})$ with the unbiased proxy $d_{t-1}=\mathbbm{1}_{[z_{t-1}<c_{t-1}]}$, gives the QPI specification:
\begin{equation}\label{eq:ind}
c_{t} = \omega + \alpha\left(\tau-d_{t-1}\right) + \beta c_{t-1}
\end{equation}
with $0<\beta<1$ and $\alpha>0$. 
Since $|\sum_{j=1}^{q}\alpha_{j}\left(\tau-d_{t-1-j}\right)|$ is bounded by $\sum_{j=1}^{q}|\alpha_{j}|$, which does not depend on $\left\{c_{t-1-j}\right\}_{j=1}^{q}$, stationarity conditions for the general $(p,q)$ specification coincide with those of the autoregressive components $\sum_{i=i}^{p}\beta_{i} c_{t-1-i}$.

\subsubsection{Violations Tracking}
Specifications with lagged quantile violations as forcing terms tend to induce oscillations in the predicted quantiles $c_{t}$: by analogy with the field of digital signal processing \citep[e.g.][]{Horowitz:Hill:2015}, we will refer to them as \textit{parasitic oscillations}. Such unappealing feature can be easily discussed in the context of the QPI model of Equation (\ref{eq:ind}). In particular, consider the case in which the true quantiles are locally constant at $c$ for $t\in[T_{1},T_{2}]$ and the filtered quantile at $T_{1}$ is $c_{T_{1}}=c$. It follows that, on average, there would be $(\tau^{-1}-1)$ consecutive periods in which $\mathbbm{1}_{[z_{t-1}<c_{t-1}]}=0$. Over such periods, the quantile dynamics reduce to:
\begin{equation*}
c_{t} = \omega +\alpha\tau + \beta c_{t-1} 
\end{equation*}
which implies the geometric reversion of $c_{t}$ to $(1-\beta)^{-1}(\omega+\alpha\tau)$ up until the occurrence of a violation.\footnote{Despite the additional smoothing provided by the logistic function, it may be shown that parasitic oscillations are also present in the CAViaR's predicted quantiles.}

We propose a Violations Tracking procedure aimed at reducing parasitic oscillations induced by the low frequency with which, under normal circumstances, predicted quantiles are violated. To begin with, let us consider the exponentially-weighted estimator $\hat{p}_{t-1}$ of the frequency with which given quantiles $\left\{c_{t-1-i}\right\}_{i=0}^{\infty}$ are violated:
\begin{equation}\label{eq:expsmoo}
\hat{p}_{t-1} = (1-\lambda)\sum_{i=0}^{\infty}\lambda^{i}d_{t-1-i} = \lambda\hat{p}_{t-2} + (1-\lambda)d_{t-1}
\end{equation}
with $\lambda\in(0,1)$. The idea behind this approach is to adjust the quantile $c_{t}$ only when the discrepancy between the violation frequency $\hat{p}_{t-1}$ and the nominal $\tau$ should be attributed to real movements in the underlying true quantile, rather than being a random fluctuation. 

The information content of $\hat{p}_{t-1}$ is exploited in the following two conditional quantiles parameterizations.

\paragraph{TT: Test Tracking}\label{tt-sec}
This approach consists of adjusting the quantile $c_{t}$ only when, for a nominal $\tau$, the observed violation frequency $\hat{p}_{t-1}$ of Equation (\ref{eq:expsmoo}) exceeds a range $[\theta_{l},\theta_{h}]$ of tolerable fluctuation. Loosely speaking, falling below $\theta_{l}$ means that the chosen quantile does not produce enough violations for the given nominal level $\tau$ and hence signals a need to increase it (\textit{low side}); by contrast, exceeding $\theta_{h}$ means that too many violations are present and hence a decrease of the quantile is in order (\textit{high side}). An intrinsic asymmetry is present, due to the nature of the problem: a long string of no violations has a different meaning than a few consecutive violations; for example, for $\tau=1\%$, ninety-nine consecutive zero violations is likely, whereas two consecutive ones are less so. 

To make this procedure operational, we treat the thresholds $\theta_{l}$ and $\theta_{h}$ as parameters and we suggest two functions for the adjustment on the low side, $\phi_l\in(0,1]$, respectively, the high side $\phi_h\in[1,+\infty)$. In principle, we could think that the size of the adjustment may depend on lagged observables at time $t-1$ i.e. $\underline{c_{t-1}}'=(c_{t-1},c_{t-2},\ldots)$, $\underline{z_{t-1}}'=(z_{t-1},z_{t-2},\ldots)$,  resulting in $\phi_{l}(\underline{c_{t-1}},\underline{z_{t-1}};\beta)$ and
$\phi_{h}(\underline{c_{t-1}},\underline{z_{t-1}};\beta)$, with a vector of parameters $\beta$. However, to reduce the number of parameters to be estimated, we formally set a baseline specification in general terms:
\begin{equation}
	c_{t} = \left\{
	\begin{array}{rcl}
		\phi_l(c_{t-1},z_{t-1},\beta_{l}) \, c_{t-1} & \text{ if } &  \hat{p}_{t-1} < \theta_{l}  \\ 
		c_{t-1} & \text{ if } & \widehat{p}_{t-1}\in[\theta_{l},\theta_{h}] \label{eq:TT} \\ 
		\phi_h(c_{t-1},z_{t-1},\beta_{h}) \, c_{t-1} & \text{ if } &  \hat{p}_{t-1} > \theta_{h} 
	\end{array}
	\right.
\end{equation}
where, specifically, we simply set $\phi_l(c_{t-1},z_{t-1},\beta_{l})=\beta_{l}$ and $\phi_h(c_{t-1},z_{t-1},\beta_{h})=\beta_{h}$, with $0<\beta_l\leq 1\leq \beta_h$. 
This implies that when $\hat{p}_{t-1} < \theta_{l}$ the quantile is increased by $\beta_{l}$, and, analogously, when $\hat{p}_{t-1} > \theta_{h}$ the quantile is decreased by $\beta_{h}$.
This specification can  be compactly rewritten as
\begin{equation}
\label{eq:TTsimp}
c_{t}=c_{t-1}\left[1+d_{t-1}^{l}(\beta_{l}-1)+d_{t-1}^{h}(\beta_{h}-1)\right]
\end{equation}
where, in general,
\begin{equation}
  \label{eq:dl-dh}
  d_{t}^{l}=\mathbbm{1}\left[\hat{p}_{t}<\theta_{l}\right] \qquad d_{t}^{h}=\mathbbm{1}\left[\hat{p}_{t}>\theta_{h}\right].
\end{equation}

One way to further motivate the approach stems from the consideration that the tolerance region $[\theta_{l},\theta_{h}]$ is closely related to the acceptance region of a test of the null hypothesis $H_{0}:\mathbb{E}[\hat{p}_{t-1}|\underline{c_{t-1}}]=\tau$ versus $H_{1}:\mathbb{E}[\hat{p}_{t-1}|\underline{c_{t-1}}]\ne\tau$. This falls within the class of \textit{unconditional coverage} tests suggested by \citet{Kupiec:1995} \citep[cf. Ch.13.2 in][]{Christoffersen:2012} with the peculiarity here that the exponential smoothing derivation of  $\hat{p}_{t-1}$ complicates the determination of the functional form of the distribution of the test statistic.\footnote{Notice that each of the $z$'s in $\widehat{p}_{t-1}$ is sampled from a potentially different distribution and the exponential smoothing in Equation (\ref{eq:expsmoo}) does not allow to appeal to a Central Limit Theorem.} Since there is a one-to-one correspondence between the critical values of the test and $\theta_{l}$ and $\theta_{h}$, working with the latter allows to treat the issue within an estimation framework.

\paragraph{MT: Multiplicative Tracking}
The Test Tracking just examined provides the dynamics of not directly observable time-varying quantities (similarly to what is done for conditional variance models), specified in terms of five parameters in its baseline specification: $\lambda$ in Equation (\ref{eq:expsmoo}), the thresholds $\theta_{l}$, $\theta_{l}$ and the adjustment parameters $\beta_{l}$, $\beta_{h}$. To hedge against the possibility of weakly identified parameters giving rise to multiple local extrema in estimation, noisy filtered values of the quantiles and poor forecasts, we suggest an alternative updating criterion of $c_{t}<0,\ \forall t$, doing away with the dependence on the parameters $\theta_{l}$ and $\theta_{h}$ in Equation (\ref{eq:TT}) and the logic behind the thresholds.  To that end we  propose a \textit{Multiplicative Tracking} specification:
\begin{equation}\label{eq:one}
c_{t} = \left[1+\alpha\ln\left(\frac{1+\hat{p}_{t-1}}{1+\tau}\right)\right]\cdot c_{t-1},
\end{equation}
where $0\leq \alpha< [\ln(1+\tau)]^{-1} $; the choice of the term in parentheses ensures the desired direction of adjustment, that is, $c_t < c_{t-1}$ {when} $\hat p_{t-1} > \tau $, and vice versa.

\section{Estimation}
\label{sec:estima}
Parameters of the quantile specifications in Section \ref{sec:ModelQDyn} are estimated from the minimization of the asymmetric Mean Absolute Deviation loss:
\begin{equation}\label{eq:ec}
\text{MAD}_{\tau}=\frac{1}{T}\sum_{t=1}^{T}\left(z_{t}-c_{t}\right)\left(\tau-\mathbbm{1}_{[z_{t}<c_{t}]}\right).
\end{equation}
Introduced by \citet{Koenker:Bassett:1978}, quantile regressions have been extended to time-series models by \citet{Bloomfield:Steiger:1983} and to autoregressions by \citet{Koul:Saleh:1995} and \citet{Koenker:Zhao:1996}. Asymptotic theory for nonlinear models with independent innovations has been derived by \citet{Oberhofer:1982} and \citet{Powell:1983}, among others, while consistency and asymptotic normality results in the case of nonlinear dynamic models are due to \citet{Weiss:1991} and \citet{Mukherjee:1999}. Building on such results, \citet{Engle:Manganelli:2004} provide conditions for the consistency and asymptotic normality of the parameter estimates of dynamic quantile models.

The discontinuities in (\ref{eq:ec}) render derivative--based optimization of the estimation criterion infeasible: in what follows, we resort to adaptive simulated annealing \citep[proposed by][]{Ingber:1993, Ingber:2000} to minimize the MAD$_{\tau}$ loss function in (\ref{eq:ec}), using several random starting values  to avoid local extrema.\footnote{In general, the adopted annealing schedule proposes candidate moves from a random-walk of the model's parameters with Gaussian increments of variance $\delta_{n}$. The evaluation of $\kappa$ candidate moves (per parameter) is carried out at the constant temperature $T_{n}$ and is followed by the assessment of the acceptance ratio $a_{n}$. In particular, if $a_{n}\ge\phi a_{n-1}$, where $\phi\in(0,1)$, the algorithm loops back, unaltered, to evaluate the successive $\kappa$ candidate moves. Instead, if $a_{n}<\phi a_{n-1}$, both the temperature and the increments' variance are adjusted: $T_{n+1}=\lambda_{1}T_{n}$ and $\delta_{n+1}=\lambda_{2}\delta_{n}$, where $\lambda_{1}\in(0,1)$ and $\lambda_{2}\in(0,1)$. In the estimation of dynamic quantile models, we set $\delta_{0}=1$, $\kappa=50$, $\phi=0.5$, $\lambda_{1}=\lambda_{2}=0.95$. The convergence criterion (stopping of the algorithm) is specified for the increments' variance as $\delta_{\min}=10^{-7}$. Thus, the initial temperature $T_{0}$ is selected case by case so that when $\delta_{n}<\delta_{\min}$ the temperature $T_{n}\gg 0$ is still well above zero.}

In view of our experience with the empirical application, discussed below, it must be said that estimation of dynamic quantile specifications is very challenging due to the step-function nature of the estimation criterion. While the adopted simulated annealing does bypass the differentiability issues of other optimizers, it too is impeded in its search by the numerous flat spots on the criterion's surface. In the end, the large number of necessary random starting values of the search algorithm blurred the boundary between simulated annealing and a fine random grid-search.

\section{Quantile Tracking Evaluation}
\label{sec:QTrackEval}
\begin{figure}[h]
\captionsetup{singlelinecheck=off}
\caption[ac]{
For given $\tau=10\%, 5\%, 1\%$, typical realizations of $z_{s}$ simulated from the data generating process ($5,000$ draws) defined by Equations (\ref{eq:zs}) and (\ref{eq:sine}), together with the corresponding true $\tau$ quantiles, $c_s$ (in blue). Different scales on the vertical axis.}\vspace{-0.25cm}
\label{fig:figure01}
\begin{center}
	\centering
\subfloat[$\tau=10\%$]{\includegraphics[width=0.32\textwidth]{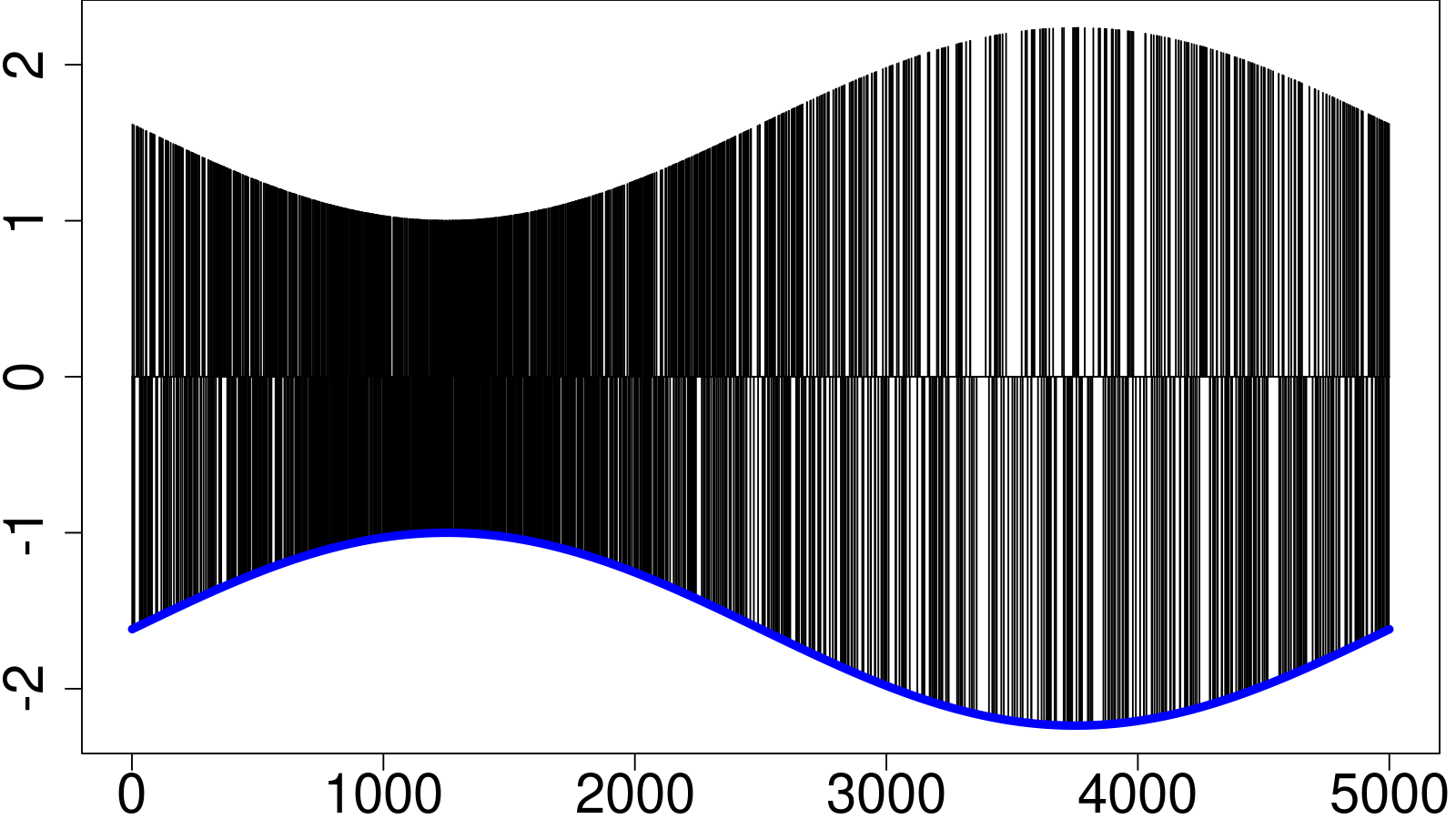}}
\hspace{0.1cm}
\subfloat[$\tau=5\%$]{\includegraphics[width=0.32\textwidth]{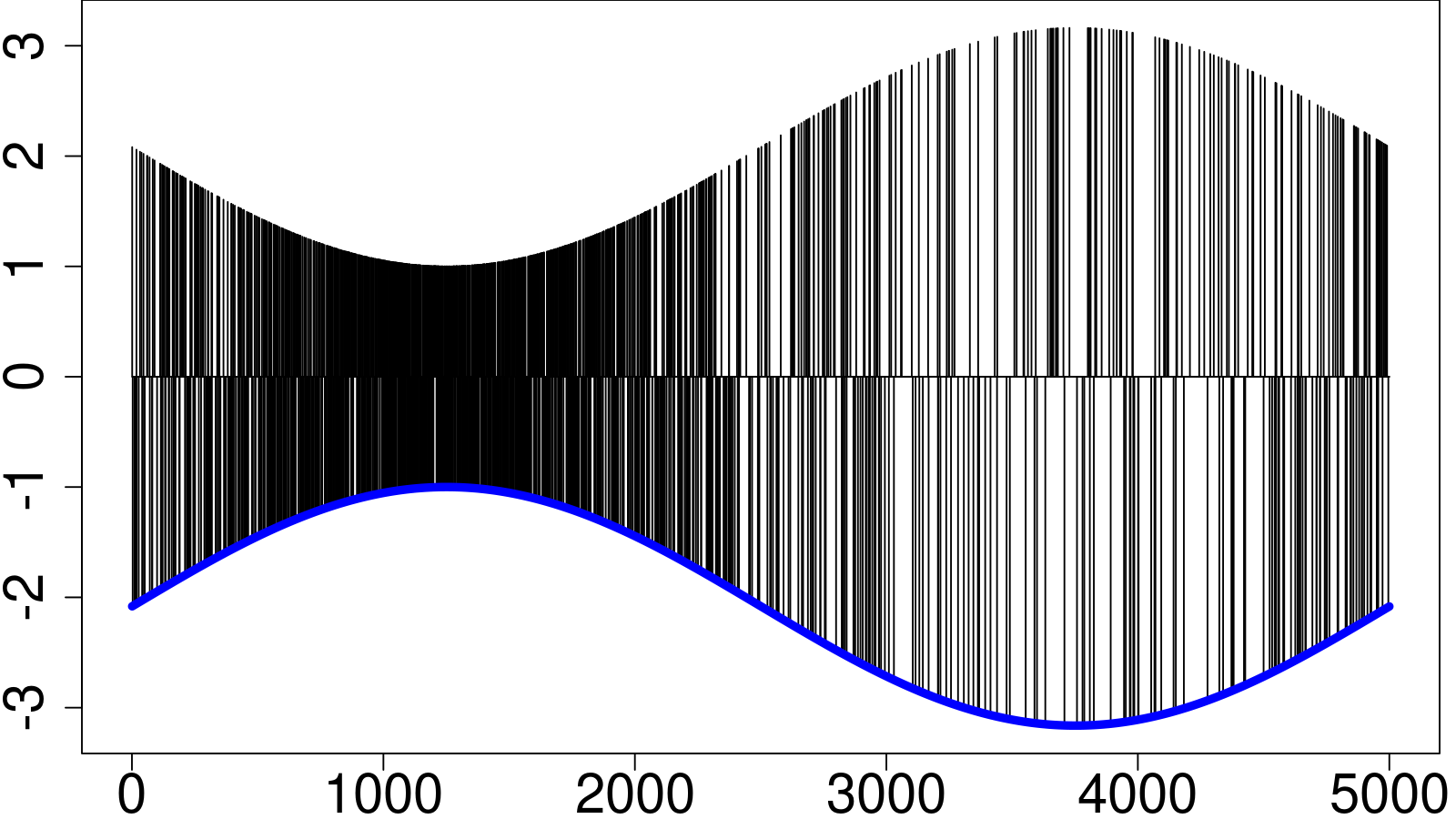}}
\hspace{0.1cm}
\subfloat[$\tau=1\%$]{\includegraphics[width=0.32\textwidth]{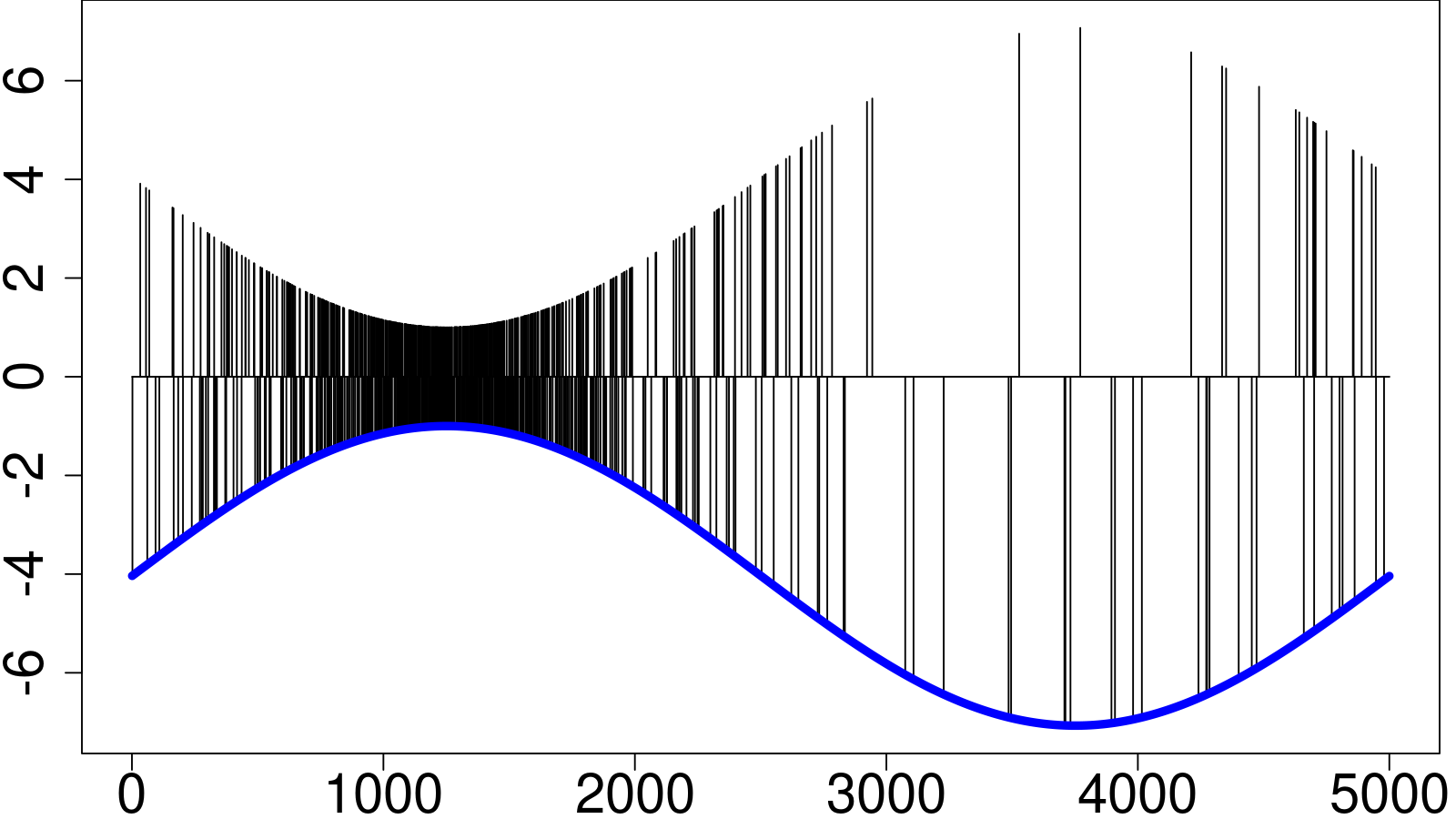}}
\end{center}
\end{figure}

The ability of the specifications in Section \ref{sec:ModelQDyn} to track time--varying $\tau$--quantiles is investigated via a long Monte Carlo simulation (full details of the data generating process are presented in Appendix \ref{appdx_setup}) of the standardized innovations $\{z_{s}\}_{s=1}^{S}$, with $S=10^{7}$. 
Figure \ref{fig:figure01} displays three subsets of $5,000$ draws for $z_{s}$, together with the underlying $\tau$-quantiles $c_s$, left--to--right corresponding to $\tau=\{10\%, 5\%, 1\%\}$ (in blue -- one cycle of the sine function each). 
Notice that, at each time $s$, the variable $z_{s}$ can take only three values.
Moreover, what may appear as clustering of the $z_{s}$ in Figure \ref{fig:figure01} does not correspond to a time-varying volatility since $\mathbb{V}(z_{s})=1$, $\forall s$.\footnote{This is confirmed empirically, as the simulated $z_{s}^{2}$ do not display significant autocorrelations nor do GARCH specifications estimated on $z_{s}$ exhibit statistically significant parameters beyond the intercept.}

\begin{figure}[h]
	\captionsetup{singlelinecheck=off}
	\caption[ac]{
			Time series of true (in blue) and predicted (in black) quantiles for all Direct- and Indirect-Dynamics approaches considered when $\tau=5\%$.}\vspace{-0.25cm}
	\label{fig:sim-fit-005}
	\begin{center}
		\subfloat[HS250]{\includegraphics[width=0.328\textwidth]{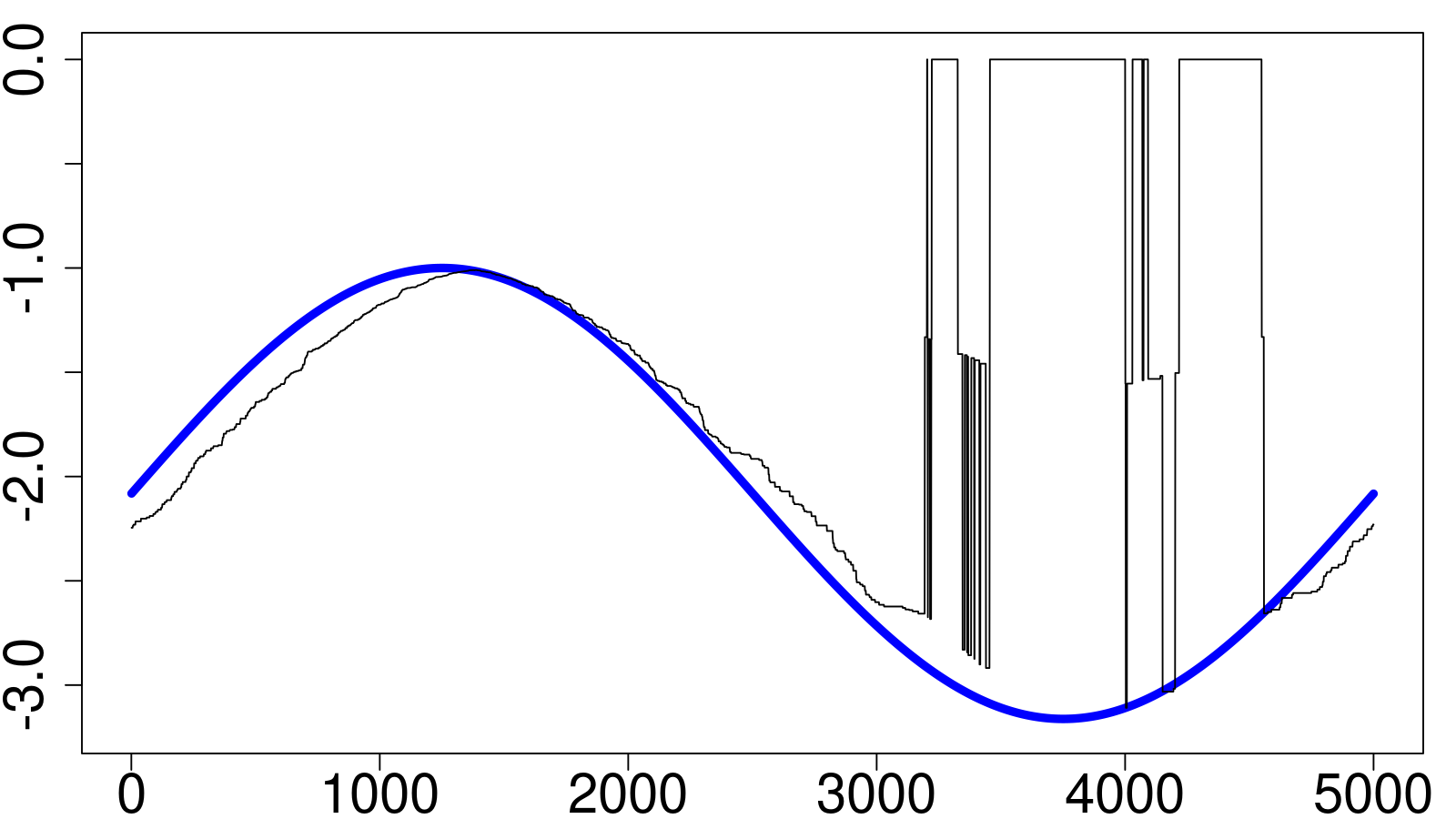}}
		\subfloat[HS1000]{\includegraphics[width=0.328\textwidth]{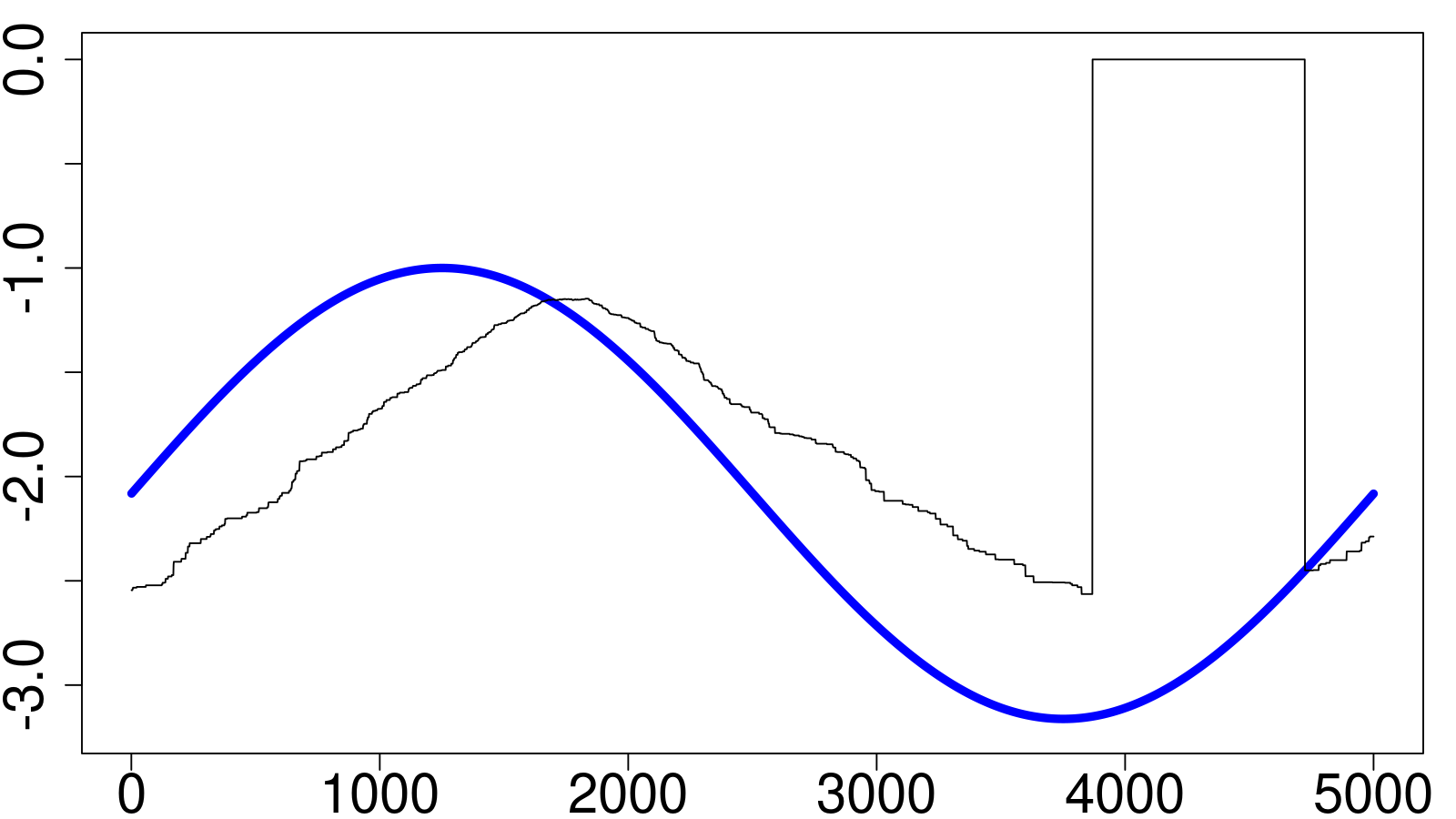}}
		\subfloat[WHS95]{\includegraphics[width=0.328\textwidth]{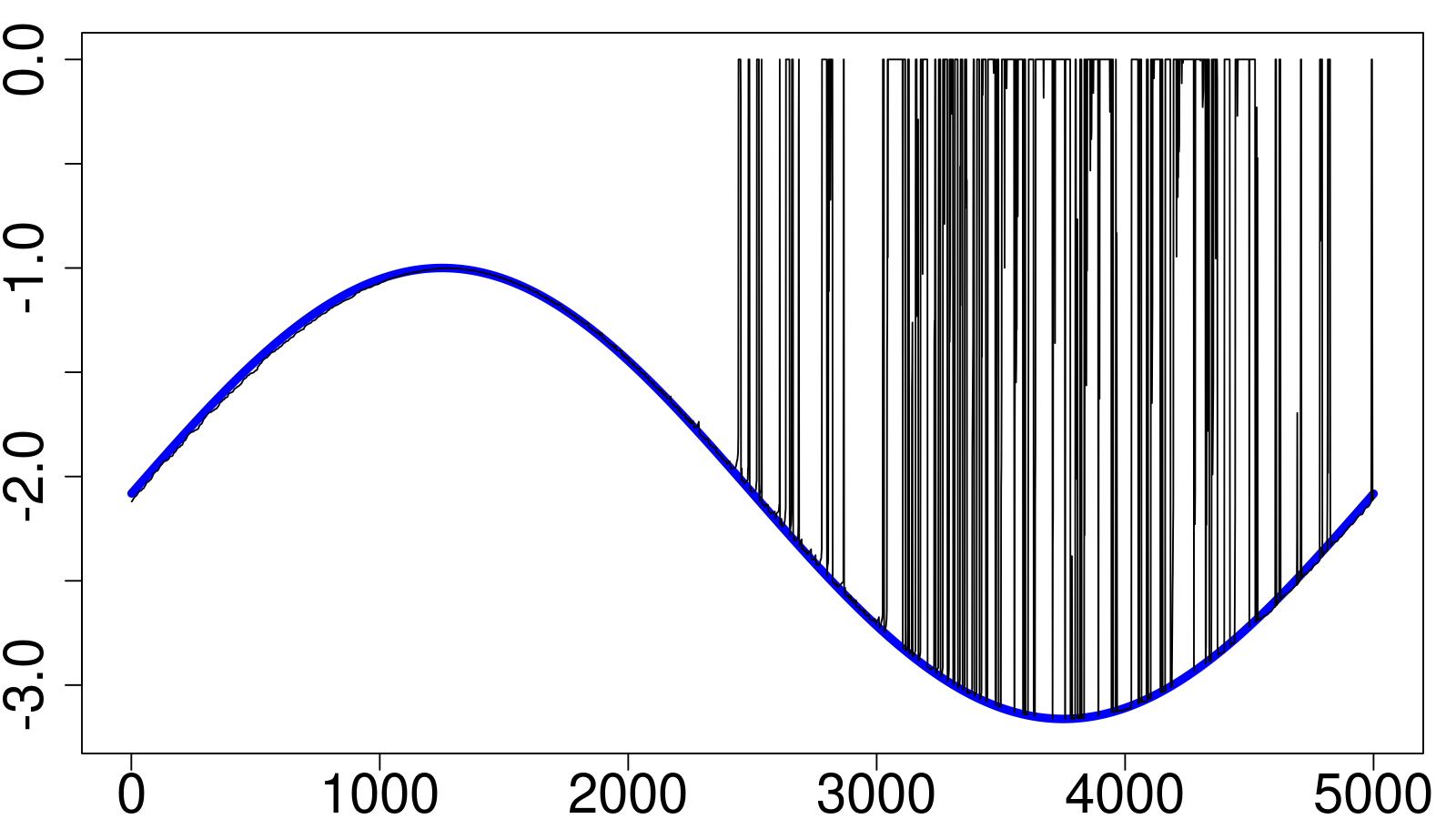}}\\
		\subfloat[WHS99]{\includegraphics[width=0.328\textwidth]{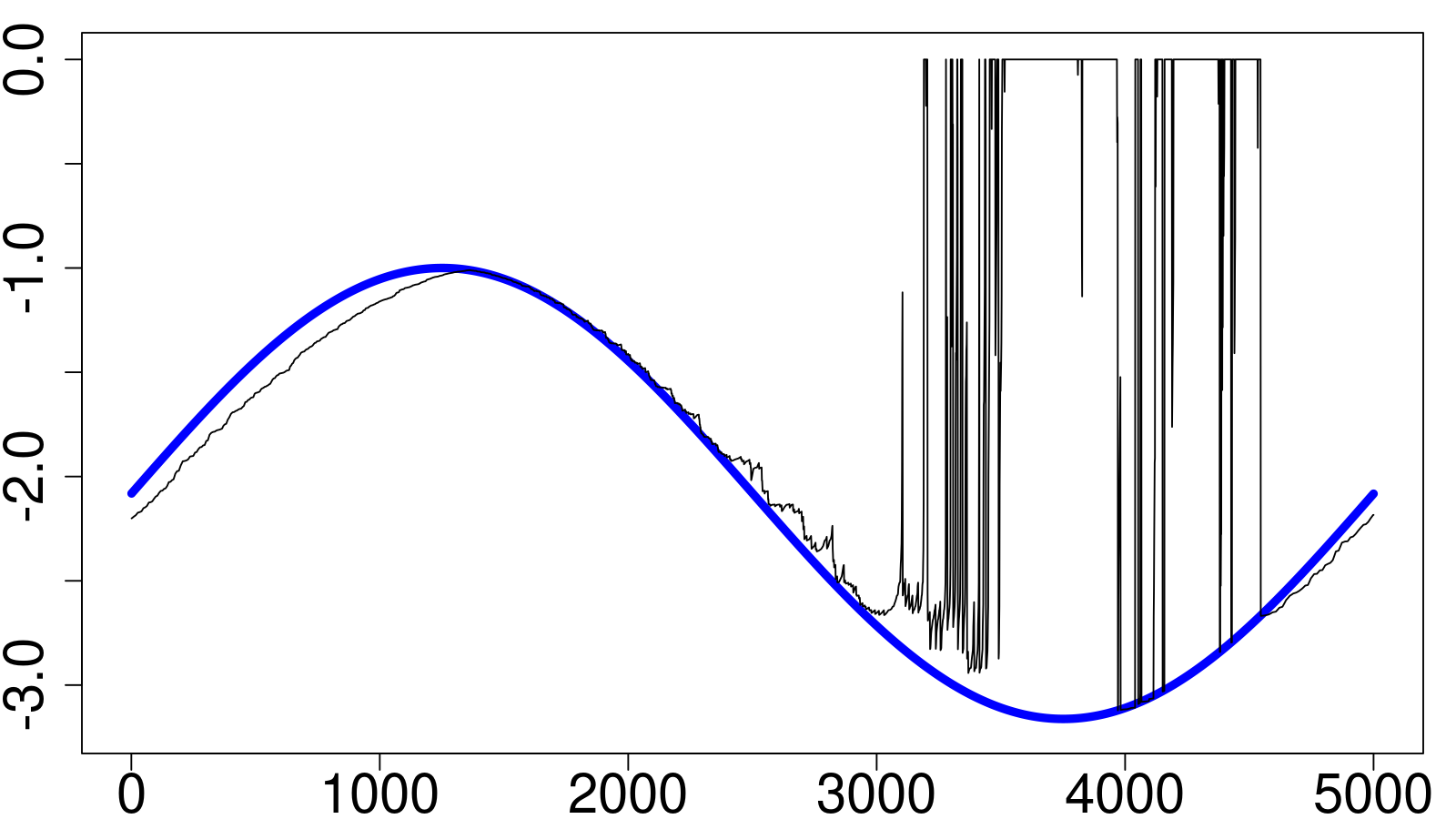}}
		\subfloat[GARCQ]{\includegraphics[width=0.328\textwidth]{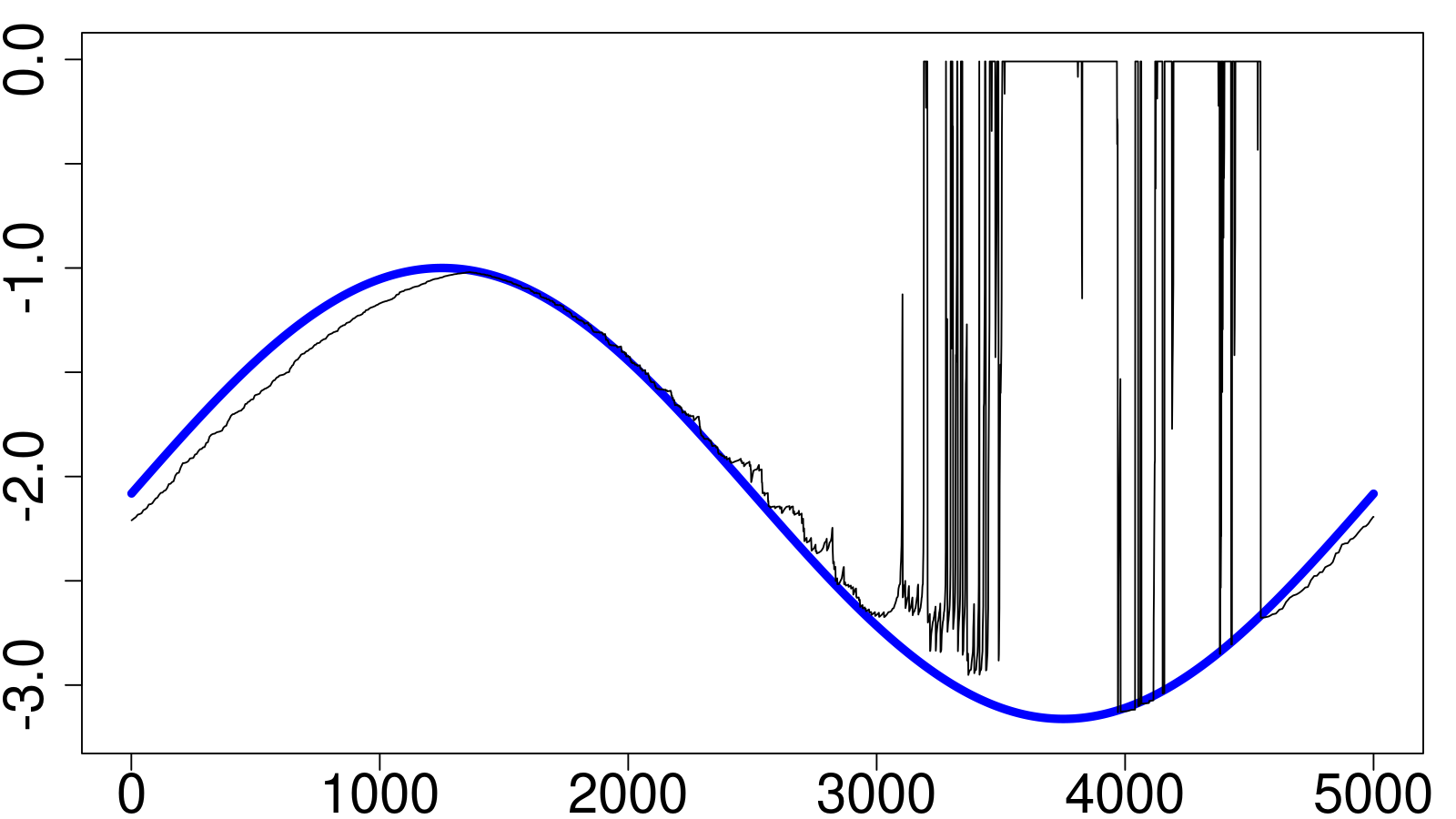}}
		\subfloat[CAViaR]{\includegraphics[width=0.328\textwidth]{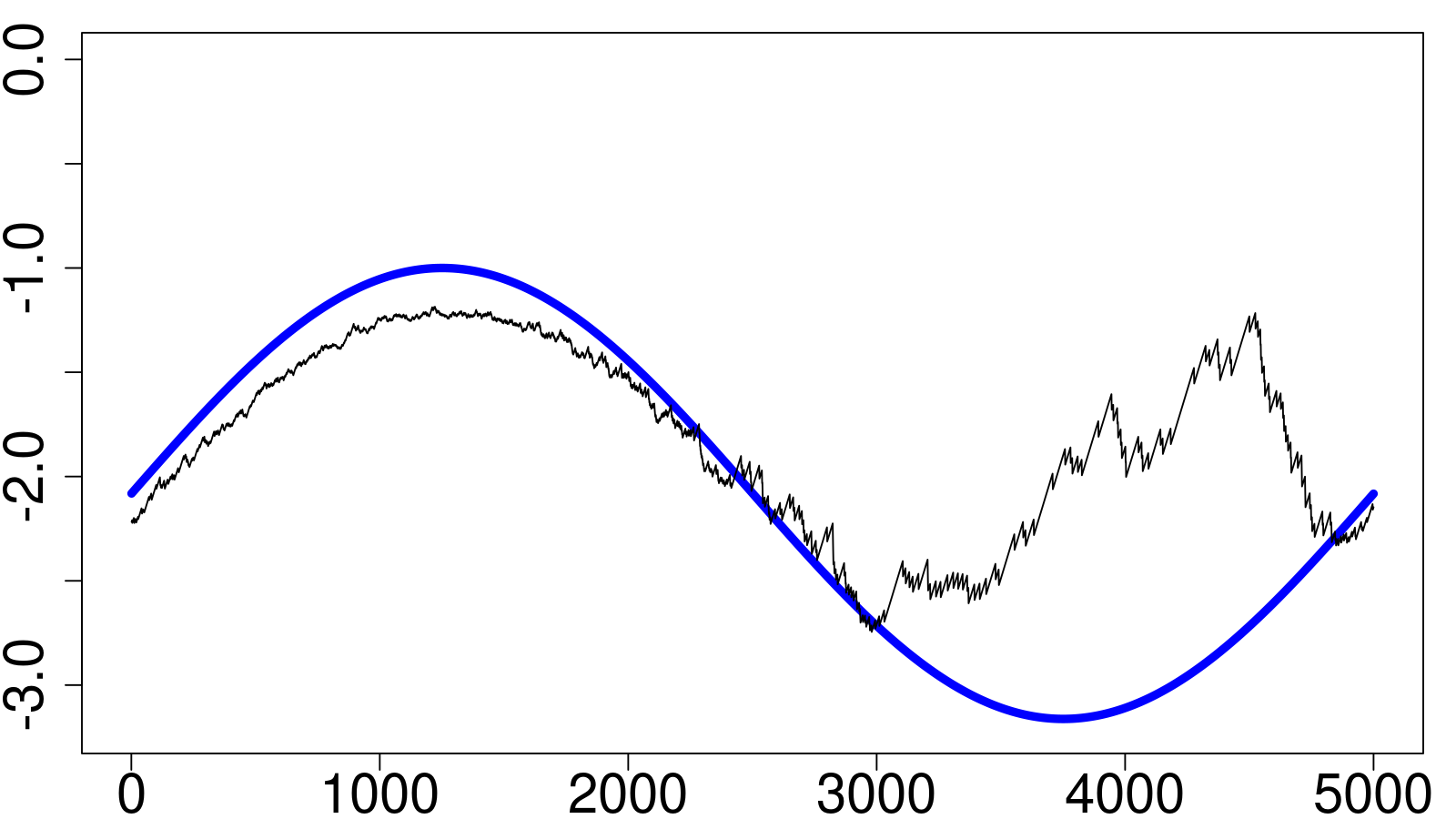}}\\
		\subfloat[QPI]{\includegraphics[width=0.328\textwidth]{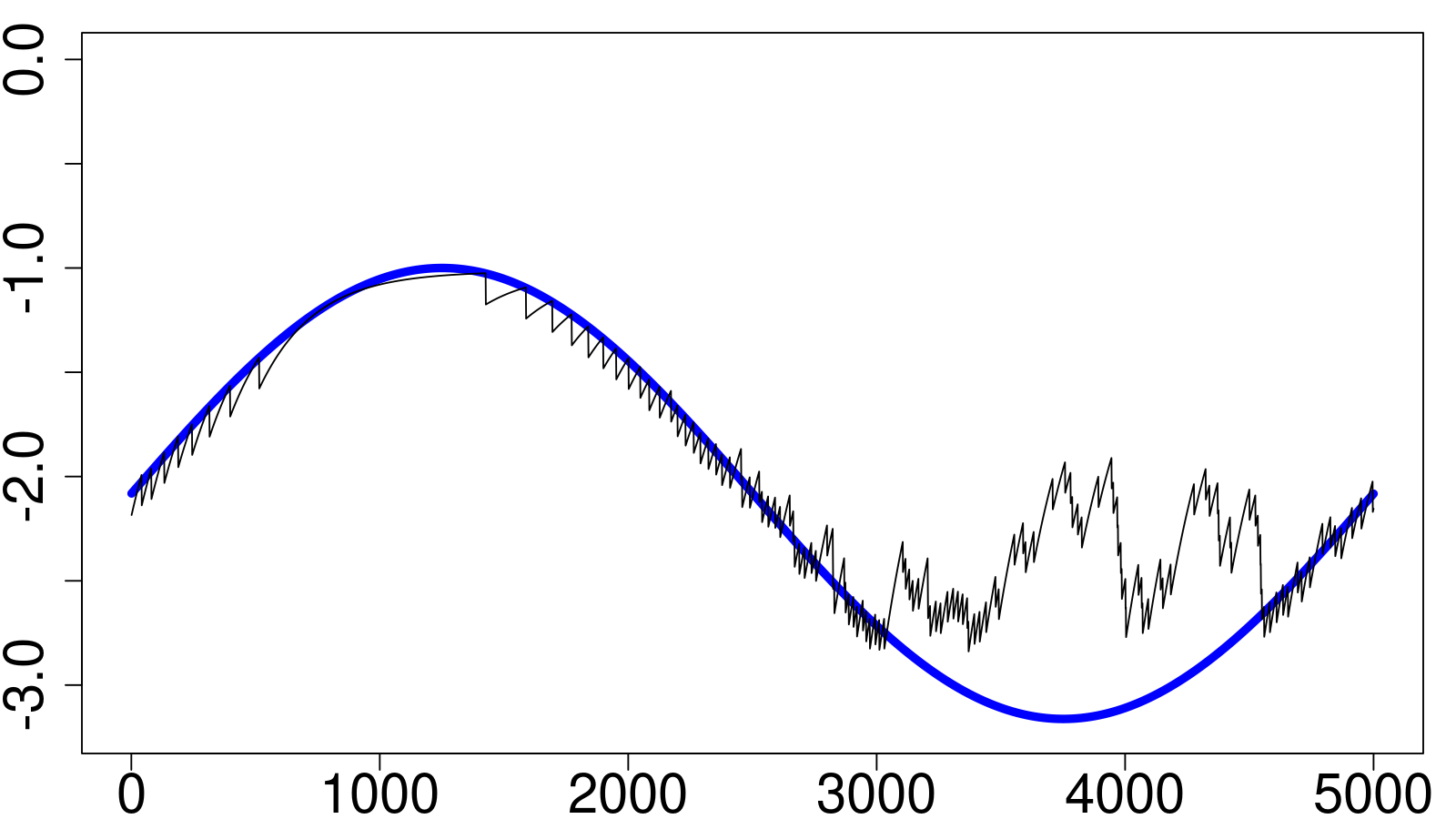}}
		\subfloat[TT]{\includegraphics[width=0.328\textwidth]{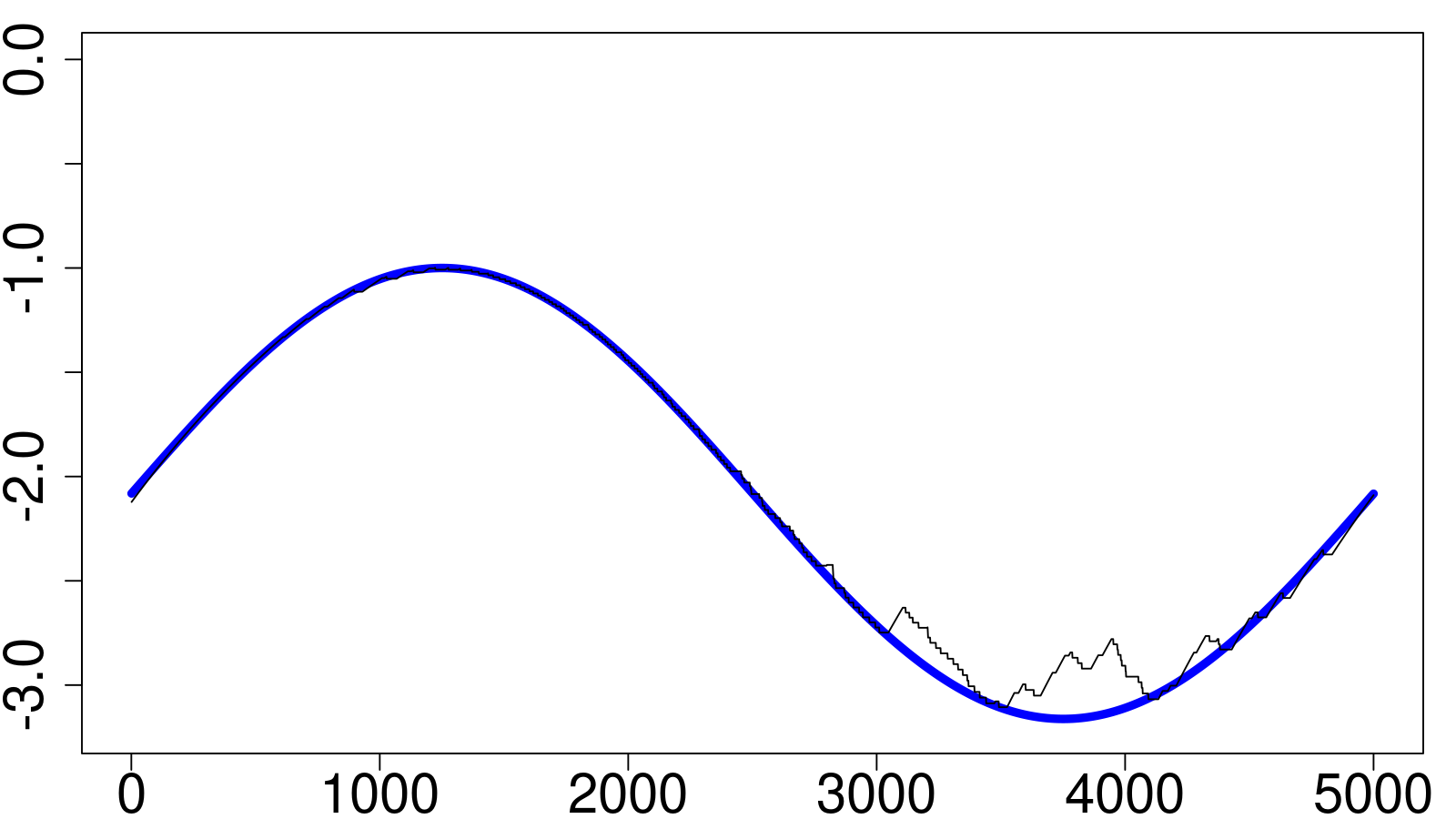}}
		\subfloat[MT]{\includegraphics[width=0.328\textwidth]{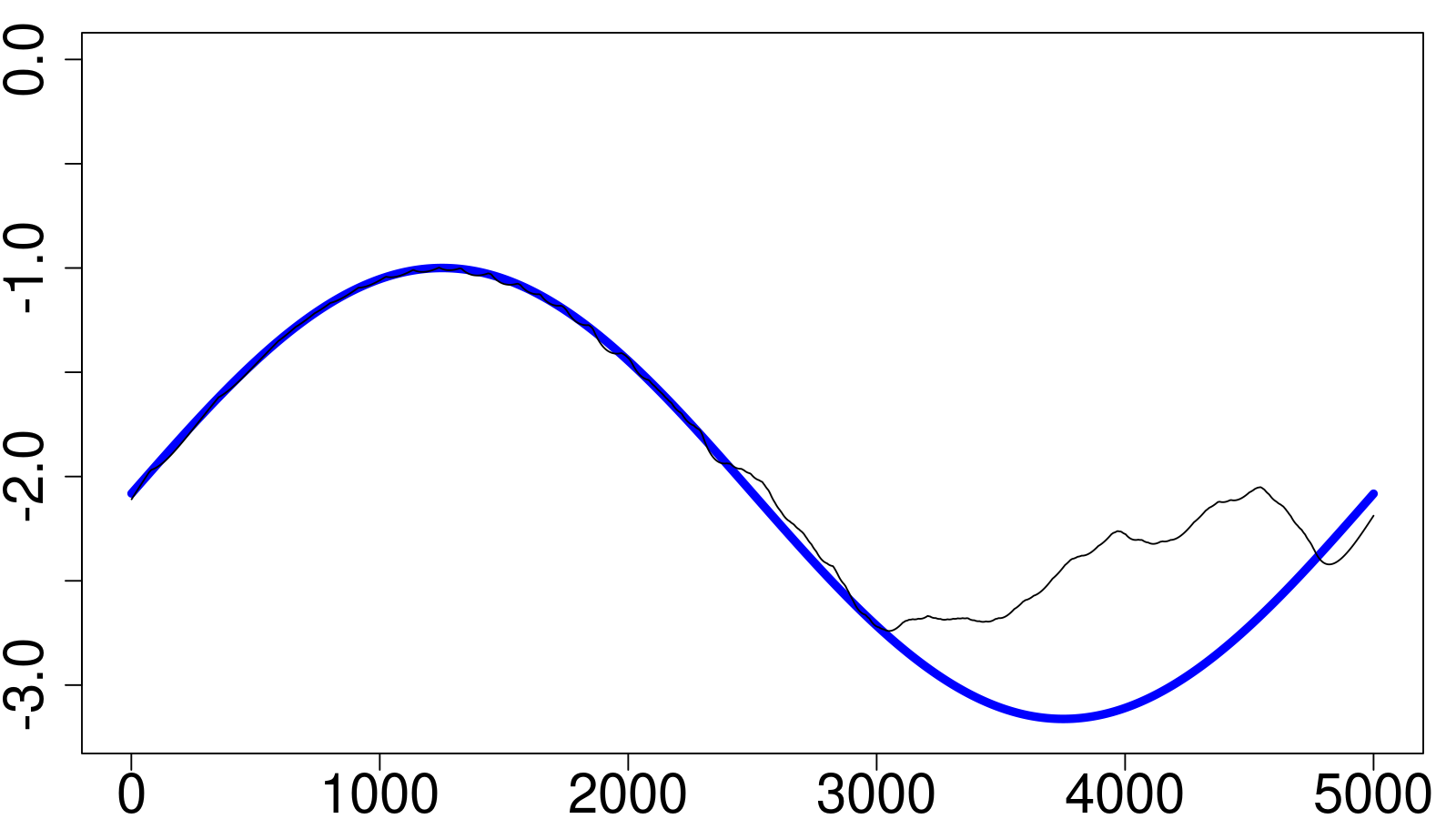}}
	\end{center}
\end{figure}

The simulated values of $z_s$ can be used to derive the corresponding one--step ahead predictions $\hat{c}_s$ for each of the nine specifications of Section \ref{sec:ModelQDyn} (estimated when needed). 
From a visual perspective, the corresponding tracking  of the $5\%$ quantile over the subset of $5,000$ simulations is plotted in Figure \ref{fig:sim-fit-005}.\footnote{The simulations are selected to exhibit a pattern in line with what is discussed below as a comment to Table \ref{tab:table_loss01}. Similar results for $\tau=10\%$ (Figure \ref{fig:sim-fit-010}) and for $\tau=1\%$ (Figure \ref{fig:sim-fit-001}) are shown in Appendix \ref{appdx_D}.} 
Notably, for the Direct-Dynamics approaches considered (panels (a) to (e)), the predictions $\hat{c}_{s}$ are often zero while the true $c_{s}\ll 0$: this is attributable to the low frequency with which $z_{s}<0$ are encountered in the right-most portion of the sinusoidal. By contrast, for the Indirect-Dynamics approaches (panels (f) to (h)) such large mispredictions are absent and the predicted quantile dynamics display a substantially less erratic behavior. 

The distance between the predicted and the true quantiles is accurately\footnote{Estimated parameter uncertainty is not a factor in the comparisons, given the size of the simulated sample.} measured by the RMSE\footnote{The RMSE satisfies the rank-invariance conditions of \citet{Hansen:Lunde:2005b} and \citet{Patton:2011}.} in the first three columns (for each $\tau$) of Table \ref{tab:table_loss01}, where the first row corresponds to a constant predicted quantile. 
All Direct-Dynamics approaches exhibit RMSE's that are larger than that of the constant quantile, confirming the visual assessment that all Historical Simulations and GARCQ generate particularly erratic predictions. 
The RMSE confirms the superior tracking ability of the Indirect-Dynamics approaches (the worst performing Indirect-Dynamics displays a RMSE that is no less than $50\%$ smaller than that of the constant quantile) and, within this group, the distinguished performance of TT's tracking.

\begin{table}[h]
\begin{center} 
\captionsetup{singlelinecheck=off}
\caption [ac] {
By column: in boldface the best performance; starred the best performance within the sub-group.} \label{tab:table_loss01}
\resizebox{!}{0.145\textheight}{
			\begin{tabular}{|l| r@{.}l r@{.}l r@{.}l || r@{.}l r@{.}l r@{.}l || r@{.}l r@{.}l r@{.}l || r@{.}l r@{.}l r@{.}l |}\toprule\toprule
				& \multicolumn{6}{c||}{RMSE$(c_{s},\hat{c}_{s})$} & \multicolumn{6}{c||}{MAD$_{\tau}(z_{s},\hat{c}_{s})$} & \multicolumn{6}{c||}{Coverage} & \multicolumn{6}{c|}{RMSE$(\hat{\pi}_{i},\tau)$} \\
				\midrule
				Model & \multicolumn{2}{r}{10\%} & \multicolumn{2}{r}{5\%} & \multicolumn{2}{r||}{1\%} & \multicolumn{2}{r}{10\%} & \multicolumn{2}{r}{5\%} & \multicolumn{2}{r||}{1\%} & \multicolumn{2}{r}{10\%} & \multicolumn{2}{r}{5\%} & \multicolumn{2}{r||}{1\%} & \multicolumn{2}{r}{10\%} & \multicolumn{2}{r}{5\%} & \multicolumn{2}{r|}{1\%} \\
				\midrule
				const   & 0&5250 & 0&8505 & 2&2335 & 0&1786 & 0&1214 & 0&0517 & 0&1000 & 0&0500 & 0&0100 & 0&0749 & 0&0460 & 0&0113 \\
				\hline                            
				HS250   & 0&7548 & 1&0834 & 2&8187 & 0&1639 & 0&1060 & 0&0394 & 0&1189 & 0&0860 & 0&0431 & 0&1335 & 0&1184 & 0&0851  \\
				HS1000  & 0&5475$^{*}$ & 0&8517$^{*}$ & 2&4247$^{*}$ & 0&1773 & 0&1201 & 0&0502 & 0&0977 & 0&0602 & 0&0191 & 0&1053$^{*}$ & 0&0769 & 0&0355  \\
				WHS95   & 0&8725 & 1&2372 & 2&9544 & \textbf{0}&\textbf{1353} & \textbf{0}&\textbf{0773} & \textbf{0}&\textbf{0255} & 0&1215 & 0&0002 & 0&0000 & 0&1500 & 0&0500$^{*}$ & 0&0100$^{*}$  \\
				WHS99   & 0&7015 & 0&9996 & 2&4726 & 0&1560 & 0&0980 & 0&0316 & 0&1185 & 0&0830 & 0&0000 & 0&1367 & 0&1173 & 0&0100$^{*}$  \\
				GARCQ   & 0&6974 & 0&9964 & 2&4723 & 0&1559 & 0&0978 & 0&0316 & \textbf{0}&\textbf{1000} & \textbf{0}&\textbf{0500} & \textbf{0}&\textbf{0100} & 0&1150 & 0&0654 & 0&0153  \\
				\midrule                                                         
				CAViaR  & 0&2427 & 0&3427 & 1&0766 & 0&1682 & 0&1093 & 0&0432 & 0&0666 & 0&0292 & 0&0060 & 0&0692 & 0&0365 & 0&0081  \\
				QPI     & 0&1923 & 0&2497 & 0&9135 & 0&1644 & 0&1065 & 0&0432 & 0&0640 & 0&0295 & 0&0062 & \textbf{0}&\textbf{0545} & 0&0287 & \textbf{0}&\textbf{0063}  \\
				TT      & \textbf{0}&\textbf{0157} & \textbf{0}&\textbf{0733} & \textbf{0}&\textbf{7365} & 0&1622$^{*}$ & 0&1046$^{*}$ & 0&0420$^{*}$ & 0&0636 & 0&0394 & 0&0101$^{*}$ & 0&0569 & \textbf{0}&\textbf{0273} & 0&0089  \\
				MT      & 0&0177 & 0&2051 & 1&1159 & 0&1624 & 0&1051 & 0&0427 & 0&1008$^{*}$ & 0&0505$^{*}$ & 0&0102 & 0&0662 & 0&0296 & 0&0079  \\
				\bottomrule\bottomrule
			\end{tabular}
		}
	\end{center}
\end{table}

Contrary to the RMSE, which is an infeasible measure in real applications due to the unobservability of the true quantiles $c_{s}$, the MAD$_{\tau}$ estimation criterion of Equation (\ref{eq:ec}) is a feasible measure based on predicted quantiles $\hat{c}_{s}$ and observables $z_{s}$.

Although the MAD$_{\tau}$ is not a consistent measure for model rankings \citep[it fails to satisfy the conditions of][]{Hansen:Lunde:2005b, Patton:2011}, when we predict quantiles for standard values of $\tau$, it may be expected to be positively correlated with the RMSE, as shown in Appendix \ref{App_MADMSE}. 
Therefore, for each model, we  calculate $40,000$  MAD$_{\tau}$ and RMSE measurements over non-overlap\-ping sub-samples of $250$ observations to mimic the forecasting window of one-year in the empirical application. The frequencies with which the two criteria agree in the pairwise comparisons are reported in Table \ref{tab:table00a}, where we observe that the rankings induced by the MAD$_{\tau}$ on the constant and the Indirect-Dynamics are very similar to those of the RMSE, while this is less true of Direct-Dynamics. 

As a point of reference for the entries therein, we set up a similar setting for competing conditional variance models\footnote{As a check--up analysis on this more familiar ground, we generate artificial returns with conditional variances that evolve according to the same sinusoidal pattern used for the quantiles. Estimated GARCH, IGARCH and TGARCH models are then ranked relative to one another in pairwise comparisons. 
}: when comparing two RMSEs, one considering the true variance, the second considering squared returns, both relative to predicted variances, the agreement frequencies are between $0.6425$ and $0.7150$. The results in Table \ref{tab:table00a} are thus encouraging, since in the conditional variance exercise both losses are RMSEs, and the second RMSE is calculated on a noisy but unbiased proxy of the conditional variance (the squared-return), while for the MAD$_{\tau}$ loss in the quantile results, an analogous proxy does not even exist.

Furthermore, given that the MAD$_{\tau}$ appears to be skewed in favor of the Direct-Dynamics and in particular of the WHS95, we  calculate the agreement frequencies in the pairwise comparisons of a Direct- {\it versus} an Indirect-Dynamics model when the MAD$_{\tau}$ favors an Indirect-Dynamics specification. As shown in Table \ref{tab:table00b}, whenever an Indirect-Dynamics model is selected by the MAD$_{\tau}$, the rate of agreement of the infeasible RMSE is generally above $90\%$. To synthesize these results, therefore, we maintain that the MAD$_{\tau}$ may be used in pairwise comparisons to rank the constant and Indirect-Dynamics specifications, but caution is needed when considering also Direct-Dynamics specifications. In such a case, in fact, while the preference provided by the MAD$_{\tau}$ in favor of an Indirect-Dynamics model over a Direct-Dynamics is generally shared by the RMSE, when it prefers a Direct-Dynamics model, the agreement with the RMSE falls to more or less half of the times.

In-sample coverages $S^{-1}\sum_{s}\mathbbm{1}_{[z_{s}<\hat{c}_{s}]}$ of the predicted quantiles are reported in the third set of columns of Table \ref{tab:table_loss01} which signal that exact coverages of the nominal $\tau$ are found for the constant quantile and the GARCQ only. This is the result of the estimator's first-order condition associated to the intercept when the dynamic specification of the quantiles does not contain an endogenous forcing variable.\footnote{For the constant, such a condition is trivially satisfied. For the GARCQ of Equation (\ref{eq:garcq}): $\partial c_{s}/\partial\omega=1+\beta\partial c_{s-1}/\partial\omega$ which, for large $s$, is such that $\partial c_{s}/\partial\omega\approx(1-\beta)^{-1}$ and in turn $\lim_{S\to\infty}S^{-1}\sum_{s}\left[\mathbbm{1}_{[z_{s}<c_{s}]}-\tau\right]=0$. As a counterexample, for the QPI of Equation (\ref{eq:ind}) which contains an endogenous forcing term, $\partial c_{s}/\partial\omega=1+\beta\partial c_{s-1}/\partial\omega -\alpha\partial d_{t-1}/\partial\omega$ which does not yield the orthogonality condition $\lim_{S\to\infty}S^{-1}\sum_{s}\left[\mathbbm{1}_{[z_{s}<c_{s}]}-\tau\right]=0$. This holds even in the limit, where the discontinuous proxy $d_{t-1}$ may be replaced with the actual distribution in $t-1$.} For the same reason, such exact in-sample coverages should not be expected for the historical simulation approaches (random-walk dynamics with no intercept) or for the Indirect-Dynamics (forcing terms contain lags of the endogenous variable).

To shed more light on coverage, we split the simulated series into $4\cdot10^4$ subsamples of $250$ observations each (mimicking roughly one year of daily data), calculate the empirical coverage $\hat{\pi}_{i}$ in each subsample $i$ and report the resulting RMSE$(\hat{\pi}_{i},\tau)$ in the last set of columns in Table \ref{tab:table_loss01}.
As a benchmark, consider the case in which the predicted quantiles coincide with the true $c_{s}$: the violations will be i.i.d. Bernoulli random variables with probability $\tau$ and $\mathbb{E}[\hat{\pi}_{i}]=\tau$ and $\mathbb{V}[\hat{\pi}_{i}]=\tau(1-\tau)/250$, so that, for nominal probabilities of $10\%$, $5\%$ and $1\%$, the reference RMSE$(\hat{\pi}_{i},\tau)$ values are equal to $0.0190$, $0.0138$ and $0.0063$, respectively. {Empirically, we find that all Direct-Dynamics specifications (including the GARCQ despite its exact unconditional coverage) exhibit more sub-sample coverage departures from nominal than those of the constant quantile: exceptions occur for the weighted historical simulations when $\tau=1\%$.} Indirect-Dynamics display the smallest RMSE$(\hat{\pi}_{i},\tau)$, in line with their superior tracking abilities. Therefore, while better tracking does translate into rejection rates that are closer to nominal at each point in time, it does not necessarily imply that rejection rates are closer to nominal over the entire period.

\section{Empirical Analysis}
\label{sec:EmpAnalysis}
The conditional quantiles of stock market investments are studied on the daily returns of the Fama-French 25 value-weighted portfolios\footnote{They are constructed, at the end of each June, from the intersection of five portfolios formed on size and five portfolios formed on the ratio of book- to market-equity considering all NYSE, AMEX and NASDAQ stocks. Data source is Kenneth French's site \texttt{https://mba.tuck.dartmouth.edu/pages/faculty/ken.french/data\_library.html}} which represent a wide spectrum of diversification, ranging from 28 to 611 firms in the 5-5 and 1-5 portfolios, respectively, with a median of 94 firms in the 5-1 portfolio. The time series of each of the 25 portfolios contain 11 years of daily returns from January 2010 to December 2020 for a total of 2769 days. We follow \citet{Andersen:Bollerslev:Huang:2011}, among others, in recognizing the need for a balance between a realistic length for the estimation window, the possible instability of the coefficients, and a long enough out-of-sample period for evaluation. We thus split the overall sample into six 5-year in-sample periods (2010-2014, 2011-2015, 2012-2016, 2013-2017, 2014-2018, 2015-2019), with about 1500 observations each. The conditional quantile specifications of Section \ref{sec:ModelQDyn} are estimated on the specified in-sample periods and the ensuing out-of-sample forecasts are computed for the following year (about 252 observations, for a total of about 1250 out-of-sample observations).

The series of log-returns are calculated from the net returns of the Fama-French portfolios. After subtracting the sample means from each series, in modeling the conditional variances we adhere to the choice of a GJR-GARCH(1,1)  \citep{Glosten:Jaganathan:Runkle:1993} made by \citet{Hansen:Lunde:2005}, due to the superior forecasting performance of the (1,1) specification, the asymmetric response of volatility to shocks and the good overall fit to the data. Thus, the standardized returns $z_{i,t}$ to be used in the quantile specifications of Section \ref{sec:ModelQDyn} are defined as $z_{i,t}\equiv \sigma_{i,t}^{-1}(r_{i,t}-\overline{r}_{i})$ where $\overline{r}_{i}$ is the in-sample average return and $\sigma_{i,t}^{2}$ are the conditional variances (fitted, in-sample and one-step-ahead forecasts, out-of-sample).

\subsection{In--sample Evidence}
\label{sec:ISEvidence}
We test for the in-sample (IS) presence of time-varying conditional quantiles by performing a signed-likelihood ratio test of independence (sLRT). Originally proposed by \citet{Christoffersen:1998} as an out-of-sample backtesting procedure, the independence test is here employed in-sample to test for {\it quantile clustering} in a manner akin to the test for volatility clustering via the Ljung-Box test statistic. The sLRT statistic assumes  a positive (respectively, negative) value when the in-sample frequency of a violation being followed by another violation is higher (respectively, lower) than the nominal $\tau$. When the statistic is negative we talk about \textit{anti--clustering}, because violations tend to be followed by a non--violation. Given that in the empirical application $\tau T$ is not particularly large, rather than relying on Gaussianity (for fixed $\tau$ and $T\to\infty$), we prefer to calculate the one--sided p-values of the sLRT over $10^8$ simulations under the null of independence, which also capture distributional asymmetry.

Results of the tests are reported in Table \ref{tab:table01} where we limit ourselves to report positive significant sLRT in red, and negative significant sLRT in blue, darker color tones indicating rejections at higher significance levels. Interestingly, the IS periods considered exhibit all possible testing outcomes: sparse {\it anti-clustering} in the first IS period (2010-2014), no-clustering in the second period (2011-2015) and clustering of the violations in the four sub-periods covering the years from 2012 to 2019. Therefore, in the IS periods exhibiting clustering ({\it anti-clustering})\footnote{In particular, letting VaR$_{t}$ be the true $\tau$-quantile and VaR$_{t}^{f}$ its $(t-1)$--conditional prediction, it follows that $\mathbb{P}_{t-1}(z_{t}<\text{VaR}_{t}^{f})>\tau$ when VaR$_{t}^{f}$ $>$ VaR$_{t}$ and $\mathbb{P}_{t-1}(z_{t}<\text{VaR}_{t}^{f})<\tau$ when VaR$_{t}^{f}$ $<$ VaR$_{t}$. Clearly, a persistent misalignment between VaR$_{t}^{f}$ and VaR$_{t}$ translates in an equally persistent probability of VaR violations that differs from the nominal $\tau$ and which is likely to lead to the rejection of independence.}, it is possible to improve upon the constant VaR by decreasing (increasing) the conditional quantile after every violation. Although less clear-cut, not rejecting independence in the (2011-2015) period does not necessarily rule out the possibility of improvements upon the constant VaR.

\subsection{Model Extensions to Capture Anti--clustering}
\label{sec:Modelext}
Since the lag-1 specifications of Section \ref{sec:ModelQDyn} cannot capture {\it anti-clustering} without giving rise to implausible OOS behaviors\footnote{A negative lag-1 coefficient implies that the forecasts of the quantiles move toward zero the more the actual quantiles (negative in sign) move away from zero for Direct Dynamics; the forecasts of the quantiles move toward zero the higher the frequency of violations for Indirect Dynamics.}, we include a second lag in the proposed parameterizations\footnote{CAViaR has been purposely left unchanged because adding lags is at odds with its parsimonious parameterization (zero intercept and unit autoregressive coefficient) and the ensuing advantages, such as ease of estimation, smaller estimator variance, relatively low probability of misinterpreting noise for signal, etc. Should more flexibility be needed, we suggest adding lags to the QPI, essentially an unconstrained CAViaR with an indicator function in place of the logistic.}: in the presence of a negative lag-1 coefficient $\alpha_{1}$, a lag-2 positive coefficient $\alpha_{2}$, with $\alpha_{1}+\alpha_{2}\ge0$, eliminates the problem. Therefore, we add the term $\alpha_{2}q_{t-2}$ to the Equation (\ref{eq:garcq}) for the GARCQ; the term  $\alpha_{2}(\tau-d_{t-2})$ to the Equation (\ref{eq:ind}) for the QPI;  the term $\alpha_{2}[\ln(1+\hat{p}_{t-2})-\ln(1+\tau)]$ within the square brackets of the Equation (\ref{eq:one}) for the MT. 

For the TT, we adopt a parsimonious parameterization to incorporate the lag-2 information. 
Beginning with the compact representation of the TT dynamics in Equations (\ref{eq:TTsimp})-(\ref{eq:dl-dh}), lag-2 information is included multiplicatively according to:
\begin{equation*}
c_{t}=c_{t-1}\left[1+d_{t-1}^{l}(\beta_{l}-1)+d_{t-1}^{h}(\beta_{h}-1)\right]\cdot\left[1+d_{t-2}^{l}(\beta_{l}^{-1}-1)+d_{t-2}^{h}(\beta_{h}^{-1}-1)\right]
\end{equation*}
where the inverse coefficients $\beta_{l}^{-1}$ and $\beta_{h}^{-1}$ capture the opposing effects induced at lags 1 and 2 as well as leaving the predicted quantile unchanged at $c_{t-1}$ in the case of conflicting signals, such as when $d_{t-1}^{l}=d_{t-2}^{l}=1$ or $d_{t-1}^{h}=d_{t-2}^{h}=1$.
While in the lag 1 specification the coefficient constraint $\beta_{l} \leq 1 \leq \beta_{h}$ is needed to preserve interpretation, in the lag 2 model the ranking between $\beta_{l} \leq \beta_{h}$ is no longer obvious, so that we remove it, leaving data to decide.

{\subsection{Model Selection}
\label{sec:ModelSelect}
Each of the nine dynamic specifications, for each estimation window (six) and portfolio (twenty-five), is first evaluated IS against the constant quantile. Specifically, expressing the log-likelihood $l$ of Asymmetric Laplace Distributed (see \citet{Keming:Zhang:2005} and \citet{Koenker:Machado:1999}, among others) observations with location $c_{t}$, scale $\sigma$ and asymmetry parameter $\tau$, as a function of MAD$_{\tau}$ gives $l=T\ln[\tau(1-\tau)]-T\ln\sigma-T\sigma^{-1}$MAD$_{\tau}$. After replacing the scale parameter with its maximum-likelihood estimator $\hat{\sigma}=$MAD$_{\tau}$, the usual Information Criteria take the form:
\begin{equation*}
IC(T,k) = 2T\ln\text{MAD}_{\tau}+p(T,k)
\end{equation*}
where $p(T,k)$ is the penalty term and $k$ the number of estimated parameters. For the Akaike Information Criterion (AIC) the penalty is $2k$ while for the Bayes (BIC), it is a more conservative $k\ln T$. 

\subsection{Model Performance Evaluation}

When the constant quantile is selected IS (in each of the 150 cases) by the AIC,\footnote{We discuss results pertaining to OOS forecasts generated either by a given dynamic specification or the constant quantile based on the outcome selected by the  AIC. When using the BIC, IS model selection  always favors the constant quantile, and hence no further analysis is performed.} 
the corresponding OOS MAD$_{\tau}$ of the model is set equal to that of the constant. The frequency with which this occurs is reported in the \textit{Tie} columns of Table \ref{tab:table_mads}, showing that most models are seldom AIC-preferred to the constant. At opposite ends of the spectrum are the WHS95, never selected, and the TT, selected 57\%, 67\% and 95\% of the instances, respectively, for the 10\%, 5\% and 1\% VaR. The \textit{Win} and \textit{Lose} columns report the frequency with which the model forecasts produce OOS MAD$_{\tau}$ that are, respectively, lower and higher than that of the constant quantile. The Win to Lose ratio \textit{W/L} is reported in each subpanel of Table \ref{tab:table_mads}. All Direct Dynamics, which include the popular historical simulations, and the MT have W/L ratios substantially lower than one, while the CAViaR and QPI have generally better ratios. The TT is the only parameterization that yields W/L greater than unity for any $\tau$: $3.25$, $3.55$ and $2.11$ for the 10\%, 5\% and 1\% VaR, respectively.

Gains connected to switching from a constant- to a dynamic-quantile specification are shown in Table \ref{tab:table_mads03} where the columns \textit{On} contain the frequency with which each model is AIC-selected IS (and hence its forecasts are used OOS in place of the constant quantile). Two other columns report the annualized returns\footnote{Consider two cost components associated to the VaR$_\tau$ of a given investment. The first, $C_V (c_{t}-z_{t})$, is the cost of a violation, encompassing all costs incurred by the investor as a result of a liquidity shortage, where $C_V$ is the marginal cost of a violation. The second, $C_{NV}(z_{t}-c_{t})$, is the cost of a non-violation, summarizing all investment opportunities foregone (e.g. larger than needed cash reserves, inadequate leverage, etc.), with $C_{NV}$ the marginal cost of a non-violation. For an agent interested in a VaR$_\tau$ it must be that $C_{NV}/(C_{V}+C_{NV})=\tau$. Hence, given that the MAD$_{\tau}$ is a return where $C_{V}=1-\tau$ and $C_{NV}=\tau$, the general loss is obtained by multiplying MAD$_{\tau}$ by $(C_{V}+C_{NV})$.} demanded by the investor to switch from a constant- to a dynamic-quantile VaR. Specifically, let:
\begin{eqnarray*}
\delta MAD_{\tau} & = & \frac{250}{25H}\sum_{i=1}^{25}\sum_{h=1}^{H}\left\{(\hat{z}_{i,h}-\hat{c}_{i,h})(\tau-\mathbbm{1}_{[\hat{z}_{i,h}<\hat{c}_{i,h}]}) -(\hat{z}_{i,h}-\overline{c}_{i,h})(\tau-\mathbbm{1}_{[\hat{z}_{i,h}<\overline{c}_{i,h}]})\right\}\\
\Delta MAD_{\tau} & = & \frac{250}{25H}\sum_{i=1}^{25}\sum_{h=1}^{H}\left\{(r_{i,h}-\hat{q}_{i,h})(\tau-\mathbbm{1}_{[r_{i,h}<\hat{q}_{i,h}]}) -(r_{i,h}-\overline{q}_{i,h})(\tau-\mathbbm{1}_{[r_{i,h}<\overline{q}_{i,h}]})\right\}
\end{eqnarray*}
where: $250$ annualizes the daily return, $H$ is the total number of days in the six OOS periods, $i$ is the portfolio index and $h$ the time index; $\hat{z}_{i,h}$ is the OOS standardized return; $\hat{c}_{i,h}$ is the model's quantile forecast and $\overline{c}_{i,h}$ is the constant quantile updated at the beginning of each year; $r_{i,h}$ is the OOS original return, $\hat{q}_{i,h}$ is the model's VaR forecast and $\overline{q}_{i,h}$ is the VaR arising from the constant quantile. The following relationships hold: $r_{i,h}=\hat{\mu}_{i,h}+\hat{\sigma}_{i,h}\hat{z}_{i,h}$; $\hat{q}_{i,h}=\hat{\mu}_{i,h}+\hat{\sigma}_{i,h}\hat{c}_{i,h}$;  $\overline{q}_{i,h}=\hat{\mu}_{i,h}+\hat{\sigma}_{i,h}\overline{c}_{i,h}$.

The first index, $\delta$MAD$_{\tau}$, reports the losses computed w.r.t. the standardized returns while in the second one, $\Delta$MAD$_{\tau}$, they are calculated for the original returns, hence jointly evaluating mean-, variance- and quantile-forecasts. Even though both $\delta$MAD$_{\tau}$ and $\Delta$MAD$_{\tau}$ may be seen as a weighted average of the other's constituents, from an investor's perspective $\Delta$MAD$_{\tau}$ is more relevant. The results in Table \ref{tab:table_mads03} confirm the poor performance of the Direct Dynamics specifications. Standard historical simulations approaches, applied to the standardized returns, bring no improvements upon the constant-quantile. The dynamic GARCQ (based on the realized quantiles from WHS99) performs worse than standard approaches, with incremental costs for the investor ranging from $10.44\%$ to $487.12\%$. In particular, the latter value is attained by the $5\%$-VaR forecasts and is due to a handful of cases, scattered across a couple of OOS periods in which, by picking the dynamics IS, the model  generates OOS rates of violations in excess of $50\%$. Within the Indirect Dynamics, MT forecasts are strongly dominated by those produced by the constant quantile. The performance of CAViaR is hardly distinguishable from a constant quantile. A similar result holds for QPI with the exception of the $1\%$-VaR for which it exhibits a substantial loss of $23.42\%$. Notice that, for the $1\%$-VaR, in spite of the fact that the QPI is selected IS by the AIC in $41\%$ of the instances against the constant, the model features are not of help OOS. Finally, the TT confirms the results of Table \ref{tab:table_mads} with annualized cost reductions of $28.50\%$ and $17.19\%$ for the $10\%$ and $5\%$ VaR, respectively. For the $1\%$-VaR, although the TT may capture some persistent IS features even with sparse data, generating a negative, albeit moderate, $\delta$MAD$_{\tau}$, the OOS gains are lost when weighted by the variance forecasts entering $\Delta$MAD$_{\tau}$.

\subsection{Coverage and Independence Tests}
\label{sec:CoverIndep}
{As a final evaluation, we analyze coverage and independence of the violations of each model's quantile forecasts. Summary statistics of coverage, over the 25 Fama-French portfolios and six OOS periods, are reported in Table \ref{tab:table_ucfs01} by model. Here, with the exception of the 5\%-VaR forecasts of the GARCQ (already discussed), no major differences can be found across the specifications considered.} 

For a given portfolio and set of quantile forecasts, we perform coverage and independence tests in each of the six OOS periods; to account for the Risk Manager's interest in quality VaR forecasts, we evaluate whether the number of rejections is compatible with the chosen significance level and the six trials associated with the available OOS periods.

To that end, let us consider a given $\tau$ level and each model. The values reported in each panel of Table \ref{tab:table_bin} are the result of a two stage procedure. For the \textit{Coverage} panel we have:
\begin{enumerate}
  \item For each portfolio and OOS year, in the first stage we test at $5\%$ whether the proportion of quantile violations in the year is different from $\tau$, providing  $25$ sets of $6$ test outcomes across the years.
  \item In the second stage, for each portfolio, we count the number of first stage rejections across the $6$ years and then we test whether that total is compatible with the $5\%$ significance level considered. The count reported in each cell is the number of rejections across such $25$ second stage tests.
\end{enumerate}

For the \textit{Independence} panel of Table \ref{tab:table_bin}, the procedure and the counts reported in the table are similar, except for the fact that in the first stage we perform the independence tests described in Section \ref{sec:ISEvidence}.

\section{Conclusions}

In this paper, we find it instructive to suggest a formal definition of the {\it unconditional}, {\it conditional} and {\it actual} quantiles as the objects of interest in the discussion about time-varying modeling of the VaR at a given coverage $\tau$. In keeping with the SDM framework, the main goal is to assess whether the distribution of the returns standardized by their conditional means and standard deviations exhibit predictable dynamics in their quantiles.

With this distinction in mind, we can classify approaches (both existing and the ones we propose) depending on the information used to update the conditional quantile: we have Direct-Dynamics, when models exploit discrepancies between past {\it actual}- and {\it conditional}-quantiles; and  Indirect-Dynamics, when the specifications make use of the differences between nominal and past empirical rejection frequencies.

We introduce some new specifications: the GARCQ (inspired by the stationary GARCH) within the Direct-Dynamics specifications; and three new Indirect-Dynamics models, the QPI (inspired by the CAViaR of \citet{Engle:Manganelli:2004}) and the {\it Violations Tracking} approaches Test Tracking (TT) and Multiplicative Tracking (MT) {which employ an exponentially-weighted average of past violations as a forcing term in order to dampen parasitic oscillations.  The TT specification further dampens such oscillations by updating the {\it conditional}-quantile only when past discrepancies may be attributed to real movements of the underlying quantile.}

Given the novelty of the focus on standardized returns, we resort to extensive simulations to assess the ability of each model to track the behavior of time--varying quantiles: this task requires the use of appropriate loss functions. Since the  RMSE, calculated over {\it true} and predicted quantiles, is consistent but operationally infeasible, we suggest to base the comparisons on the feasible MAD$_{\tau}$. We assess the overall superior quantile-tracking abilities of the Indirect-Dynamics specifications and the adequacy of the rankings induced by the MAD$_{\tau}$ relative to those induced by the RMSE, which can be computed only in this simulated framework.

The empirical analysis, conducted on the daily returns of the Fama-French 25 value-weighted portfolios, shows that, for all approaches, the violations (resulting from the quantile forecasts) exhibit empirical coverage in line with the nominal $\tau$ and no statistically significant clustering. However, when interpreting the MAD$_{\tau}$  as the cost of mispredicting the quantile in returns terms, and taking the constant quantile as the benchmark, only the Test Tracking forecasts produce substantial reductions  quantifiable in annualized returns of $28.50\%$ and $17.19\%$ for $\tau=10\%$ and $\tau=5\%$, respectively. All other approaches, both existing and new, generate forecasts that are dominated by the constant conditional-quantile.

{We find that using conditional variances to standardize returns, the distribution of the resulting innovations may still exhibit some relevant time dependence.} Hence, the effort of modeling their conditional quantiles {may result} in substantial forecast improvements: the evidence is in favor of a new updating technique which changes the predicted quantile only when the empirical frequency of violations falls outside a data-driven interval around the nominal $\tau$. We deem this to be a reasonable trade--off between capturing the underlying dynamics, while not indulging in detrimental adjustments driven by noise.

\label{sec:concl}


\newpage 
\clearpage
\begin{table}[h!]
\begin{center} 
\captionsetup{singlelinecheck=off}
\caption [ac] {
For each $\tau=10\%, 5\%, 1\%$, the panels  report the frequency with which the feasible MAD$_{\tau}(z_{s},\hat{c}_{s})$ loss and the infeasible 	RMSE$(c_{s},\hat{c}_{s})$ agree by pairwise comparison of model rankings. MAD$_{\tau}$ and RMSE are calculated $40,000$ times over simulated sub-samples of $250$ observations.} \label{tab:table00a}
\resizebox{!}{0.47\textheight}{
\begin{tabular}{|l| r@{.}l r@{.}l r@{.}l r@{.}l r@{.}l | r@{.}l r@{.}l r@{.}l r@{.}l |}\toprule\toprule
$\tau=10\%$ & \multicolumn{2}{l}{HS250} & \multicolumn{2}{l}{HS1000} & \multicolumn{2}{l}{WHS95} & \multicolumn{2}{l}{WHS99} & \multicolumn{2}{l|}{GARCQ} & \multicolumn{2}{l}{CAViaR} & \multicolumn{2}{l}{QPI} & \multicolumn{2}{l}{TT} & \multicolumn{2}{l|}{MT} \\
\midrule
const   & 0&8234 & 0&8032 & 0&4855 & 0&7604 & 0&7687 & 0&8568 & 0&8685 & 0&8831 & 0&8833 \\
\midrule
HS250   & \multicolumn{2}{c}{-} & 0&7774 & 0&5499 & 0&8390 & 0&8404 & 0&7020 & 0&6562 & 0&8107 & 0&8088 \\
HS1000  & \multicolumn{2}{c}{-} & \multicolumn{2}{c}{-} & 0&5290 & 0&7634 & 0&7639 & 0&8306 & 0&7950 & 0&8664 & 0&8637 \\
WHS95   & \multicolumn{2}{c}{-} & \multicolumn{2}{c}{-} & \multicolumn{2}{c}{-} & 0&5059 & 0&5036 & 0&4247 & 0&4234 & 0&3746 & 0&3503 \\
WHS99   & \multicolumn{2}{c}{-} & \multicolumn{2}{c}{-} & \multicolumn{2}{c}{-} & \multicolumn{2}{c}{-} & 0&6753 & 0&6411 & 0&5331 & 0&6872 & 0&6815 \\
GARCQ   & \multicolumn{2}{c}{-} & \multicolumn{2}{c}{-} & \multicolumn{2}{c}{-} & \multicolumn{2}{c}{-} & \multicolumn{2}{c|}{-} & 0&6525 & 0&5741 & 0&6868 & 0&6744 \\
\midrule
CAViaR  & \multicolumn{2}{c}{-} & \multicolumn{2}{c}{-} & \multicolumn{2}{c}{-} & \multicolumn{2}{c}{-} & \multicolumn{2}{c|}{-} & \multicolumn{2}{c}{-} & 0&8200 & 0&9041 & 0&8932 \\    
QPI     & \multicolumn{2}{c}{-} & \multicolumn{2}{c}{-} & \multicolumn{2}{c}{-} & \multicolumn{2}{c}{-} & \multicolumn{2}{c|}{-} & \multicolumn{2}{c}{-} & \multicolumn{2}{c}{-} & 0&9061 & 0&8927 \\
TT      & \multicolumn{2}{c}{-} & \multicolumn{2}{c}{-} & \multicolumn{2}{c}{-} & \multicolumn{2}{c}{-} & \multicolumn{2}{c|}{-} & \multicolumn{2}{c}{-} & \multicolumn{2}{c}{-} & \multicolumn{2}{c}{-} & 0&8114 \\
\toprule \toprule
$\tau=5\%$ & \multicolumn{2}{l}{HS250} & \multicolumn{2}{l}{HS1000} & \multicolumn{2}{l}{WHS95} & \multicolumn{2}{l}{WHS99} & \multicolumn{2}{l|}{GARCQ} & \multicolumn{2}{l}{CAViaR} & \multicolumn{2}{l}{QPI} & \multicolumn{2}{l}{TT} & \multicolumn{2}{l|}{MT} \\
\midrule
const   & 0&7672 & 0&7775 & 0&4768 & 0&7630 & 0&7696 & 0&8484 & 0&8564 & 0&8803 & 0&8803 \\
\midrule
HS250   & \multicolumn{2}{c}{-} & 0&7720 & 0&5561 & 0&8387 & 0&7793 & 0&6489 & 0&6196 & 0&7817 & 0&7631 \\
HS1000  & \multicolumn{2}{c}{-} & \multicolumn{2}{c}{-} & 0&5403 & 0&7638& 0&7634 & 0&7986 & 0&8307 & 0&8537 & 0&8438 \\
WHS95   & \multicolumn{2}{c}{-} & \multicolumn{2}{c}{-} & \multicolumn{2}{c}{-} & 0&5046 & 0&5038 & 0&4214 & 0&4179 & 0&3442 & 0&3653 \\
WHS99   & \multicolumn{2}{c}{-} & \multicolumn{2}{c}{-} & \multicolumn{2}{c}{-} & \multicolumn{2}{c}{-} & 0&7342 & 0&5999 & 0&4765& 0&6523 & 0&6238 \\
GARCQ   & \multicolumn{2}{c}{-} & \multicolumn{2}{c}{-} & \multicolumn{2}{c}{-} & \multicolumn{2}{c}{-} & \multicolumn{2}{c|}{-} & 0&6223 & 0&5832 & 0&6237 & 0&6100 \\
\midrule
CAViaR  & \multicolumn{2}{c}{-} & \multicolumn{2}{c}{-} & \multicolumn{2}{c}{-} & \multicolumn{2}{c}{-} & \multicolumn{2}{c|}{-} & \multicolumn{2}{c}{-} & 0&8021& 0&8940 & 0&8907 \\
QPI     & \multicolumn{2}{c}{-} & \multicolumn{2}{c}{-} & \multicolumn{2}{c}{-} & \multicolumn{2}{c}{-} & \multicolumn{2}{c|}{-} & \multicolumn{2}{c}{-} & \multicolumn{2}{c}{-} & 0&9023 & 0&8521 \\
TT      & \multicolumn{2}{c}{-} & \multicolumn{2}{c}{-} & \multicolumn{2}{c}{-} & \multicolumn{2}{c}{-} & \multicolumn{2}{c|}{-} & \multicolumn{2}{c}{-} & \multicolumn{2}{c}{-} & \multicolumn{2}{c}{-} & 0&8178 \\
\toprule \toprule
$\tau=1\%$ & \multicolumn{2}{l}{HS250} & \multicolumn{2}{l}{HS1000} & \multicolumn{2}{l}{WHS95} & \multicolumn{2}{l}{WHS99} & \multicolumn{2}{l|}{GARCQ} & \multicolumn{2}{l}{CAViaR} & \multicolumn{2}{l}{QPI} & \multicolumn{2}{l}{TT} & \multicolumn{2}{l|}{MT} \\
\midrule
const   & 0&6885 & 0&6803 & 0&4471 & 0&6983& 0&6984 & 0&7819 & 0&7399 & 0&7950 & 0&7780 \\
\midrule
HS250   & \multicolumn{2}{c}{-} & 0&7366 & 0&5475 & 0&7772 & 0&7771 & 0&5060 & 0&4417 & 0&6209 & 0&5101 \\
HS1000  & \multicolumn{2}{c}{-} & \multicolumn{2}{c}{-} & 0&5266 & 0&7351 & 0&7351 & 0&7201 & 0&6832 & 0&7853 & 0&7636 \\   
WHS95   & \multicolumn{2}{c}{-} & \multicolumn{2}{c}{-} & \multicolumn{2}{c}{-} & 0&2584 & 0&2581 & 0&3644 & 0&3745 & 0&3536 & 0&3642 \\    
WHS99   & \multicolumn{2}{c}{-} & \multicolumn{2}{c}{-} & \multicolumn{2}{c}{-} & \multicolumn{2}{c}{-} & 0&5184 & 0&5642 & 0&5817 & 0&4639 & 0&5792 \\
GARCQ   & \multicolumn{2}{c}{-} & \multicolumn{2}{c}{-} & \multicolumn{2}{c}{-} & \multicolumn{2}{c}{-} & \multicolumn{2}{c|}{-} & 0&5642 & 0&5817 & 0&4640 & 0&5792 \\    
\midrule
CAViaR  & \multicolumn{2}{c}{-} & \multicolumn{2}{c}{-} & \multicolumn{2}{c}{-} & \multicolumn{2}{c}{-} & \multicolumn{2}{c|}{-} & \multicolumn{2}{c}{-} & 0&7662 & 0&8159 & 0&7477 \\
QPI     & \multicolumn{2}{c}{-} & \multicolumn{2}{c}{-} & \multicolumn{2}{c}{-} & \multicolumn{2}{c}{-} & \multicolumn{2}{c|}{-} & \multicolumn{2}{c}{-} & \multicolumn{2}{c}{-} & 0&8196 & 0&7638 \\
TT      & \multicolumn{2}{c}{-} & \multicolumn{2}{c}{-} & \multicolumn{2}{c}{-} & \multicolumn{2}{c}{-} & \multicolumn{2}{c|}{-} & \multicolumn{2}{c}{-} & \multicolumn{2}{c}{-} & \multicolumn{2}{c}{-} & 0&7965 \\
\bottomrule\bottomrule
\end{tabular}
}
\end{center}
\end{table}

\clearpage 

\begin{table}[h!]
\begin{center} 
\captionsetup{singlelinecheck=off}
\caption [ac] {
Frequency with which the feasible MAD$_{\tau}(z_{s},\hat{c}_{s})$ loss agrees with the infeasible RMSE$(c_{s},\hat{c}_{s})$, when the former prefers an Indirect Dynamics (by column) over a Direct Dynamics (by row) specification. MAD$_{\tau}$ and RMSE are calculated $40,000$ times over simulated sub-samples of $250$ observations.} \label{tab:table00b}
\resizebox{!}{0.11\textheight}{
\begin{tabular}{|l| r@{.}l r@{.}l r@{.}l r@{.}l | r@{.}l r@{.}l r@{.}l r@{.}l | r@{.}l r@{.}l r@{.}l r@{.}l | }\toprule\toprule
 & \multicolumn{8}{c|}{$\tau=10\%$} & \multicolumn{8}{c|}{$\tau=5\%$} & \multicolumn{8}{c|}{$\tau=1\%$} \\
 & \multicolumn{2}{c}{\footnotesize{CAViaR}} & \multicolumn{2}{c}{\footnotesize{QPI}} & \multicolumn{2}{c}{\footnotesize{TT}} & \multicolumn{2}{c|}{\footnotesize{MT}} & \multicolumn{2}{c}{\footnotesize{CAViaR}} & \multicolumn{2}{c}{\footnotesize{QPI}} & \multicolumn{2}{c}{\footnotesize{TT}} & \multicolumn{2}{c|}{\footnotesize{MT}} & \multicolumn{2}{c}{\footnotesize{\footnotesize{CAViaR}}} & \multicolumn{2}{c}{\footnotesize{QPI}} & \multicolumn{2}{c}{\footnotesize{TT}} & \multicolumn{2}{c|}{\footnotesize{MT}} \\
\midrule
\footnotesize{const}   & 0&9800 & 1&0000 & 1&0000 & 0&9999 & 0&9979 & 1&0000 & 1&0000 & 1&0000 & 0&9746 & 0&9266 & 0&9739 & 0&9718 \\
\midrule                                                                                                    
\footnotesize{HS250}   & 0&7880 & 0&8044 & 0&9990 & 0&9991 & 0&8068 & 0&8057 & 0&9985 & 0&9960 & 0&7435 & 0&6348 & 0&9212 & 0&7833 \\   
\footnotesize{HS1000}  & 0&9369 & 0&8885 & 1&0000 & 0&9999 & 0&9347 & 0&9704 & 0&9996 & 0&9946 & 0&9303 & 0&8612 & 0&9922 & 0&9746 \\   
\footnotesize{WHS95}   & 1&0000 & 1&0000 & 0&9791 & 0&9831 & 1&0000 & 0&9985 & 0&9936 & 0&9932 & 0&3333 & 0&9855 & 0&9639 & 0&9850 \\   
\footnotesize{WHS99}   & 0&5925 & 0&7392 & 0&9999 & 0&9964 & 0&7150 & 0&7161 & 0&9991 & 0&9795 & 0&9986 & 0&9451 & 0&9904 & 0&9953 \\   
\footnotesize{GARCQ}   & 0&6419 & 0&8112 & 0&9994 & 0&9960 & 0&8514 & 0&9495 & 0&9595 & 0&9682 & 0&9986 & 0&9452 & 0&9905 & 0&9953 \\   
\bottomrule\bottomrule
\end{tabular}
}
\end{center}
\end{table}

\clearpage 
\begin{table}[h!]
\begin{center} 
\captionsetup{singlelinecheck=off}
\caption [ac] {
 Independence tests of the violations of the in-sample unconditional 10\%, 5\% and 1\% VaRs. The F-F column lists the Fama-French 25 portfolios formed from the intersection of quintile sorts on size (first entry) and book-to-market (second entry). Signed-likelihood ratio tests, with p-values calculated over $10^{8}$ simulations, are performed for six 5-year rolling in-sample periods ranging from 2010-2014 to 2015-2019. Rejection of independence with a negative value of the test statistic (fewer consecutive rejections than expected) is reported at the significance levels of $0.10$, $0.05$ and $0.01$, respectively, as \colorbox{blue!25}{\textcolor{blue!25}{a}}, \colorbox{blue!50}{\textcolor{blue!50}{a}} and \colorbox{blue!100}{\textcolor{blue!100}{a}}. Rejection of the null with a positive value of the test statistic (more consecutive rejections than expected) is reported  at the significance levels of $0.10$, $0.05$ and $0.01$, respectively, as \colorbox{red!25}{\textcolor{red!25}{a}}, \colorbox{red!50}{\textcolor{red!50}{a}} and \colorbox{red!100}{\textcolor{red!100}{a}}. 
} \label{tab:table01}
\resizebox{!}{0.39\textheight}{
\begin{tabular}{|l| c c c| c c c| c c c| c c c| c c c| c c c|}\toprule\toprule
& \multicolumn{3}{c|}{2010-2014} & \multicolumn{3}{c|}{2011-2015} & \multicolumn{3}{c|}{2012-2016} & \multicolumn{3}{c|}{2013-2017} & \multicolumn{3}{c|}{2014-2018} & \multicolumn{3}{c|}{2015-2019} \\
F-F & 10\% & \enskip 5\% & \enskip 1\% & 10\% & \enskip 5\% & \enskip 1\% & 10\% & \enskip 5\% & \enskip 1\% & 10\% & \enskip 5\% & \enskip 1\% & 10\% & \enskip 5\% & \enskip 1\% & 10\% & \enskip 5\% & \enskip 1\% \\
\toprule 
1-1 & & \colorbox{blue!25}{\textcolor{blue!25}{A}} & & & & & \colorbox{red!25}{\textcolor{red!25}{A}} & & & \colorbox{red!25}{\textcolor{red!25}{A}} & & & \colorbox{red!25}{\textcolor{red!25}{A}} & & & \colorbox{red!50}{\textcolor{red!50}{A}} & & \\ \midrule
1-2 & \colorbox{blue!50}{\textcolor{blue!50}{A}} & & & & & & & & & & & & & & & & & \\ \midrule
1-3 & & & & & & & & & & & & & & & & & & \\ \midrule 
1-4 & \colorbox{blue!50}{\textcolor{blue!50}{A}} & & & & & & & \colorbox{red!50}{\textcolor{red!50}{A}} & & & \colorbox{red!25}{\textcolor{red!25}{A}} & & & \colorbox{red!25}{\textcolor{red!25}{A}} & & \colorbox{red!25}{\textcolor{red!25}{A}} & \colorbox{red!25}{\textcolor{red!25}{A}} & \\ \midrule
1-5 & & \colorbox{blue!25}{\textcolor{blue!25}{A}} & & & \colorbox{red!25}{\textcolor{red!25}{A}} & & \colorbox{red!25}{\textcolor{red!25}{A}}& \colorbox{red!25}{\textcolor{red!25}{A}}& & & \colorbox{red!25}{\textcolor{red!25}{A}} & & \colorbox{red!25}{\textcolor{red!25}{A}} & \colorbox{red!50}{\textcolor{red!50}{A}} & & \colorbox{red!50}{\textcolor{red!50}{A}} & \colorbox{red!50}{\textcolor{red!50}{A}} & \\ \midrule
2-1 & & & & & & & & \colorbox{red!25}{\textcolor{red!25}{A}} & & & & & & & & & & \\ \midrule
2-2 & & \colorbox{blue!25}{\textcolor{blue!25}{A}} & & & & & & & & & & & & & & & & \\ \midrule
2-3 & & \colorbox{blue!25}{\textcolor{blue!25}{A}} & & & & & & & & & & & & & & & & \\ \midrule
2-4 & & \colorbox{blue!25}{\textcolor{blue!25}{A}} & & & & & & \colorbox{red!25}{\textcolor{red!25}{A}} & & & \colorbox{red!50}{\textcolor{red!50}{A}} & & & \colorbox{red!25}{\textcolor{red!25}{A}} & & \colorbox{red!25}{\textcolor{red!25}{A}} & \colorbox{red!50}{\textcolor{red!50}{A}} & \\ \midrule
2-5 & \colorbox{blue!25}{\textcolor{blue!25}{A}} & & & & & & & & & & & & & & & & & \\ \midrule
3-1 & & & & & & & \colorbox{red!50}{\textcolor{red!50}{A}} & & & \colorbox{red!50}{\textcolor{red!50}{A}} & & & \colorbox{red!50}{\textcolor{red!50}{A}} & \colorbox{red!25}{\textcolor{red!25}{A}} & & \colorbox{red!100}{\textcolor{red!100}{A}} & \colorbox{red!25}{\textcolor{red!25}{A}} & \\ \midrule
3-2 & & & & & & & & & & & & & & & & \colorbox{red!100}{\textcolor{red!100}{A}} & & \\ \midrule
3-3 & & \colorbox{blue!25}{\textcolor{blue!25}{A}} & & & & & \colorbox{red!50}{\textcolor{red!50}{A}} & & & \colorbox{red!50}{\textcolor{red!50}{A}} & & & \colorbox{red!50}{\textcolor{red!50}{A}} & & & \colorbox{red!100}{\textcolor{red!100}{A}} & & \\ \midrule
3-4 & & \colorbox{blue!25}{\textcolor{blue!25}{A}} & & & & & & & & \colorbox{red!25}{\textcolor{red!25}{A}} & & & \colorbox{red!25}{\textcolor{red!25}{A}} & & & \colorbox{red!50}{\textcolor{red!50}{A}} & & \\ \midrule
3-5 & & \colorbox{blue!25}{\textcolor{blue!25}{A}} & & & & & & \colorbox{red!25}{\textcolor{red!25}{A}} & & & \colorbox{red!25}{\textcolor{red!25}{A}} & & & \colorbox{red!25}{\textcolor{red!25}{A}} & & & \colorbox{red!25}{\textcolor{red!25}{A}} & \\ \midrule
4-1 & & & & & & & & & & & & & & & & \colorbox{red!25}{\textcolor{red!25}{A}} & \colorbox{red!50}{\textcolor{red!50}{A}} & \\ \midrule
4-2 & & & & & & & & & & & & & & & & & & \\ \midrule
4-3 & & & & & & & & & & & & & & & & \colorbox{red!50}{\textcolor{red!50}{A}} & & \\ \midrule
4-4 & & \colorbox{blue!25}{\textcolor{blue!25}{A}} & & & & & & & & & & & & & & & \colorbox{red!50}{\textcolor{red!50}{A}} & \\ \midrule
4-5 & & & & & & & & & & & & & & & & & & \\ \midrule
5-1 & & & & & & & & & & & & & & & & & & \\ \midrule
5-2 & & & & & & & & & & & & & & & & & & \\ \midrule
5-3 & & \colorbox{blue!25}{\textcolor{blue!25}{A}} & & & & & & & & & & & & & & & & \\ \midrule
5-4 & & \colorbox{blue!25}{\textcolor{blue!25}{A}} & & & & & & & & & & & & & & & & \\ \midrule
5-5 & & & & & & & \colorbox{red!25}{\textcolor{red!25}{A}} & \colorbox{red!100}{\textcolor{red!100}{A}} & & \colorbox{red!25}{\textcolor{red!25}{A}} & \colorbox{red!25}{\textcolor{red!25}{A}} & & \colorbox{red!50}{\textcolor{red!50}{A}} & \colorbox{red!25}{\textcolor{red!25}{A}} & & \colorbox{red!50}{\textcolor{red!50}{A}} & \colorbox{red!25}{\textcolor{red!25}{A}} & \\ \midrule
\bottomrule
\end{tabular}
}
\end{center}
\end{table}


\begin{table}[ht]
\begin{center}
\captionsetup{singlelinecheck=off}
\caption [ac] {
For each model, performance is measured by the MAD$_{\tau}$ over 150 instances (25 portfolios over 6 OOS periods) and compared to that of the constant VaR. The columns report the frequencies with which the models Win and Lose against the constant VaR. When, for a given portfolio and time period, the AIC leads to discard the IS estimated model in favor of the constant VaR, the OOS performance results in a Tie with the constant VaR. The W/L column summarizes the ratio of Win/Lose instances for each model.} \label{tab:table_mads}
\resizebox{!}{0.15\textheight}{
\begin{tabular}{|l| r@{.}l r@{.}l r@{.}l r@{.}l| r@{.}l r@{.}l r@{.}l r@{.}l| r@{.}l r@{.}l r@{.}l r@{.}l|}\toprule\toprule
\multirow{2}{*}{Model} & \multicolumn{8}{c|}{$\tau=10\%$} & \multicolumn{8}{c|}{$\tau=5\%$} & \multicolumn{8}{c|}{$\tau=1\%$} \\
\cline{2-25}
& \multicolumn{2}{c}{Win} & \multicolumn{2}{c}{Tie} & \multicolumn{2}{c}{Lose} & \multicolumn{2}{c|}{W/L} & \multicolumn{2}{c}{Win} & \multicolumn{2}{c}{Tie} & \multicolumn{2}{c}{Lose} & \multicolumn{2}{c|}{W/L} & \multicolumn{2}{c}{Win} & \multicolumn{2}{c}{Tie} & \multicolumn{2}{c}{Lose} & \multicolumn{2}{c|}{W/L}\\
\midrule
HS250             & 0&0000 & 0&9333 & 0&0667 & 0&0000 & 0&0000 & 0&9333 & 0&0667 & 0&0000 & 0&0133 & 0&9467 & 0&0400 & 0&3333\\
HS1000            & 0&0400 & 0&7600 & 0&2000 & 0&2000 & 0&0200 & 0&8000 & 0&1800 & 0&1111 & 0&0267 & 0&9267 & 0&0467 & 0&5714\\
WHS95             & 0&0000 & 1&0000 & 0&0000 & \multicolumn{2}{c|}{--} & 0&0000 & 1&0000 & 0&0000 & \multicolumn{2}{c|}{--} & 0&0000 & 1&0000 & 0&0000 & \multicolumn{2}{c|}{--}\\
WHS99             & 0&0067 & 0&9667 & 0&0267 & 0&2500 & 0&0000 & 0&9867 & 0&0133 & 0&0000 & 0&0000 & 0&9867 & 0&0133 & 0&0000\\
GARCQ             & 0&0267 & 0&7867 & 0&1867 & 0&1429 & 0&1133 & 0&6600 & 0&2267 & 0&5000 & 0&1467 & 0&5067 & 0&3467 & 0&4231\\
\midrule
CAViaR            & 0&0933 & 0&8200 & 0&0867 & 1&0769 & 0&0800 & 0&8267 & 0&0933 & 0&8571 & 0&0000 & 0&9933 & 0&0067 & 0&0000\\
QPI               & 0&0533 & 0&9000 & 0&0467 & 1&1429 & 0&0733 & 0&8133 & 0&1133 & 0&6471 & 0&1867 & 0&5933 & 0&2200 & 0&8485\\
TT                & 0&4333 & 0&4333 & 0&1333 & 3&2500 & 0&5200 & 0&3333 & 0&1467 & 3&5455 & 0&6467 & 0&0467 & 0&3067 & 2&1087\\
MT                & 0&0267 & 0&7867 & 0&1867 & 0&1429 & 0&0200 & 0&8733 & 0&1067 & 0&1875 & 0&0467 & 0&8600 & 0&0933 & 0&5000\\
\bottomrule\bottomrule
\end{tabular}
}
\end{center}
\end{table}

\begin{table}[ht]
\begin{center}
\captionsetup{singlelinecheck=off}
\caption [ac] {

The \textit{On} column reports the frequency, in the 150 (25 portfolios over 6 OOS periods) instances, with which each model is selected IS by the AIC against the constant VaR. The column $\delta$MAD$_{\tau}$ reports the resulting OOS variation of the loss: expressed in annualized return. $\Delta$MAD$_{\tau}$ is the OOS variation of the MAD$_{\tau}$ calculated w.r.t. the original returns and the corresponding VaR which combines mean, variance and quantile forecasts: expressed in annualized return.} \label{tab:table_mads03}
\resizebox{!}{0.18\textheight}{
\begin{tabular}{|l| r@{.}l r@{.}l r@{.}l| r@{.}l r@{.}l r@{.}l| r@{.}l r@{.}l r@{.}l|}\toprule\toprule
\multirow{2}{*}{Model} & \multicolumn{6}{c|}{$\tau=10\%$} & \multicolumn{6}{c|}{$\tau=5\%$} & \multicolumn{6}{c|}{$\tau=1\%$} \\
\cline{2-19}
& \multicolumn{2}{c}{On} & \multicolumn{2}{c}{$\delta$MAD$_{\tau}$} & \multicolumn{2}{c|}{$\Delta$MAD$_{\tau}$} & \multicolumn{2}{c}{On} & \multicolumn{2}{c}{$\delta$MAD$_{\tau}$} & \multicolumn{2}{c|}{$\Delta$MAD$_{\tau}$} & \multicolumn{2}{c}{On} & \multicolumn{2}{c}{$\delta$MAD$_{\tau}$} & \multicolumn{2}{c|}{$\Delta$MAD$_{\tau}$} \\
\midrule
HS250             & 0&07 &  10&18\% & 10&27\% & 0&07 &    9&62\% & 10&87\% & 0&05 &   1&82\% & 1&27\%\\
HS1000            & 0&24 &   2&82\% &  2&81\% & 0&20 &    2&52\% &  2&32\% & 0&08 &   0&37\% & 0&04\%\\
WHS95             & 0&00 & \multicolumn{2}{c}{--} & \multicolumn{2}{c|}{--} & 0&00 & \multicolumn{2}{c}{--} & \multicolumn{2}{c|}{--} & 0&00 & \multicolumn{2}{c}{--} & \multicolumn{2}{c|}{--} \\
WHS99             & 0&04 &   0&41\% & -3&44\% & 0&01 &    2&54\% &  2&86\% & 0&01 &   0&55\% & 0&42\%\\
GARCQ             & 0&22 &  24&79\% & 24&23\% & 0&34 &  487&12\% &476&63\% & 0&50 &  10&44\% & 9&97\%\\
\midrule                                                                                       
CAViaR            & 0&18 &   0&46\% & -2&04\% & 0&17 &    2&04\% &  1&95\% & 0&01 &   0&40\% & 0&61\%\\
QPI               & 0&10 &  -1&17\% & -4&09\% & 0&19 &    1&51\% & -0&60\% & 0&41 &  13&85\% &23&42\% \\
TT                & 0&56 & -15&24\% &-28&50\% & 0&67 &   -9&70\% &-17&19\% & 0&95 &  -1&47\% & 0&49\%\\
MT                & 0&22 &  22&84\% & 27&84\% & 0&13 &   21&62\% & 24&89\% & 0&14 &   3&33\% & 3&36\%\\
\bottomrule\bottomrule
\end{tabular}
}
\end{center}
\end{table}

\begin{table}[ht]
	\begin{center}
		\captionsetup{singlelinecheck=off}
		\caption [ac] {Unconditional Coverage of the 10\%, 5\% and 1\% VaR forecasts. For every model, a total of 150 empirical rejection rates is calculated by considering all 25 Fama-French portfolios over the six OOS periods. Descriptive statistics of min, mean, max and std.dev. are reported in the corresponding columns.} \label{tab:table_ucfs01}
		\resizebox{!}{0.16\textheight}{
			\begin{tabular}{|l| r@{.}l r@{.}l r@{.}l r@{.}l| r@{.}l r@{.}l r@{.}l r@{.}l| r@{.}l r@{.}l r@{.}l r@{.}l|}\toprule\toprule
				\multirow{2}{*}{Model} & \multicolumn{8}{c|}{$\tau=10\%$} & \multicolumn{8}{c|}{$\tau=5\%$} & \multicolumn{8}{c|}{$\tau=1\%$} \\
				\cline{2-25}
				& \multicolumn{2}{c}{min} & \multicolumn{2}{c}{mean} & \multicolumn{2}{c}{max} & \multicolumn{2}{c|}{std.dev.} & \multicolumn{2}{c}{min} & \multicolumn{2}{c}{mean} & \multicolumn{2}{c}{max} & \multicolumn{2}{c|}{std.dev.} & \multicolumn{2}{c}{min} & \multicolumn{2}{c}{mean} & \multicolumn{2}{c}{max} & \multicolumn{2}{c|}{std.dev.}\\
				\midrule
				const             & 0&0319 & 0&1003 & 0&1818 & 0&0310 & 0&0120 & 0&0517 & 0&0988 & 0&0198 & 0&0000 & 0&0131 & 0&0319 & 0&0063 \\
				\midrule
				HS250             & 0&0319 & 0&1017 & 0&1818 & 0&0326 & 0&0120 & 0&0527 & 0&0988 & 0&0206 & 0&0000 & 0&0132 & 0&0319 & 0&0063 \\
				HS1000            & 0&0319 & 0&1014 & 0&1818 & 0&0309 & 0&0120 & 0&0526 & 0&0988 & 0&0198 & 0&0000 & 0&0134 & 0&0319 & 0&0062 \\
				WHS95             & 0&0319 & 0&1003 & 0&1818 & 0&0310 & 0&0120 & 0&0517 & 0&0988 & 0&0198 & 0&0000 & 0&0131 & 0&0319 & 0&0063 \\
				WHS99             & 0&0319 & 0&0997 & 0&1818 & 0&0302 & 0&0120 & 0&0519 & 0&0988 & 0&0200 & 0&0000 & 0&0132 & 0&0319 & 0&0063 \\
				GARCQ             & 0&0319 & 0&1025 & 0&2143 & 0&0354 & 0&0120 & 0&0705 & 0&8294 & 0&1068 & 0&0000 & 0&0133 & 0&0432 & 0&0076 \\
				\midrule
				CAViaR            & 0&0319 & 0&0995 & 0&1700 & 0&0297 & 0&0120 & 0&0520 & 0&0988 & 0&0200 & 0&0000 & 0&0130 & 0&0319 & 0&0063 \\
				QPI               & 0&0319 & 0&1001 & 0&1700 & 0&0302 & 0&0120 & 0&0518 & 0&0988 & 0&0193 & 0&0000 & 0&0138 & 0&0319 & 0&0065 \\
				TT                & 0&0319 & 0&0996 & 0&1660 & 0&0279 & 0&0120 & 0&0521 & 0&0992 & 0&0184 & 0&0000 & 0&0133 & 0&0435 & 0&0081 \\ 
				MT                & 0&0319 & 0&1000 & 0&2151 & 0&0320 & 0&0120 & 0&0527 & 0&1434 & 0&0219 & 0&0000 & 0&0133 & 0&0319 & 0&0066 \\
				\bottomrule\bottomrule
			\end{tabular}
		}
	\end{center}
\end{table}

\clearpage

\begin{table}[ht]
	\begin{center}
		\captionsetup{singlelinecheck=off}
		\caption [ac] {
			\textit{Coverage} panel: for each portfolio and OOS year, we first perform a standard coverage test at $5\%$ and then, for each portfolio, we count the number of first stage rejections across the $6$ years, testing whether that total is compatible with the $5\%$ significance level considered. The count reported in each cell is the number of rejections across such $25$ second stage tests. \textit{Independence} panel: same procedure but independence tests are performed at the first stage.}.
			\label{tab:table_bin}
		\resizebox{!}{0.17\textheight}{
			\begin{tabular}{|l|ccc|ccc|}\toprule\toprule
				\multirow{2}{*}{Model} & \multicolumn{3}{c|}{Coverage} & \multicolumn{3}{c|}{Independence} \\
				\cline{2-7}
				& \multicolumn{1}{c|}{~~~~~$\tau=$10\%~~~~~} & \multicolumn{1}{c|}{~~~~~~$\tau=$5\%~~~~~} & \multicolumn{1}{c|}{~~~~~~$\tau=$1\%~~~~~} & \multicolumn{1}{c|}{~~~~~$\tau=$10\%~~~~~} & \multicolumn{1}{c|}{~~~~~~$\tau=$5\%~~~~~} & \multicolumn{1}{c|}{~~~~~~$\tau=$1\%~~~~~} \\
				\midrule
				const       & 0 & 1 & 0 & 0 & 0 & 0\\
				\midrule
				HS250       & 0 & 3 & 3 & 0 & 0 & 0\\
				HS1000      & 0 & 1 & 0 & 0 & 0 & 0\\      
				WHS95       & 0 & 1 & 0 & 0 & 0 & 0\\       
				WHS99       & 0 & 2 & 0 & 1 & 0 & 0\\      
				GARCQ       & 1 & 6 & 8 & 0 & 0 & 0\\      
				\midrule
				CAViaR      & 0 & 2 & 0 & 0 & 0 & 0\\      
				QPI         & 1 & 1 & 0 & 0 & 0 & 0\\      
				TT          & 0 & 1 & 0 & 0 & 0 & 0\\      
				MT          & 0 & 3 & 1 & 0 & 0 & 0\\      
				\bottomrule\bottomrule
			\end{tabular}
		}
	\end{center}
\end{table}

\newpage 

\appendix
\section{Mixture of Distributions \label{example}}
The $F(z|\omega_{t})$ distribution of Figure \ref{fig:figure00} is the mixture of a standardized exponential distribution and its mirror image:
\begin{equation*}
F(z|\omega_{t}) = \omega_{t}\mathbbm{1}_{[z\ge-1]}\left(1-e^{-1-z}\right) + (1-\omega_{t})\mathbbm{1}_{[z\le 1]}e^{-1+z}
\end{equation*}
where the random weight $\omega_{t}$ is given by:
\begin{equation*}
\omega_{t} = 0.2 + 0.6x_{t-1}+\left(2x_{t-1}-1\right)\left(-0.6+0.75\nu_{t}\right)
\end{equation*}
with $x_{t-1}\sim B(1,0.5)$ and $\nu_{t}\sim B(1,0.8)$. From $\mathbb{E}_{t-1}[\omega_{t}]=0.2+0.6x_{t-1}$ and $\mathbb{E}[\omega_{t}]=0.5$ it follows that the unconditional distribution $F(z)$ depicted in the third panel of Figure \ref{fig:figure00} is given by:
\begin{equation*}
F(z) = 0.5\cdot\mathbbm{1}_{[z\ge-1]}\left(1-e^{-1-z}\right) + 0.5\cdot\mathbbm{1}_{[z\le 1]}e^{-1+z}
\end{equation*}
The conditional distribution $F(z|x_{t-1})$ in the second panel of Figure \ref{fig:figure00} exemplifies the case in which the value of the conditioning Bernoulli random variable is $x_{t-1}=1$:
\begin{equation*}
F(z|x_{t-1}=1) = 0.8\cdot\mathbbm{1}_{[z\ge-1]}\left(1-e^{-1-z}\right) + 0.2\cdot\mathbbm{1}_{[z\le 1]}e^{-1+z}
\end{equation*}
Finally, the actual distribution $F(z|\omega_{t})$ plotted in the first panel of Figure \ref{fig:figure00} is the result of the draw $\nu_{t}=0$ for the Bernoulli random variable $\nu_{t}$, to which corresponds $\omega_{t}=0.2$:
\begin{equation*}
F(z|\omega_{t}=0.2) = 0.2\cdot\mathbbm{1}_{[z\ge-1]}\left(1-e^{-1-z}\right) + 0.8\cdot\mathbbm{1}_{[z\le 1]}e^{-1+z}
\end{equation*}

\section{Simulations Set-Up\label{appdx_setup}}
\label{setup}
The ability of the specifications in Section \ref{sec:ModelQDyn} to track time--varying $\tau$--quantiles is investigated via a long Monte Carlo simulation of the innovations $\{z_{s}\}_{s=1}^{S}$, with $S=10^{7}$. The distribution from which to sample cannot be the {\it standardized} Gaussian random variable, as it has constant quantiles. By the same token, a heavy-tailed distribution does not entail to  appreciably decrease and increase the value of the $\tau$-quantile: as a matter of fact, as shown in Appendix \ref{appdx_A} for the tails of the Generalized Pareto distribution, {\it standardized} heavy-tailed random variables are likely to produce too narrow a range of quantiles, for it to be of interest when comparing performances. 

To maximize such a range, we can refer to Chebyshev's inequality bounds that envisage, for any $|c_{s}|>1$ (to represent here the $\tau$-quantile at time $s$), the probability $\mathbb{P}(|z_{s}|\ge |c_{s}|)$ to be less than or equal to $c_{s}^{-2}$, and take that probability to be the largest. Therefore, we sample the innovations $z_{s}$ from the symmetric three-valued distribution for which Chebyshev holds with equality:\footnote{Another possibility is to employ the two-valued distribution for which Cantelli holds with equality and for which either $\mathbb{P}(z_{s}\ge c_{s})=(1+c_{s}^{2})^{-1}$ or $\mathbb{P}(z_{s}\le -c_{s})=(1+c_{s}^{2})^{-1}$ but $\mathbb{P}(|z_{s}|\ge c_{s})<c_{s}^{-2}$.}
\begin{equation}\label{eq:zs}
z_{s}=\left\{
\begin{array}{rll}
-(2p_{s})^{-1/2} & \text{w.p.} & p_{s}\\
0 & \text{w.p.} & 1-2p_{s}\\          
(2p_{s})^{-1/2} & \text{w.p.} & p_{s}\\     
\end{array}
\right.
\end{equation}
with $p_{s}\in[0,\frac{1}{2}]$, with the required feature that $\mathbb{E}(z_s)=0$ and $\mathbb{V}(z_s)=1$. Since for $\tau\le p_{s}$, $c_{s}$ is defined from $\mathbb{P}(z_{s}\le c_{s})=\tau$, it follows that $c_{s}=-(2p_{s})^{-1/2}$ and, conversely, $p_{s}=\frac{1}{2}\left[-c_{s}\right]^{-2}$. Thus, $p_{s} \in[\tau,\frac{1}{2}]$ has a one--to--one correspondence with $c_{s}\in[-(2\tau)^{-1/2},-1]$, indeed the largest possible range for the quantiles. 

In order to mimic a smooth evolution of the quantiles, for a given $\tau$ we specify a sinusoidal pattern of $c_{s}$ with a chosen frequency $f=2000$:
\begin{equation}\label{eq:sine}
c_{s} = -\frac{1}{2}\left[(2\tau)^{-1/2}+1\right]+\frac{1}{2}\left[(2\tau)^{-1/2}-1\right]\cdot\sin\left(\frac{2\pi f s}{S}\right) \quad s=1,\ldots,S,
\end{equation}
which varies between the required minimum of $-(2\tau)^{-1/2}$ and the maximum of $-1$. Accordingly, this implies a $p_{s}$ which determines the distribution  in Expression (\ref{eq:zs}) from which $z_s$ are generated. Thus, setting a specific $\tau$ in Equation (\ref{eq:sine}) allows us to generate a corresponding sinusoidal trajectory for $c_s$, maximizing the excursion of the corresponding $\tau$-quantile.

Figure \ref{fig:figure01} displays three subsets of $5,000$ draws for $z_{s}$, together with the underlying $\tau$-quantiles $c_s$, left--to--right corresponding to $\tau=\{10\%, 5\%, 1\%\}$ (in blue -- one cycle of the sine function each). The lighter regions in the time series of $z_{s}$ appear when there is a longer persistence of the observations at $0$: given that $z_{s}$ has mean zero and unit variance, realizations with relatively larger modulus $(2p_{s})^{-1/2}$ occur with a smaller probability $p_{s}$, according to Equation (\ref{eq:zs}).\footnote{This explains why smaller values of $c_s$ in modulus  correspond to fewer occurrences at zero (cf. panel (a) in the Figure for $\tau=10\%$), and vice versa (cf. panel (c) in the Figure for $\tau=1\%$).} Furthermore, what looks like clustering of the $z_{s}$ in Figure \ref{fig:figure01} does not correspond to a time-varying volatility since the data generating process in (\ref{eq:zs}) has $\mathbb{V}(z_{s})=1$, $\forall s$.\footnote{This is confirmed empirically, as the simulated $z_{s}^{2}$ do not display significant autocorrelations nor do GARCH specifications estimated on $z_{s}$ exhibit statistically significant parameters beyond the intercept.}

\newpage
\section{Tracking for $\tau=10\%$ and $\tau=1\%$  \label{appdx_D}}
\begin{figure}[H]
\captionsetup{singlelinecheck=off}
\caption[ac]{
Time series of true (in blue) and predicted (in black) $10\%$ quantiles for all Direct- and Indirect-Dynamics approaches considered.}\vspace{-0.25cm}\label{fig:sim-fit-010}
\begin{center}
\subfloat[HS250]{\includegraphics[width=0.328\textwidth, height=0.14\textwidth]{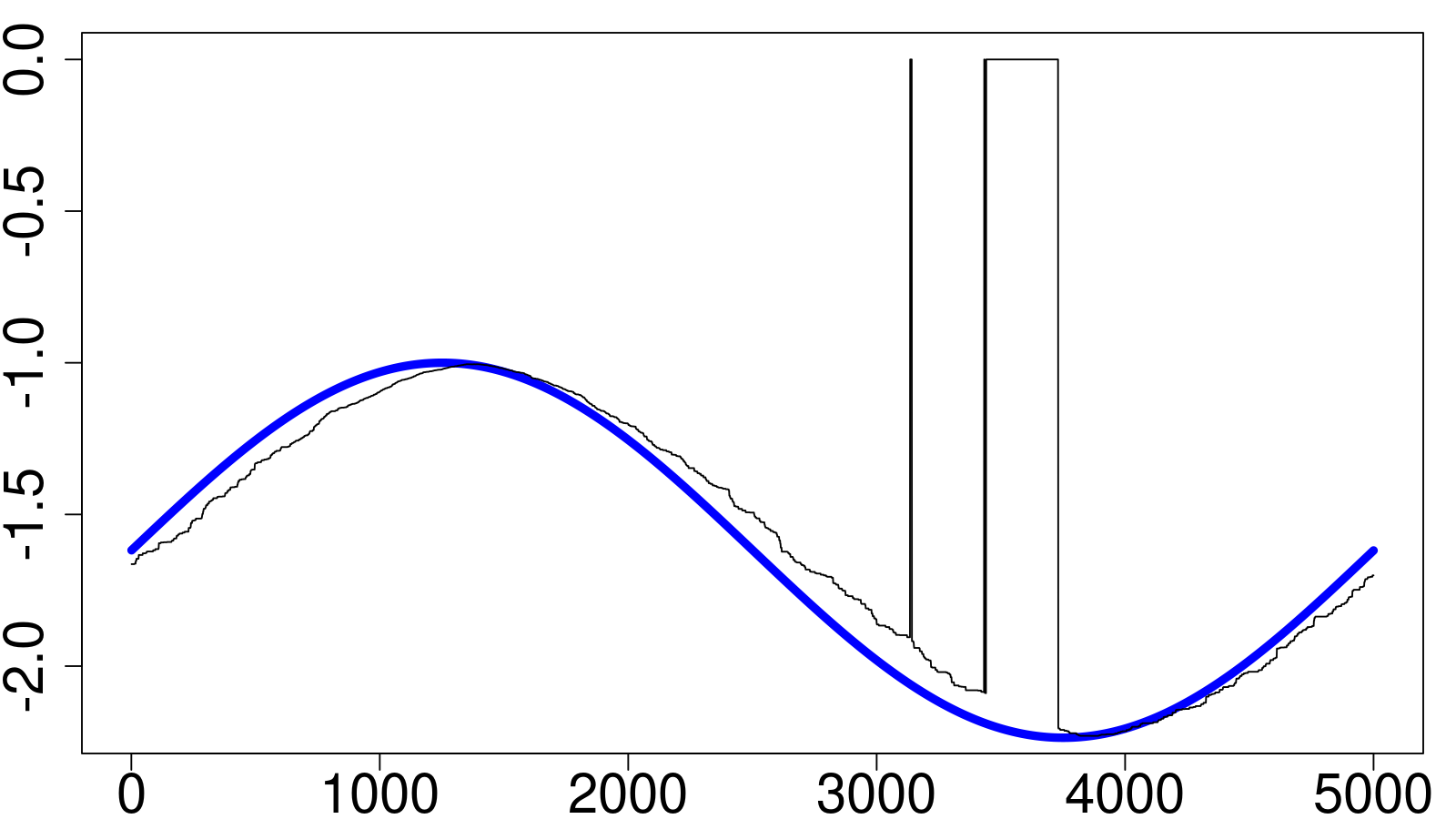}}
\subfloat[HS1000]{\includegraphics[width=0.328\textwidth, height=0.14\textwidth]{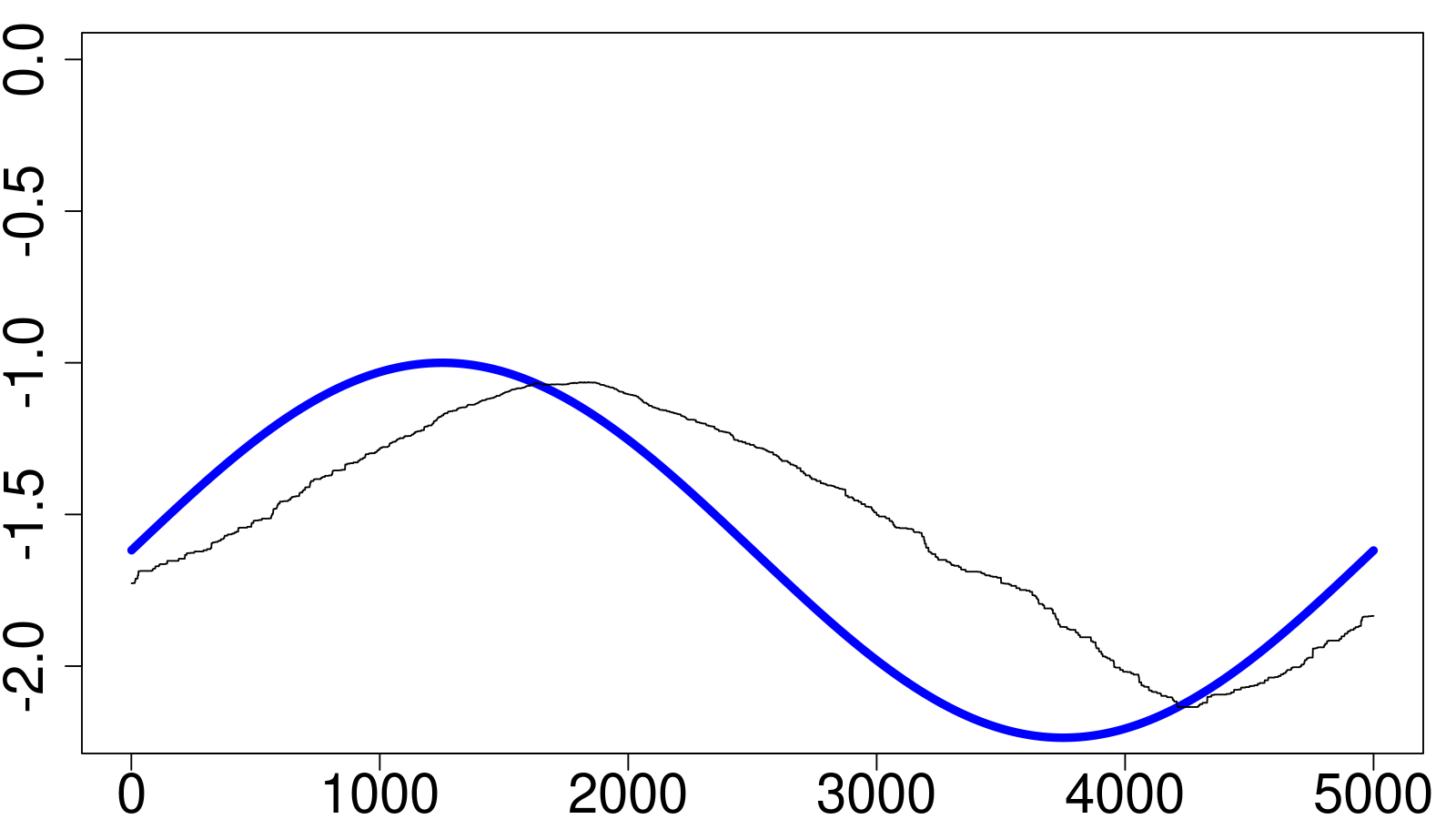}}
\subfloat[WHS95]{\includegraphics[width=0.328\textwidth, height=0.14\textwidth]{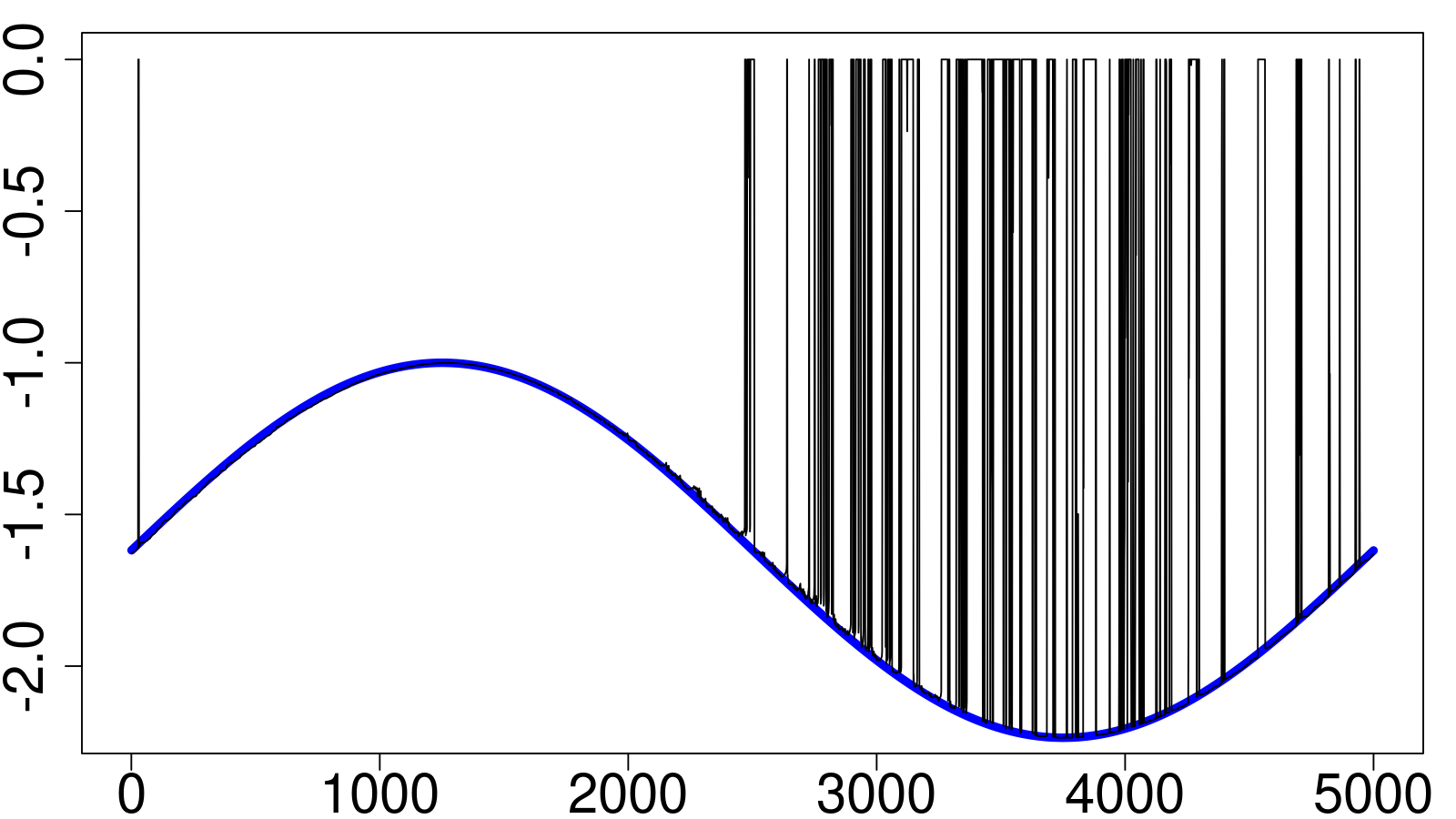}}\\
\subfloat[WHS99]{\includegraphics[width=0.328\textwidth, height=0.14\textwidth]{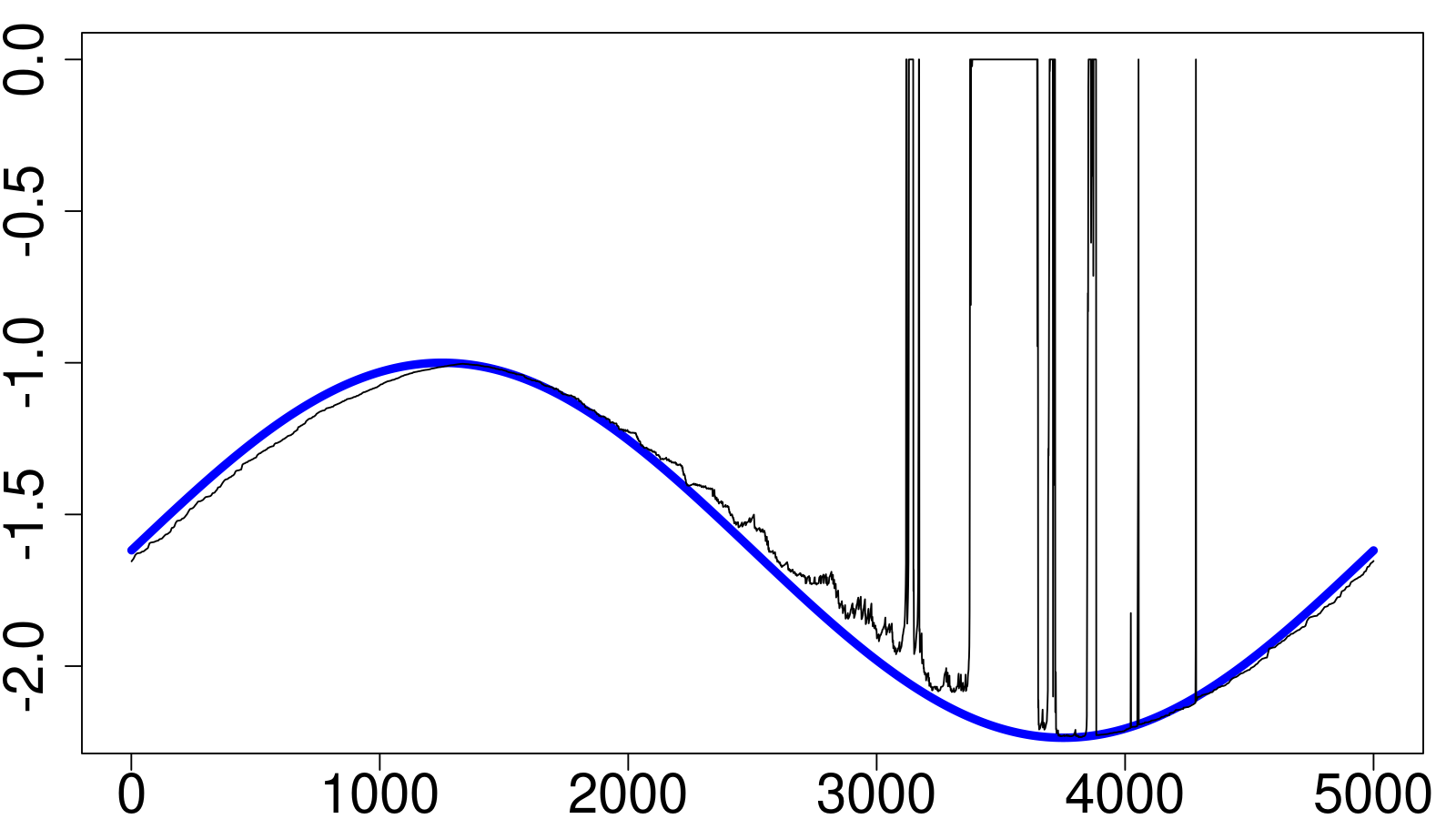}}
\subfloat[GARCQ]{\includegraphics[width=0.328\textwidth, height=0.14\textwidth]{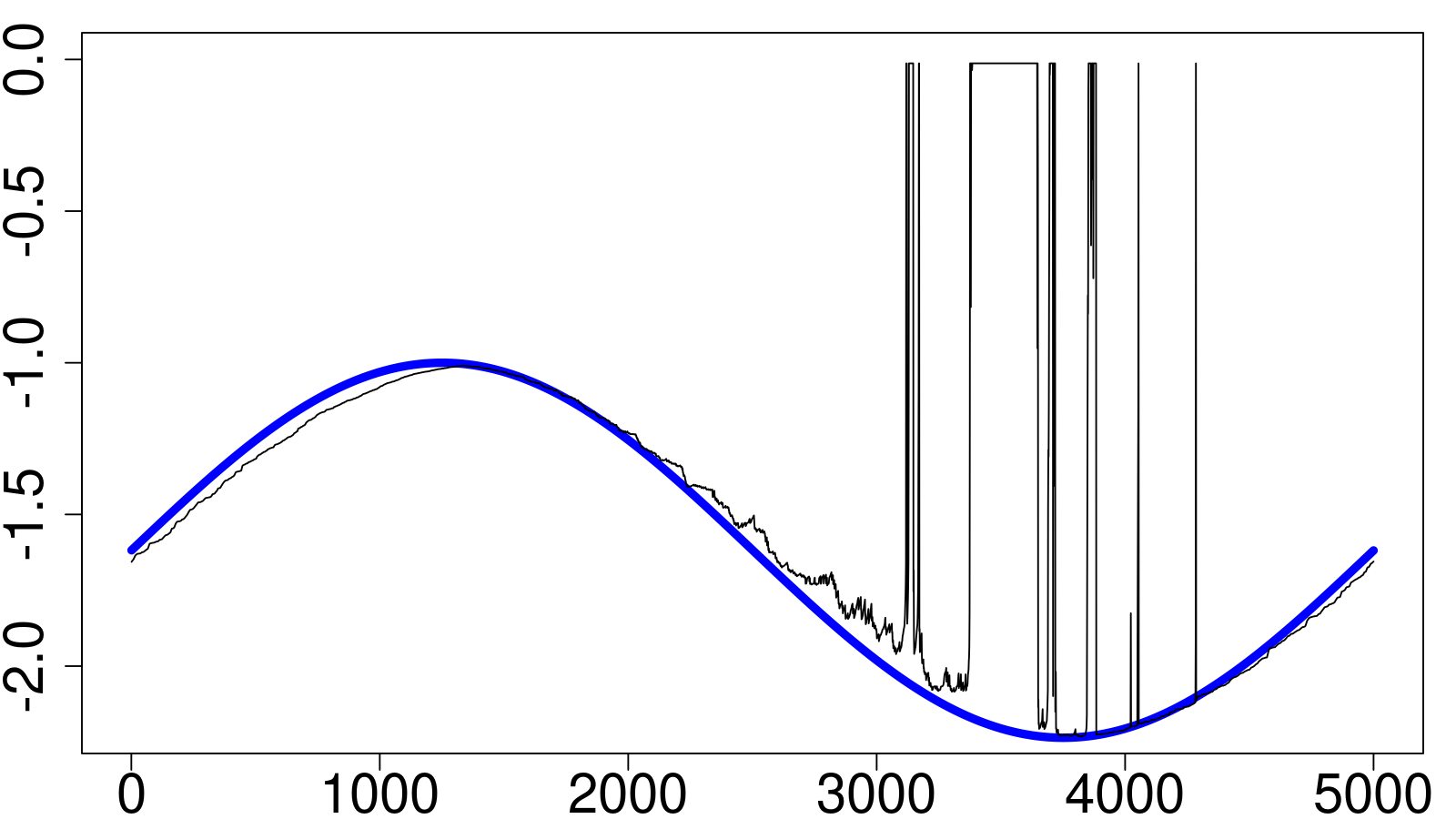}}
\subfloat[CAViaR]{\includegraphics[width=0.328\textwidth, height=0.14\textwidth]{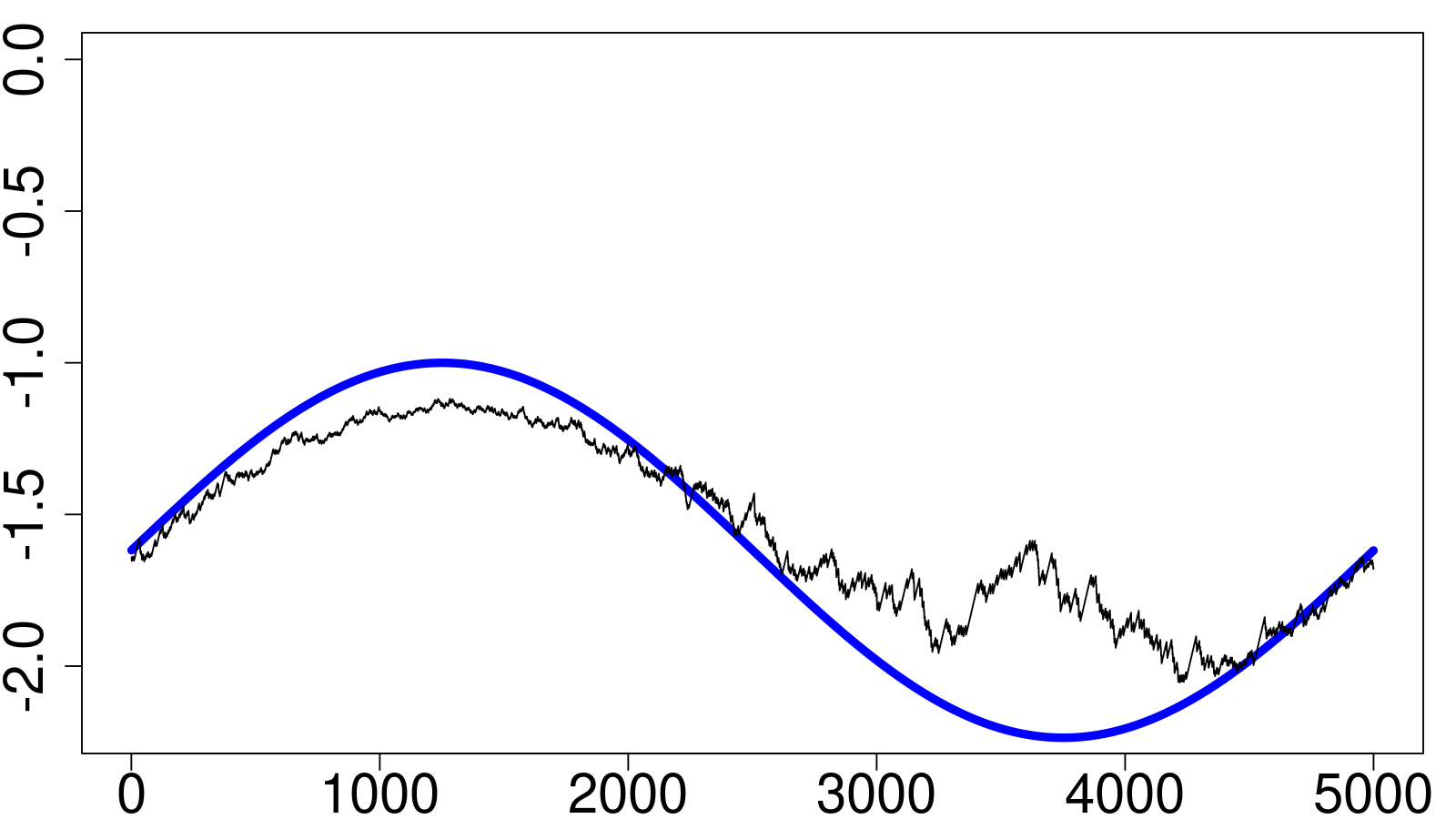}}\\
\subfloat[QPI]{\includegraphics[width=0.328\textwidth, height=0.14\textwidth]{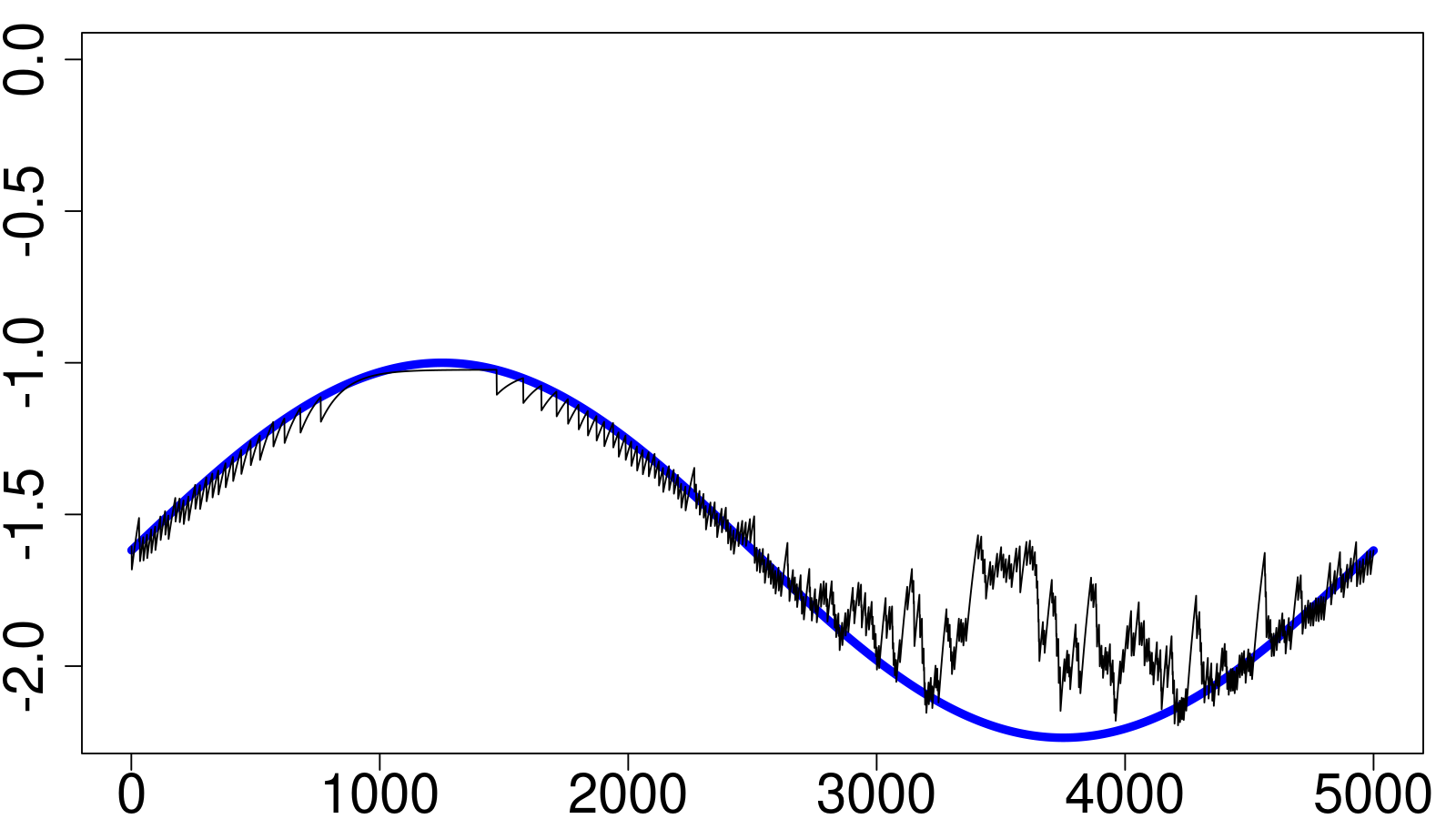}}
\subfloat[TT]{\includegraphics[width=0.328\textwidth, height=0.14\textwidth]{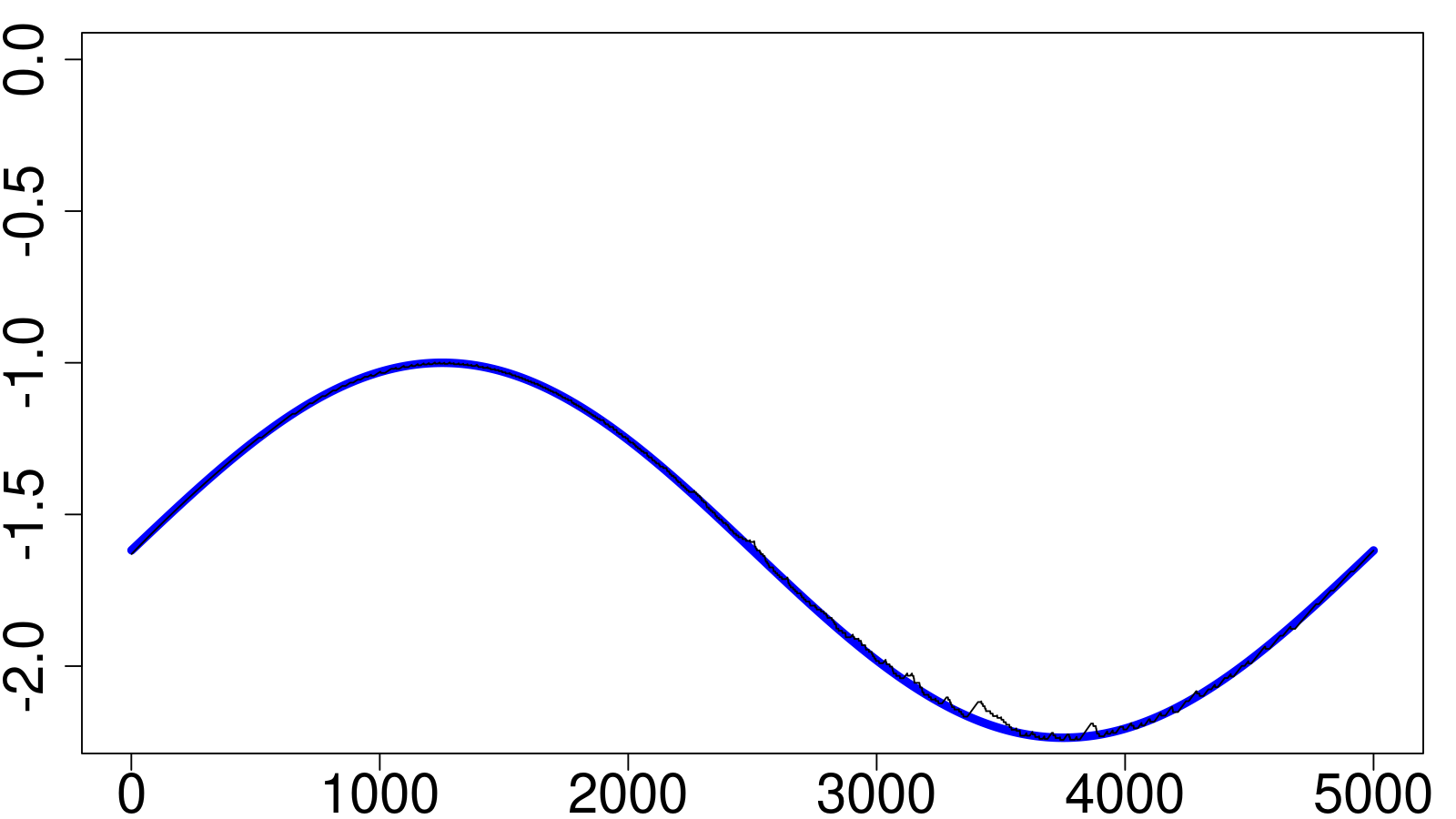}}
\subfloat[MT]{\includegraphics[width=0.328\textwidth, height=0.14\textwidth]{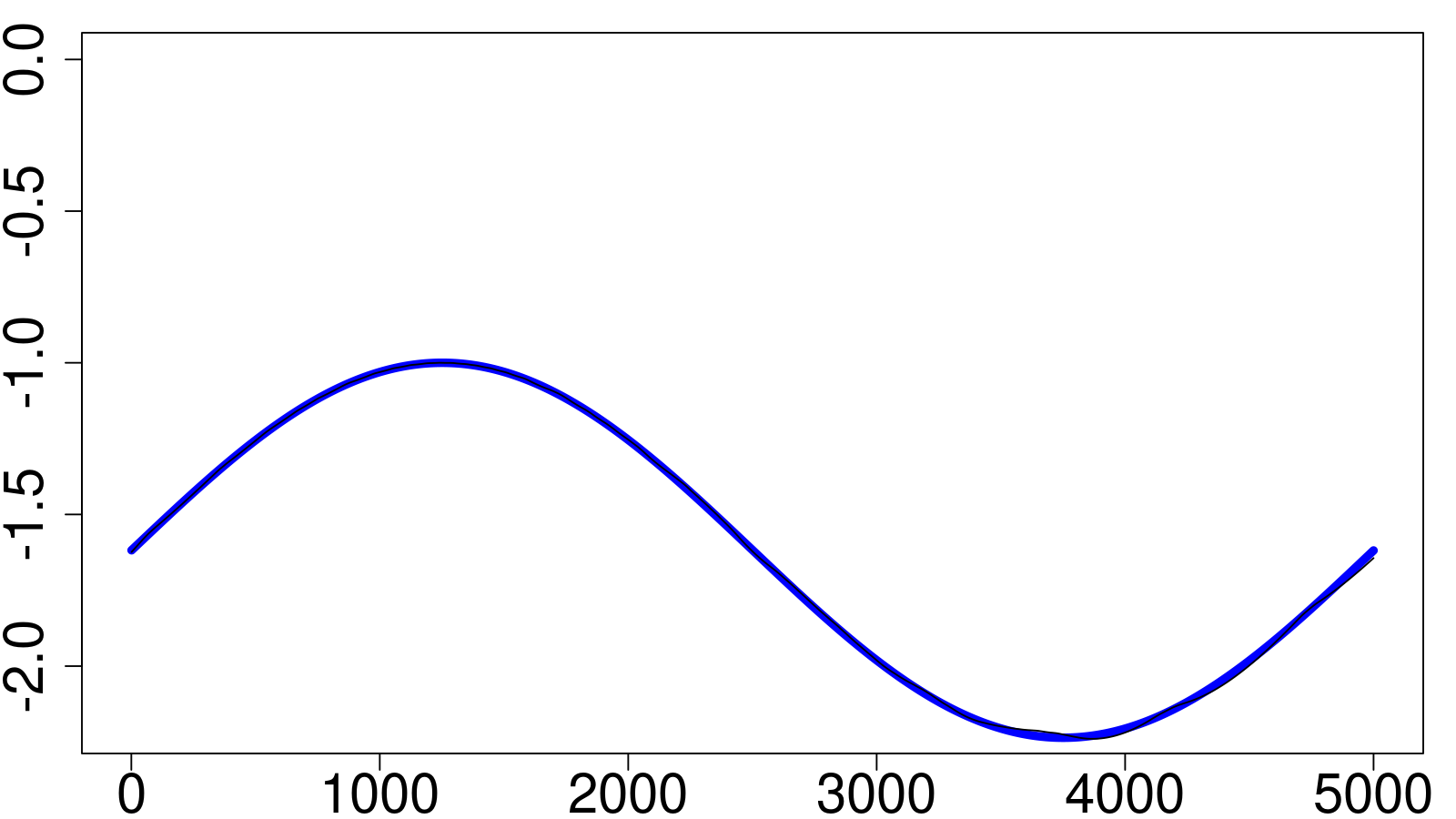}}
\end{center}

\caption[ac]{
Time series of true (in blue) and predicted (in black)  $1\%$  quantiles for all Direct- and Indirect-Dynamics approaches considered when $\tau=1\%$.}\vspace{-0.25cm}\label{fig:sim-fit-001}
\begin{center}
\subfloat[HS250]{\includegraphics[width=0.328\textwidth, height=0.14\textwidth]{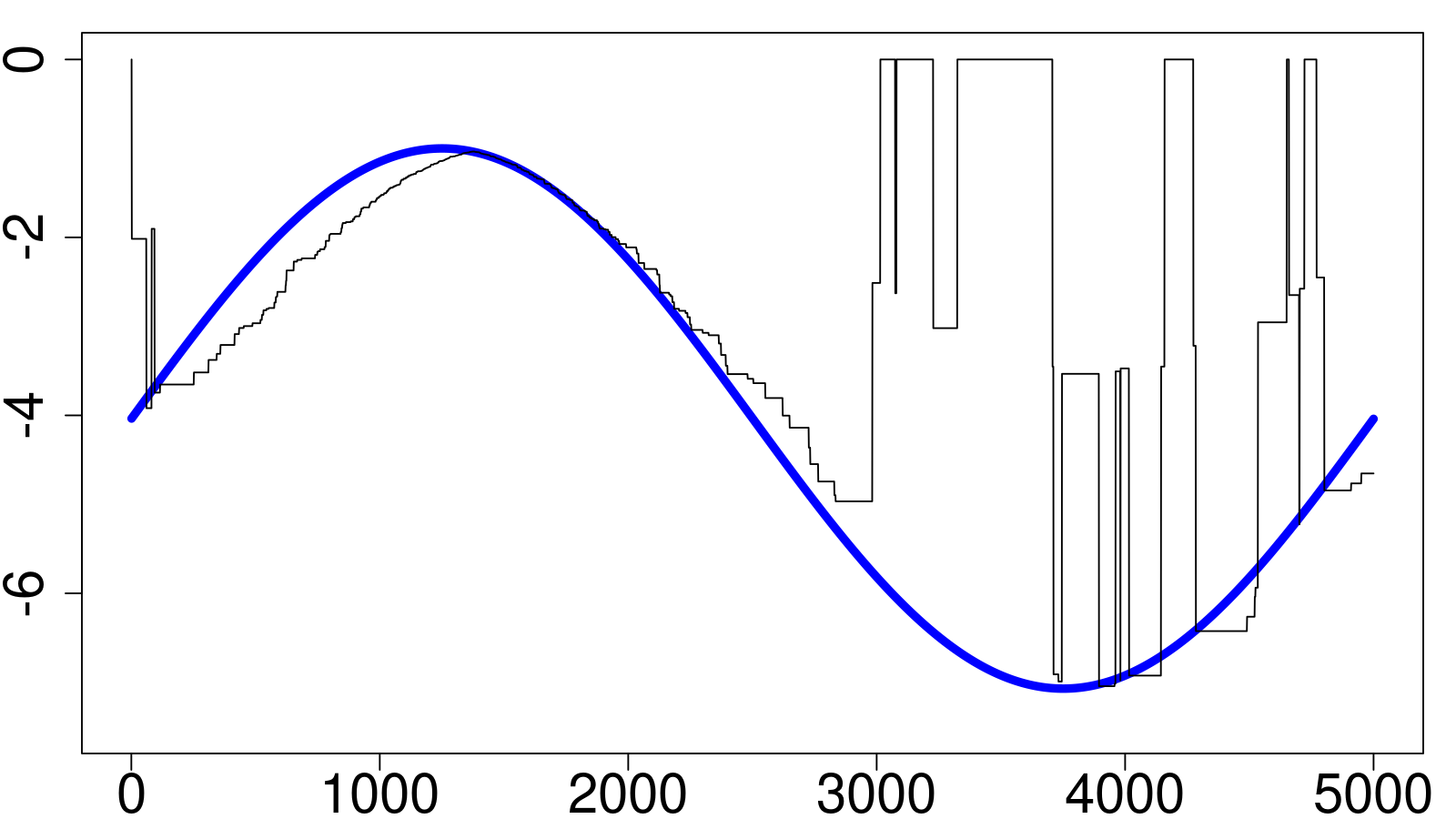}}
\subfloat[HS1000]{\includegraphics[width=0.328\textwidth, height=0.14\textwidth]{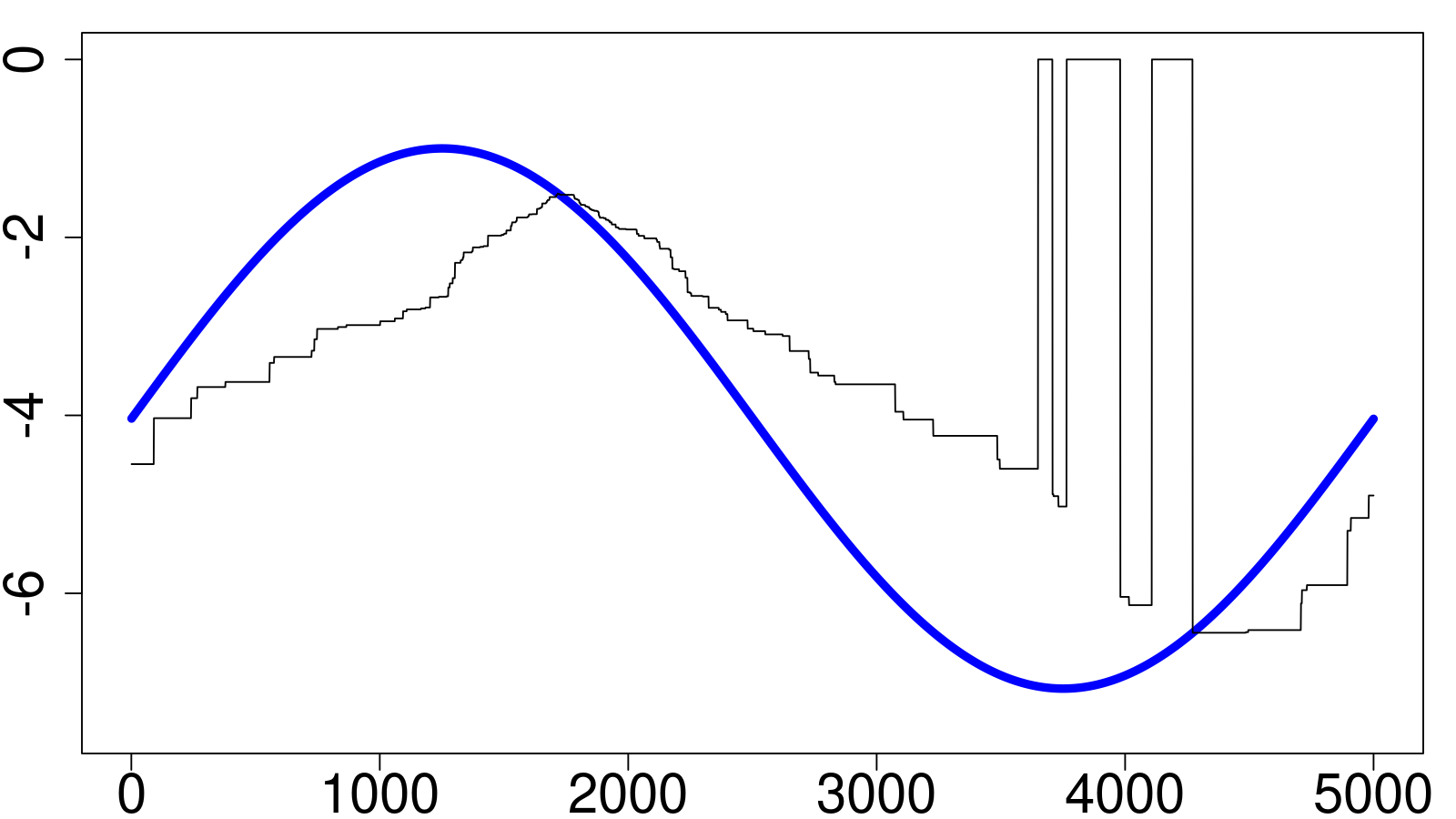}}
\subfloat[WHS95]{\includegraphics[width=0.328\textwidth, height=0.14\textwidth]{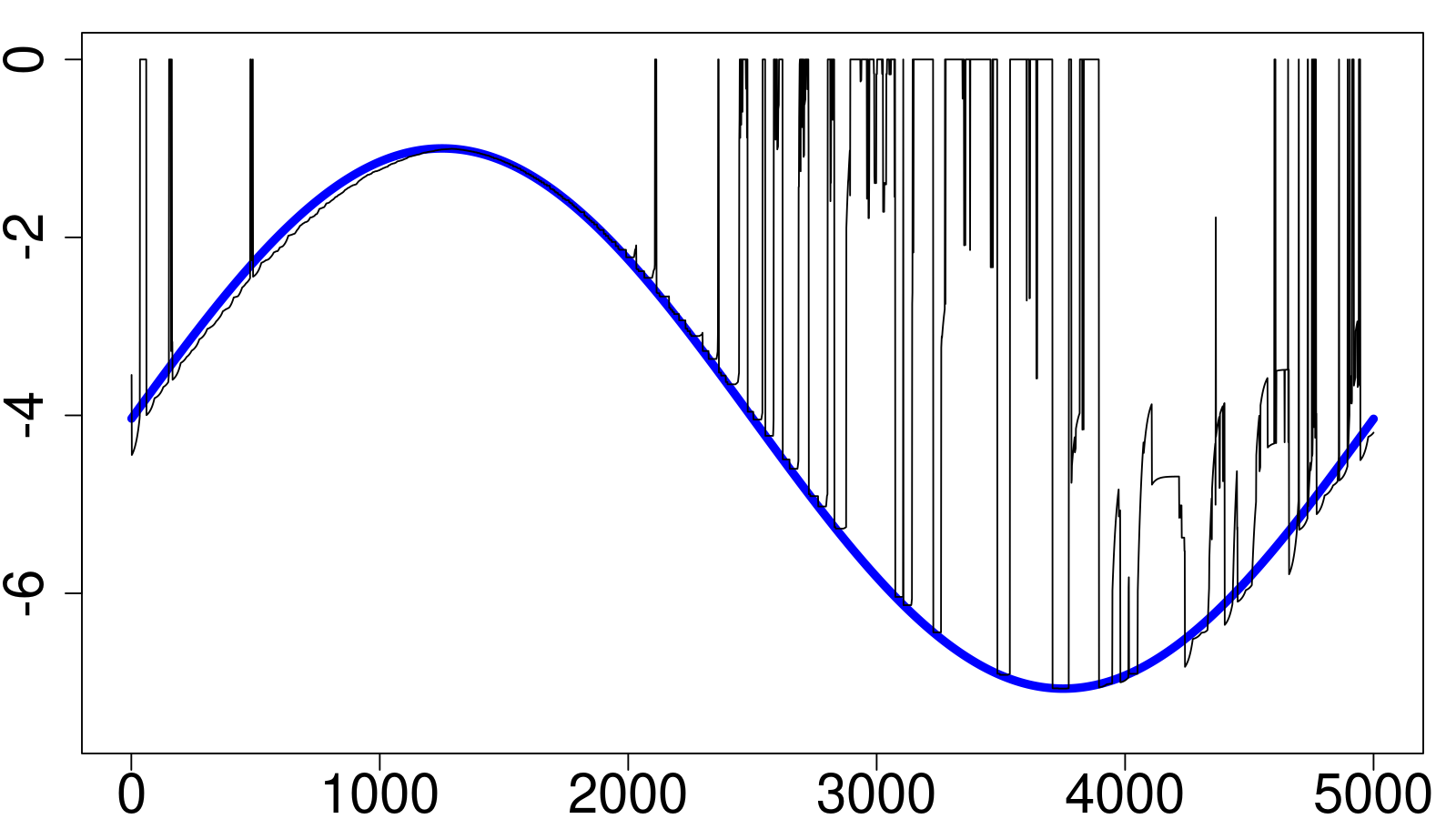}}\\
\subfloat[WHS99]{\includegraphics[width=0.328\textwidth, height=0.14\textwidth]{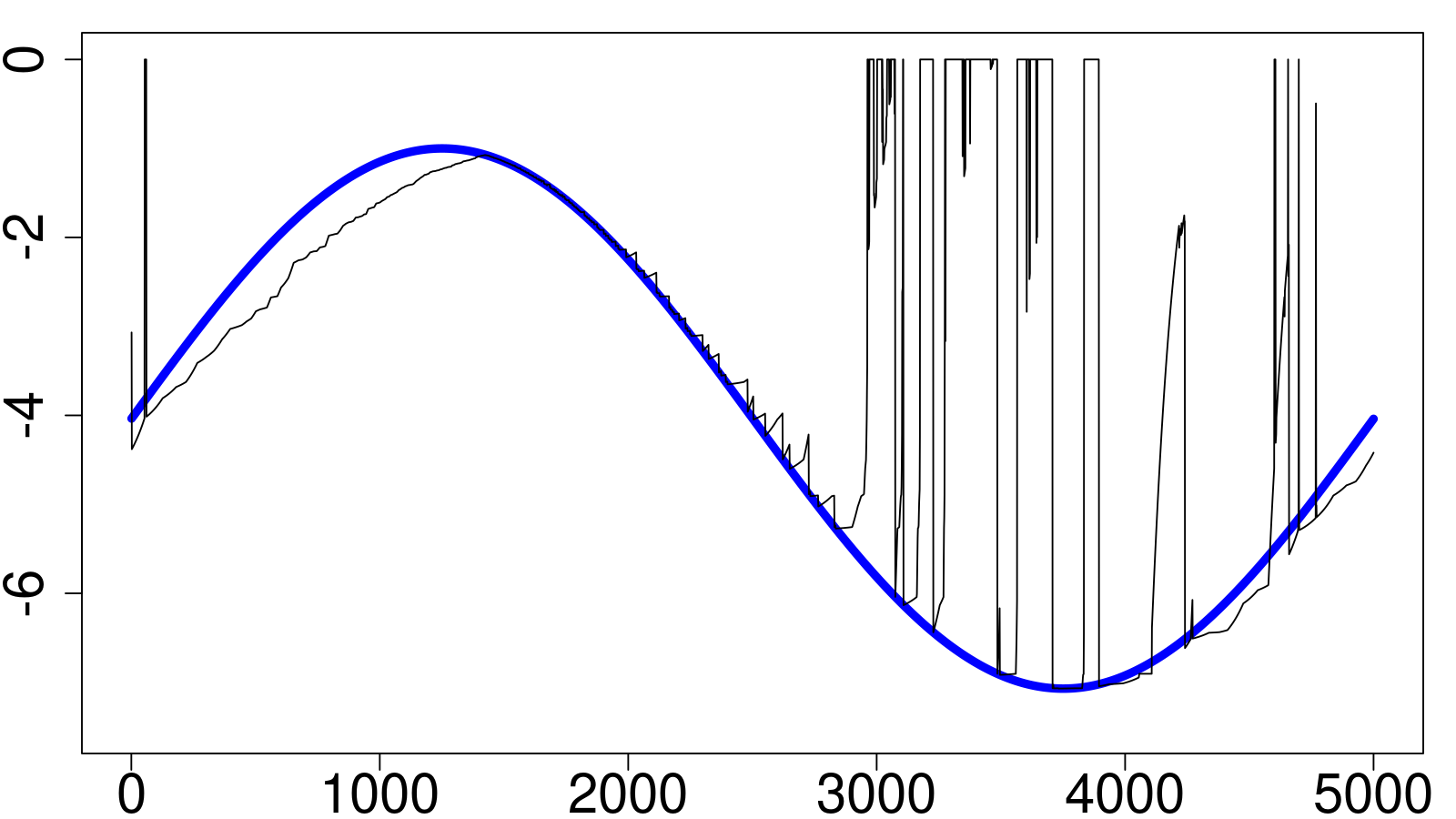}}
\subfloat[GARCQ]{\includegraphics[width=0.328\textwidth, height=0.14\textwidth]{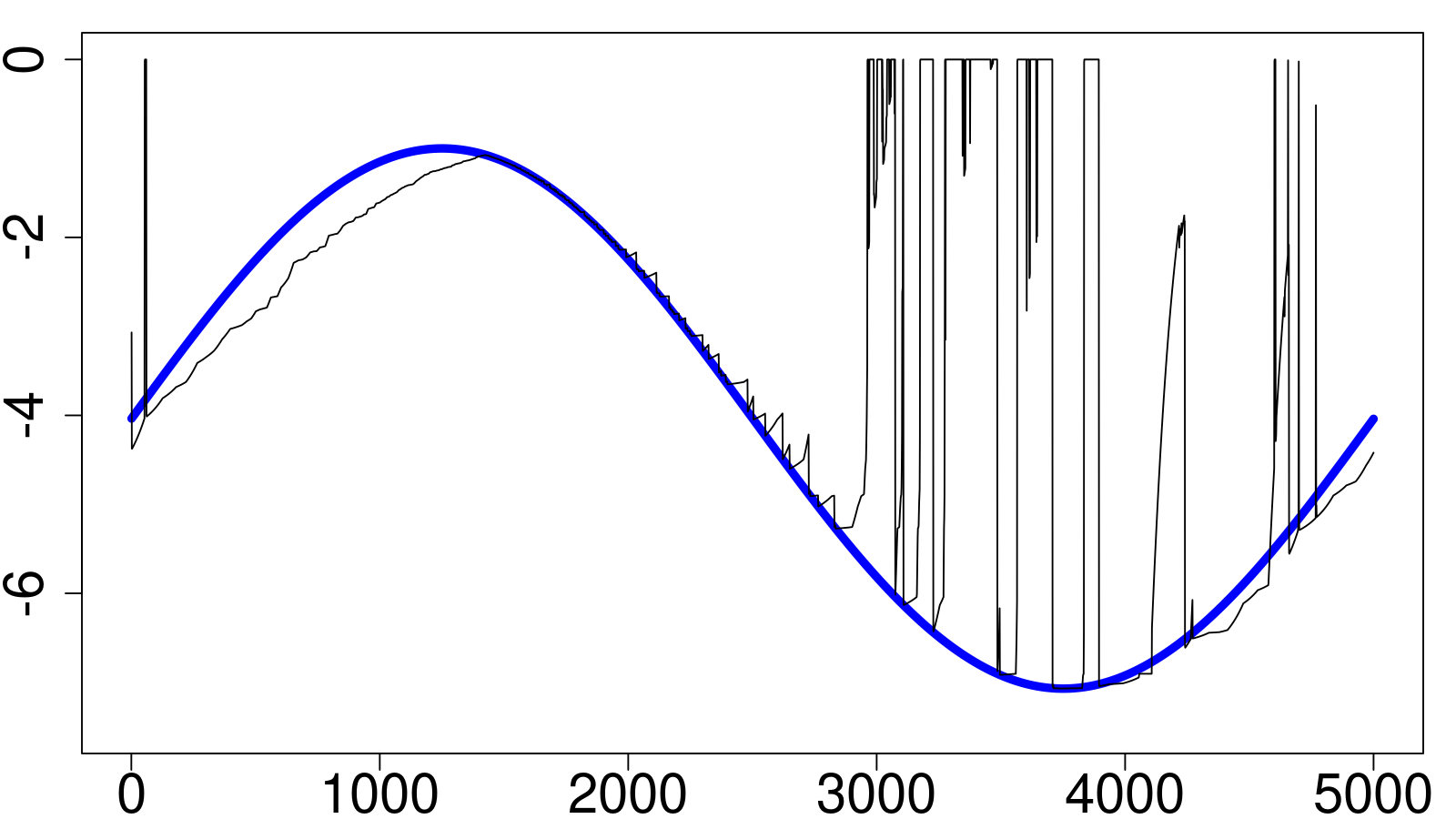}}
\subfloat[CAViaR]{\includegraphics[width=0.328\textwidth, height=0.14\textwidth]{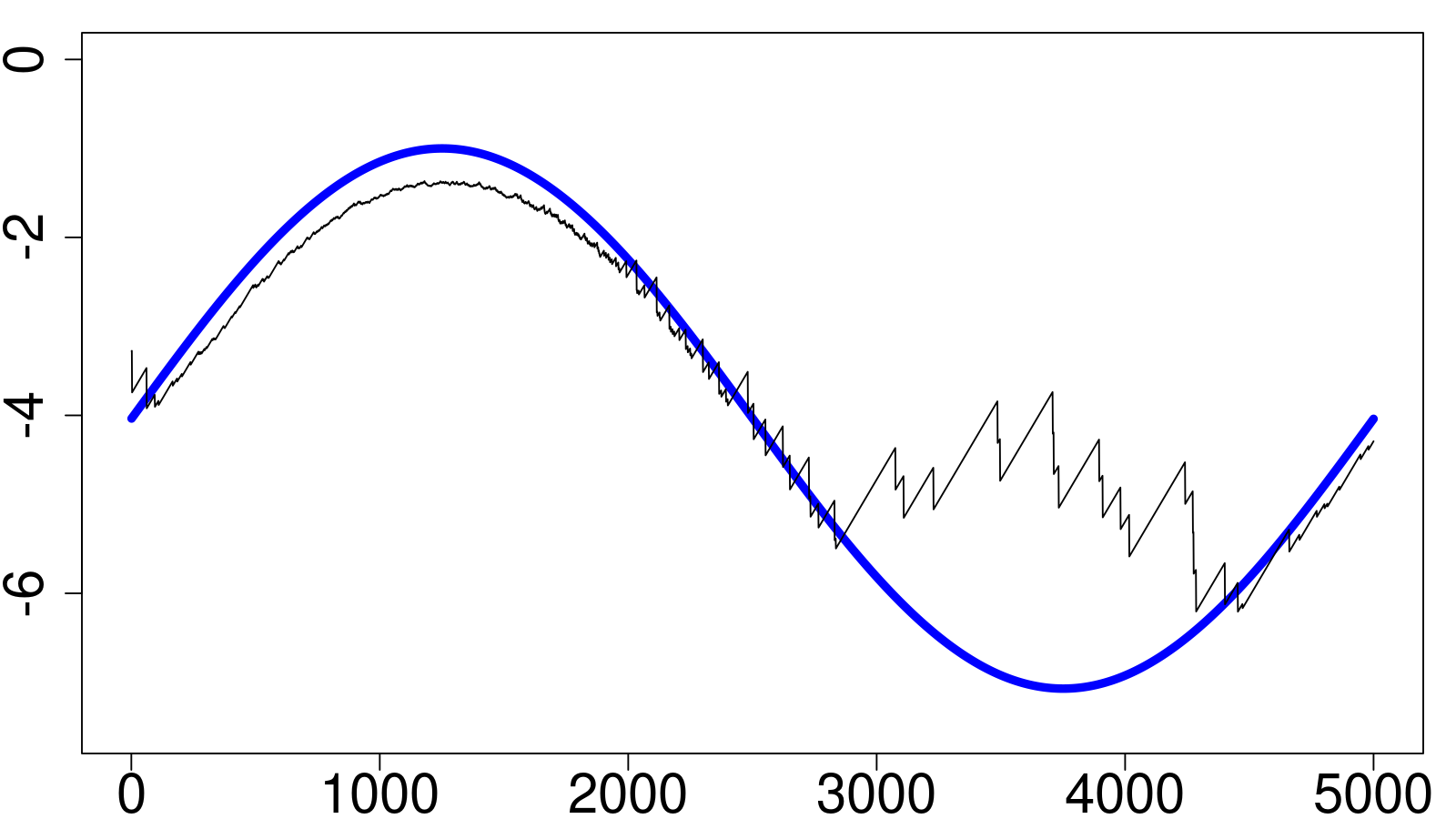}}\\
\subfloat[QPI]{\includegraphics[width=0.328\textwidth, height=0.14\textwidth]{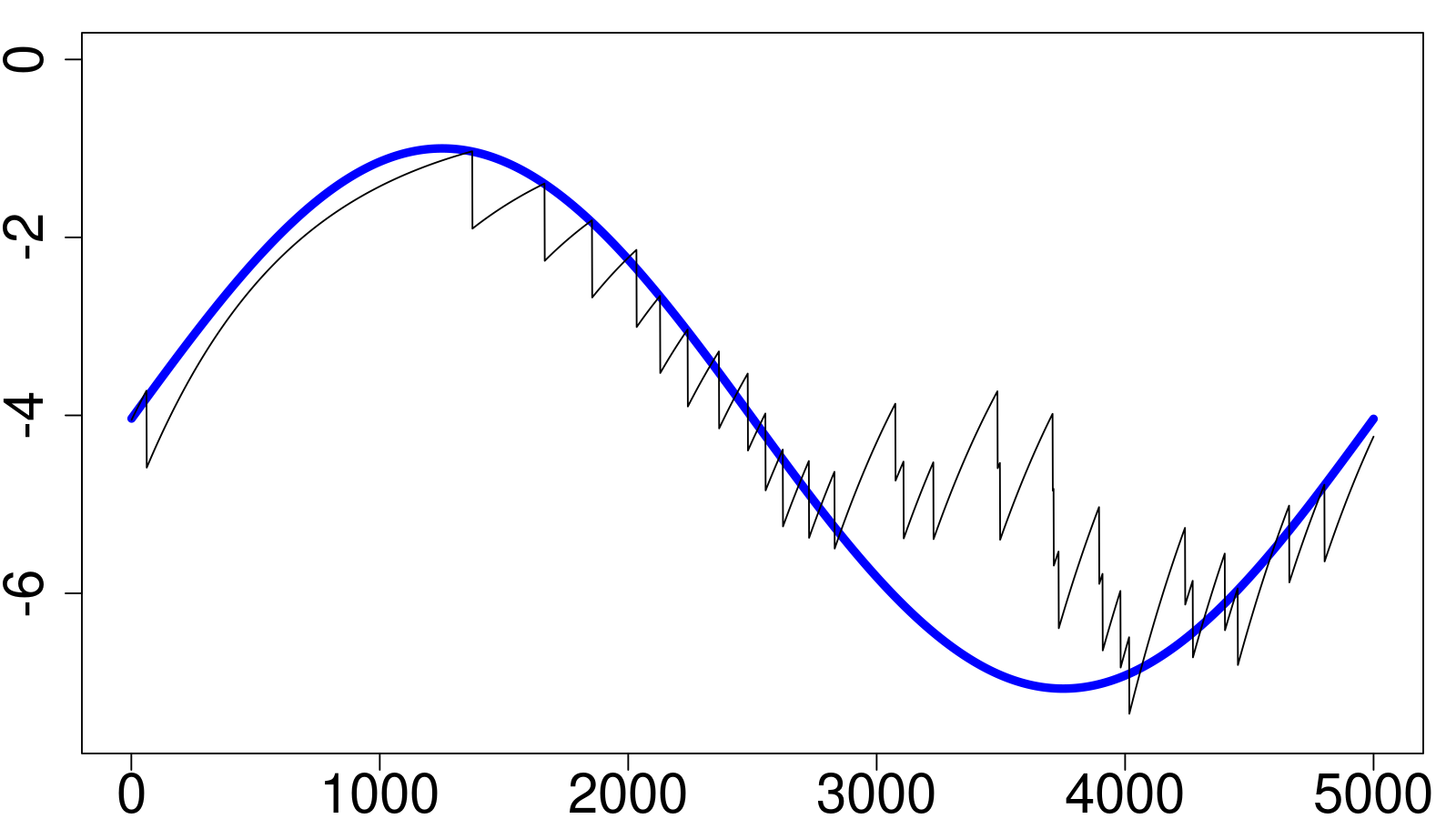}}
\subfloat[TT]{\includegraphics[width=0.328\textwidth, height=0.14\textwidth]{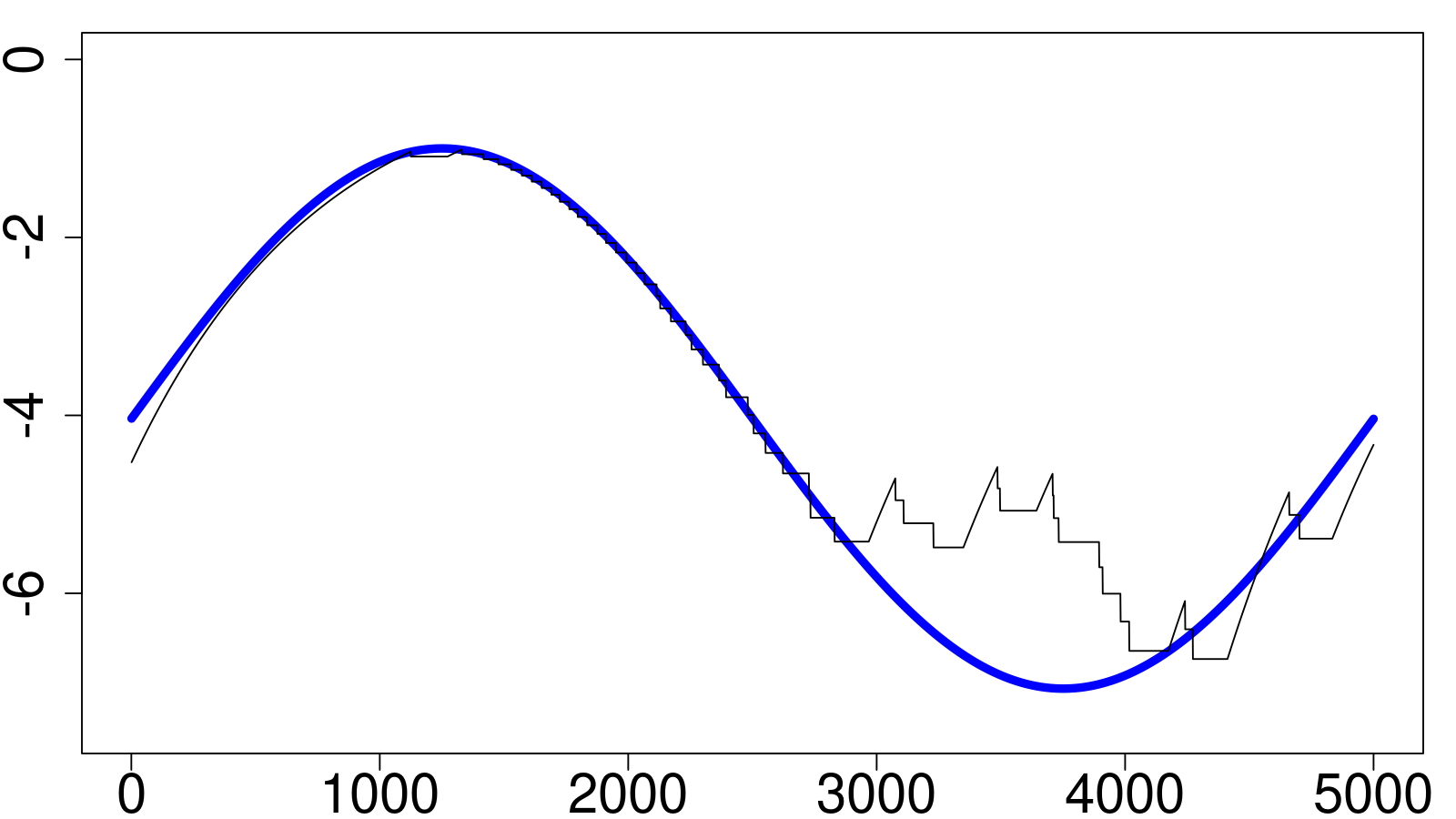}}
\subfloat[MT]{\includegraphics[width=0.328\textwidth, height=0.14\textwidth]{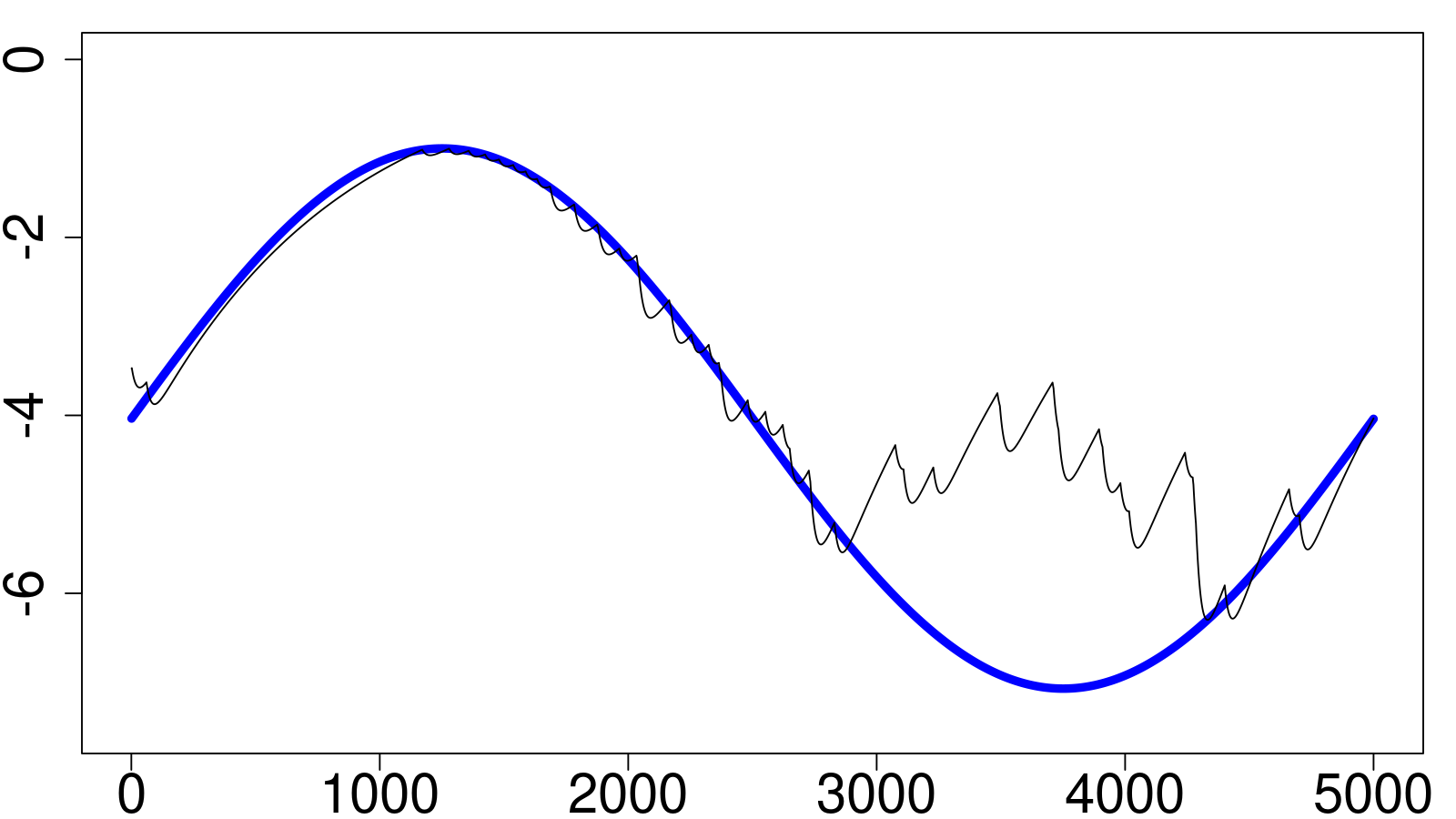}}
\end{center}
\end{figure}

\section{MAD$_{\tau}$ and MSE Relation}\label{App_MADMSE}
Given that the evaluation of the performance of the predictions is made for values in the tail of the conditional distribution $f_{t-1}(\cdot)$, it is worth noting that for small values of $\tau$, the value of $f_{t-1}(\cdot)$ evaluated at a quantile is fairly flat, i.e. it does not vary much with changing values of the quantile.  This has a bearing in establishing a relationship between the rankings of the results obtained by the MAD$_\tau$ and those obtained by the MSE.

From Markov's Law of Large Numbers for independent but not identically distributed random variables it follows that:
\begin{eqnarray*}
	\lim_{T\to\infty}\text{MAD}_{\tau}(z_{t},\hat{c}_{t}) & = & \lim_{T\to\infty}\frac{1}{T}\sum_{t=1}^{T} \mathbb{E}_{t-1}\left[(z_{t}-\hat{c}_{t})(\tau-\mathbbm{1}_{[z_{t}<\hat{c}_{t}]})\right]\\
	& = & \lim_{T\to\infty}\frac{1}{T}\sum_{t=1}^{T} \left\{\hat{c}_{t}\left[F_{t-1}(\hat{c}_{t})-\tau\right]-\int_{-\infty}^{\hat{c}_{t}}zf_{t-1}(z)dz\right\}.
\end{eqnarray*}
Adopting a second-order expansion of the terms in curly brackets around the true quantile $c_t$ (such that $F_{t-1}(c_{t})=\tau$) we have:
\begin{eqnarray*}
	& & c_{t}\left[F_{t-1}(c_{t})-\tau\right]-\int_{-\infty}^{c_{t}}zf_{t-1}(z)dz \\ && + \left\{F_{t-1}(c_{t})-\tau+c_{t}f_{t-1}(c_{t})-c_{t}f_{t-1}(c_{t})\right\}(\hat{c}_{t}-c_{t})
 + \frac{1}{2}f_{t-1}(\hat{c}_{t}^{*})\cdot(\hat{c}_{t}-c_{t})^{2}\\
	& = & -\int_{-\infty}^{c_{t}}zf_{t-1}(z)dz + \frac{1}{2}f_{t-1}(\hat{c}_{t}^{*})\cdot(\hat{c}_{t}-c_{t})^{2}
\end{eqnarray*}
where $\hat{c}_{t}^{*}$ is between $c_{t}$ and $\hat{c}_{t}$. Hence, the expression for the asymptotic limit of the MAD$_{\tau}$ becomes:
\begin{equation}
	\lim_{T\to\infty}\text{MAD}_{\tau}(z_{t},\hat{c}_{t}) = \lim_{T\to\infty}\text{MAD}_{\tau}(z_{t},c_{t}) + \frac{1}{2}\lim_{T\to\infty}\frac{1}{T}\sum_{t=1}^{T}f_{t-1}(\hat{c}_{t}^{*})\cdot(\hat{c}_{t}-c_{t})^{2}
	\label{eq:limMAD}
\end{equation}
The argument of the second limit on the right-hand-side is an arithmetic average which may be rewritten as:
\begin{equation*}
	\left[\frac{1}{T}\sum_{t=1}^{T}f_{t-1}(\hat{c}_{t}^{*})\right]\cdot\left[\frac{1}{T}\sum_{t=1}^{T}(\hat{c}_{t}-c_{t})^{2}\right] 
\end{equation*}
plus a term representing the sample covariance between $f_{t-1}(\hat{c}_{t}^{*})$ and $(\hat{c}_{t}-c_{t})^{2}$. Given that, as noted, for the usual values of $\tau$, $f_{t-1}(\hat{c}_{t}^{*})$ have small variability for changing $\hat{c}_{t}^{*}$, assuming them to be a constant  $\kappa$ allows us to set the covariance term equal to zero. Noting that $T^{-1}\sum_{t=1}^{T}(\hat{c}_{t}-c_{t})^{2}$ is equal to the MSE$(\hat{c}_{t},c_{t})$, the limiting MAD$_{\tau}$ in Equation (\ref{eq:limMAD}) becomes:
\begin{equation*}
	\lim_{T\to\infty}\text{MAD}_{\tau}(z_{t},\hat{c}_{t}) = \lim_{T\to\infty}\text{MAD}_{\tau}(z_{t},c_{t}) + \frac{\kappa}{2}\lim_{T\to\infty}\text{MSE}(\hat{c}_{t},c_{t}).
\end{equation*}
It follows that, for competing predictions/forecasts $\hat{c}_{t}$ and $\tilde{c}_{t}$, if the $\text{MAD}_\tau(z_t,\hat{c}_{t}) < \text{MAD}_\tau(z_t,\tilde{c}_{t})$, it must be that $\text{MSE}(\hat{c}_{t},c_{t}) < \text{MSE}(\tilde{c}_{t},c_{t})$. Therefore, in general, for $f_{t-1}(\hat{c}_{t}^{*})$ with small variability, the predictions/forecasts rankings induced by the MAD$_{\tau}$ are positively correlated with the rankings of the MSE.

\section{Heavy-Tailed Distributions \label{appdx_A}}
Consider a symmetric density function with Generalized Pareto distribution tails only: $\mathbb{P}(-u<z<u)=0$ but $\mathbb{P}(z<-u)=\beta (-u)^{-\alpha}$ and $\mathbb{P}(z>u)=\beta u^{-\alpha}$. Focusing on the right tail and imposing $\mathbb{P}(-\infty<z<+\infty)=1$ yields that $\mathbb{P}(z>y)=\frac{1}{2} u^{\alpha}y^{-\alpha}$, $\forall y>u$. While $\mathbb{E}(z)=0$ by construction from the distribution's symmetry, $\mathbb{E}(z^{2})=\frac{\alpha}{\alpha-2}u^{2}$, for $\alpha>2$. Therefore, the tails of a standardized random variable $z$ must start at $u=\alpha^{-1}(\alpha-2)$. Letting $q$ be the right $\tau$-quantile of the distribution, such that $\mathbb{P}(z>q)=\tau$, it follows that $q=(2\tau)^{-1/\alpha}\alpha^{-1/2}(\alpha-2)^{1/2}$. The value of $\alpha$ that maximizes the modulus of the quantile $q$ is $\alpha^{*}=2\ln(2\tau)\left[1+\ln(2\tau)\right]^{-1}$ and the corresponding value of the quantile is:
\begin{equation*}
q^{*} = (2\tau)^{-\frac{1+\ln(2\tau)}{2\ln(2\tau)}}\left[-\ln(2\tau)\right]^{-1/2}
\end{equation*}
Therefore, at the usual levels $\tau=\{10\%, 5\%, 1\%\}$, the largest quantiles attainable with the tails-only of a Generalized Pareto distribution are $q_{10\%}^{*}=1.0691$, $q_{5\%}^{*}=1.2640$ and $q_{1\%}^{*}=2.1684$, all smaller than the corresponding quantiles of a standardized Gaussian distribution: $q_{10\%}=1.2816$, $q_{5\%}=1.6449$ and $q_{1\%}=2.3264$.

\end{document}